\documentclass[aps,pra,superscriptaddress,12pt,longbibliography]{revtex4-2}

\renewcommand{\thesection}{\arabic{section}}
\renewcommand{\thesubsection}{\thesection.\arabic{subsection}}

\makeatletter
\renewcommand{\p@subsection}{}
\renewcommand{\p@subsubsection}{}
\makeatother

\usepackage{amsmath,amsfonts,amssymb,amsthm,mathtools,dsfont}
\usepackage{setspace}
\usepackage[normalem]{ulem}
\usepackage{graphicx,xcolor}
\usepackage{hyperref}
\hypersetup{pdfstartview=}
\usepackage{url}
\usepackage{bm}
\usepackage{bbold}
\usepackage{suffix} 

\numberwithin{equation}{section}
\newtheorem{thm}{Theorem}
\newtheorem{prop}[equation]{Proposition}
\newtheorem{lem}[equation]{Lemma}
\newtheorem{cor}[equation]{Corollary}

\theoremstyle{definition}
\newtheorem{defn}[equation]{Definition}

\providecommand{\vpd}[0]{\vphantom{\dagger}}
\providecommand{\vps}[0]{\vphantom{*}}
\providecommand{\vpp}[0]{\vphantom{\prime}}

\newcommand{\id}{\mathds{1}}
\newcommand{\ket}[1]{\left| #1 \right\rangle}
\newcommand{\bra}[1]{\left\langle #1 \right|}
\DeclareMathOperator*{\trace}{tr}

\providecommand{\CFrame}{C_{\mathrm F}}
\providecommand{\ZProj}[1]{\left| #1 \middle\rangle \hspace{-0.4mm} \middle\langle #1 \right|}
\providecommand{\KetBra}[2]{\left| #1 \middle\rangle \hspace{-0.4mm} \middle\langle #2 \right|}\WithSuffix\providecommand\KetBra*[2]{| #2 \rangle \hspace{-0.4mm}\langle #2 |}
\providecommand{\MatEl}[3]{\left\langle #1 \middle| #2 \middle| #3 \right\rangle}\WithSuffix\providecommand\MatEl*[3]{\langle #1 | #2 | #3 \rangle}

\usepackage{tikz}
\usetikzlibrary{quantikz}
\begin{document}

\title{Simplifying models of noisy stabilizer QEC on leakage qubits by Pauli twirling}
\begin{abstract}
The growing capabilities of quantum error-correction (QEC) experiments make it increasingly important to develop methods that predict or assess logical performance over many QEC cycles. 
Prior work showed that a stabilizer QEC cycle subject to Pauli-stochastic noise can be modeled by a classical Markov process over syndrome states,
with a logical Pauli channel attached to each syndrome transition~\cite{kwiatkowski2025approximate}. 
Here, we extend this framework to systems in which physical qubits are subject to leakage. 
We first develop the theory of leakage-dephased Pauli twirling and show that twirled processes can be expressed in terms of Pauli maps on the computational subspace, conditioned on transitions between leakage patterns. 
We apply this theory to noisy stabilizer QEC with leakage, demonstrating that a QEC cycle with leakage can be modeled by a classical Markov process over the joint syndrome and leakage patterns, with a logical Pauli channel attached to each joint transition. 
Finally, we derive a transfer-matrix description of repeated QEC cycles and an exact transfer-matrix expression for the success probability of a QEC memory experiment. 
This framework connects leaky error models of the individual components of the QEC cycle to many-cycle logical performance and provides a foundation for logical characterization and benchmarking methods that can account for leakage. 
\end{abstract}

\author{Alex Kwiatkowski}
\affiliation{National Institute of Standards and Technology, Boulder, Colorado 80305, USA}

\author{Aaron J. Friedman}
\affiliation{National Institute of Standards and Technology, Boulder, Colorado 80305, USA}
\affiliation{Department of Electrical Engineering, University of Colorado, Denver, Colorado 80204, USA}

\date{\today}
\maketitle

\section{Introduction}

As quantum error-correction (QEC) experiments continue to improve, it becomes increasingly important to develop methods to predict and assess logical
performance over many QEC cycles~\cite{kwiatkowski2025approximate, ziyad2025emergent, miao2023overcoming, google2025belowthreshold, lacroix2025scaling,caune2026realtime, atomcomputing2026toric, postler2024steane, paetznick2026improved, xin2026improved, berthusen2026experimental}.
One approach is to predict logical performance via measurements or models of the individual, low-level physical components involved in the QEC cycle~\cite{google2021exponential, iyer2022diagnostics, google2023suppressing, harper2023correlated, fazio2026characterizing}.
Alternatively, logical performance can be characterized directly at the logical level, without consideration of the low-level components~\cite{combes2017logical, xu2018integrity, pal2022relaxation,gidney2022stability, wagner2023learning}. 
However, both approaches are complicated by the fact that the effective logical dynamics under repeated QEC cycles can be non-Markovian (i.e., retain memory) even when the low-level error model is Markovian~\cite{ziyad2025emergent}.
Importantly, Ref.~\cite{kwiatkowski2025approximate} has established that a stabilizer QEC cycle subject to Pauli-stochastic noise can be modeled by a classical Markov process over syndrome states, with a logical Pauli channel attached to each syndrome transition.
As discussed in Ref.~\cite{kwiatkowski2025approximate}, studying the family of logical Pauli channels of a QEC cycle provides a direct means of characterizing the logical performance of repeated applications of that QEC cycle. 
However, these results do not hold in systems in which qubits experience leakage.

Leakage substantially complicates both the prediction and the assessment of logical performance in stabilizer QEC.
A Pauli error acts within the computational qubit space, whereas leakage transfers population into noncomputational levels and can leave information there from one QEC cycle to the next.  
As a result, modeling a general leakage process could require consideration of the full quantum dynamics, rather than the much simpler stochastic description available for Pauli errors.  
Furthermore, in general the full quantum state, including the leaked degrees of freedom, can influence subsequent gates and thereby affect the errors that occur in later cycles.  
Because these effects can substantially degrade logical performance, accurately accounting for them is essential.  
Consequently, modeling repeated QEC in the presence of leakage errors can be considerably more complicated than modeling QEC with Pauli errors.
The key simplification in this work comes from Pauli twirling.  We develop a
version of Pauli twirling for leakage-extended qubits such that twirled error processes can be modeled by Pauli error channels on the computational subspace, conditioned on transitions between leakage patterns.  This is formalized in Theorem~\ref{prop:leakage-transition-Pauli-map-decomposition}. In this work we apply it to error processes of QEC cycles, but we hope it will be of independent interest in other contexts as well.

Our second main result is obtained by applying Theorem~\ref{prop:leakage-transition-Pauli-map-decomposition} to the mathematical model developed in Section~\ref{sec:qec-cycle-model} for a noisy stabilizer-code QEC cycle in the presence of leakage.  
We show that the complete-cycle channel decomposes into subnormalized logical Pauli maps indexed by joint transitions of the syndrome and the
data-leakage pattern.  
This is formalized in Theorem~\ref{thm:single-cycle}.
One interpretation of this result is that the syndrome and the data-leakage pattern jointly function as a hidden-memory state carried from one QEC cycle to the next.  
Thus, a single QEC cycle can be viewed as a classical Markov transition rule on the hidden-memory state space, together with a logical Pauli channel conditioned on each transition. 
In the setting of Ref.~\cite{kwiatkowski2025approximate}, where the error model consists of Pauli noise and no leakage, the analogous
hidden-memory state is the syndrome alone.  
Consequently, Theorem~\ref{thm:single-cycle} can be interpreted as a generalization of the formalism of Ref.~\cite{kwiatkowski2025approximate} to account for leakage.

Our third main result comes from applying Theorem~\ref{thm:single-cycle} to QEC memory experiments.  
We define a QEC memory experiment to consist of state preparation, \(k\) repeated QEC cycles, and a final measurement for which one outcome is designated ``success.''  
For each logical Pauli component, we express its contribution to the success probability in terms of a transfer matrix, a state-preparation vector, and a measurement vector.  
Each transfer matrix is determined by the dynamics of a single QEC cycle and the dependence on \(k\) enters only through the \(k\)th power of that matrix.  
The state-preparation vector depends only on the input state, while the measurement vector depends only on the measurement process and its distinguished success outcome; neither depends on the details of the QEC cycle.
This is formalized in Theorem~\ref{thm:memory-logical-Pauli-transfer-formula}.  
Altogether, this formulation of memory experiment success probabilities provides a concrete connection between a single-cycle error model and predictions of many-cycle QEC performance.  
Furthermore, if the action of the transfer matrices can be well approximated by their dominant left and right eigenspaces, this representation could also lead to approximate Markovian models of the kind developed in Ref.~\cite{kwiatkowski2025approximate}.
Although we do not address this direction in detail here, it is an important matter for future work.

Leakage has long been recognized as an important challenge for fault-tolerant QEC.  
There has been substantial effort to develop methods for analyzing, suppressing, or removing leakage~\cite{aliferis2007faulttolerant,fowler2013coping,
ghosh2013understanding, ghosh2015leakageresilient, suchara2015leakage,battistel2021hardware,mcewen2021removing}; detecting leakage and incorporating leakage information into error correction~\cite{bultink2020protecting,varbanov2020leakage,ziad2025local}; and obtaining tractable representations or simulations of multilevel leakage noise~\cite{marshall2025incoherent,manabe2025efficient,
vendrell2026plaquette}.
Our contribution is complementary to these efforts: for any leakage-handling scheme described by our QEC-cycle model, our framework provides a general way to understand how its dynamics contribute to logical errors.

Pauli twirling has long been used to map general qubit noise to Pauli noise, both for simpler noise characterization and as a tractable approximation in simulations of fault-tolerant circuits and QEC~\cite{dur2005standard,geller2013efficient,cai2019constructing}.
For a noisy implementation of a Clifford operation, a random input Pauli can be propagated through the target Clifford to determine a compensating output Pauli, thereby preserving the target operation while Pauli twirling its residual noise. 
This construction---which we refer to as Pauli twirling with respect to a Clifford---underlies Pauli-frame randomization, randomized compiling, cycle benchmarking, and other noise-characterization protocols~\cite{wallman2016noise,erhard2019characterizing,flammia2022averaged}.
We note that prior work has generalized some of these ideas to scenarios involving leakage. 
Randomized-benchmarking protocols have been developed that account for leakage, with some constructions using random relative phases between the computational and leakage subspaces to suppress coherence between them~\cite{epstein2014investigating,chasseur2015complete,wallman2016robust}.
The subspace-twirling approximation has been used to obtain a tractable incoherent model of leakage in QEC~\cite{marshall2025incoherent}, while, under gate-independent, time-independent Markovian-noise assumptions, independent Pauli twirling has been shown to depolarize each leakage-pattern sector, leading to a classical Markov process on the leakage patterns \cite{wu2024leakage}.

This work is organized as follows.  
Section~\ref{sec:summary-results-discussion} gives a high-level summary of the results and discusses their implications, limitations, and directions for future work.
The appendices provide a detailed, largely self-contained mathematical presentation of the results. 
Appendix~\ref{sec:twirling} introduces standard Pauli twirling and the structure of Pauli maps in the absence of leakage.
Appendix~\ref{sec:Clifford-twisted-Pauli-twirl} develops the theory of Pauli twirling with respect to a Clifford.
Appendix~\ref{sec:total-Hilbert-space-and-leakage-subspaces} introduces leakage-extended qubits, leakage-extended qubit registers, and other mathematical concepts related to leakage, and then proves Theorem~\ref{prop:leakage-transition-Pauli-map-decomposition}.
Appendix~\ref{sec:setup} provides the mathematical formulation of a QEC cycle on leakage-extended qubit registers.
Appendix~\ref{sec:leakage-aware-cycle} formulates the model of a QEC cycle in terms of logical Pauli maps conditioned on transitions between leakage patterns and syndrome states, and proves Theorem~\ref{thm:single-cycle}.
Appendix~\ref{sec:single-cycle} develops the joint syndrome-leakage hidden-memory representation and the transfer-matrix description of repeated
QEC experiments, and proves Theorem~\ref{thm:memory-logical-Pauli-transfer-formula}.
Finally, Appendix~\ref{app:elementary-operator-space-lemmas} contains lemmas related to operator spaces.

\section{Summary of results and discussion}
\label{sec:summary-results-discussion}

\subsection{Leakage-extended qubits and leakage-dephased Pauli twirling}
\label{sec:twirling-main}

Our first main result is a structure theorem for leakage-dephased Pauli twirling. 
This is formalized in Theorem~\ref{prop:leakage-transition-Pauli-map-decomposition}, and is described in detail in the rest of this subsection.

Throughout this paper, a leakage-extended qubit space is a Hilbert space \(\mathfrak{h}=\mathfrak{c}\oplus\mathfrak{l}\), where \(\mathfrak{c}\) is a two-dimensional computational qubit subspace and \(\mathfrak{l}\) is a one-dimensional leakage subspace (see Definition~\ref{defn:leakage-extended-qubit}).  
Theorem~\ref{prop:leakage-transition-Pauli-map-decomposition} is formulated in terms of a leakage-extended qubit register \(\mathsf{Q}\), which consists of a set \(\mathsf{Q}\) of system (qubit) labels and a leakage-extended qubit space \(\mathfrak{h}_i\) for each label \(i \in \mathsf{Q}\) (see Definition~\ref{defn:register-of-leakage-extended-qubits}). 
The register's total Hilbert space is \(\mathcal{H}_{\mathsf{Q}} =\bigotimes_{i\in\mathsf{Q}}\mathfrak{h}_i\).
A leakage pattern \(S\subseteq\mathsf{Q}\) specifies which labels correspond to leaked qubits (see Definition~\ref{defn:leakage-pattern}) and determines a leakage subspace \(\mathcal{H}_{\mathsf{Q}}^{S}\) (see Definition~\ref{defn:leakage-subspace}).  
These subspaces give the decomposition \(\mathcal{H}_{\mathsf{Q}}=\bigoplus_{S\subseteq\mathsf{Q}}\mathcal{H}_{\mathsf{Q}}^{S}\)~\eqref{eq:leakage-subspace-direct-sum}.

Denote by \(\mathsf{K}_{\mathsf{Q}}\) the computational qubit register associated with \(\mathsf{Q}\), obtained by considering only the computational
subspace \(\mathfrak{c}_i\) for each leakage-extended qubit in \( \mathsf{Q}\) (see Definition~\ref{defn:computational-qubit-register}).  
The Hilbert space of \(\mathsf{K}_{\mathsf{Q}}\) is denoted \(\mathcal{K}_{\mathsf{Q}}\), and it embeds naturally as a subspace of \(\mathcal{H}_{\mathsf{Q}}\). Pauli operators on
\(\mathcal{K}_{\mathsf{Q}}\) are indexed by the Pauli-label space \(V_{\mathsf{K}_{\mathsf{Q}}}\), with binary \(X\) and \(Z\) labels for each qubit (see Definition~\ref{defn:Pauli-label-space}). 

For \(v\in V_{\mathsf{K}_{\mathsf{Q}}}\), denote by \(W^{\mathsf{K}_{\mathsf{Q}}}(v)\) the corresponding Pauli operator (see Definition~\ref{defn:Pauli-operator-computational}), acting on \(\mathcal{K}_{\mathsf{Q}}\). 
Denote by \(W^{\mathsf{Q}}(v)\) the corrresponding leakage-extended Pauli operator, which acts as the Pauli labeled by \(v\) on the computational subspace of \(\mathcal{H}_{\mathsf{Q}}\) and as the identity on the leakage subspaces (see Definition~\ref{defn:leakage-extended-Pauli-family}). 
We denote by \(\mathcal{W}^{\mathsf{N}}(v) = \operatorname{Ad}_{W^{\mathsf{N}}(v)}\) the map associated with conjugation by the Pauli \(W^{\mathsf{N}}(v)\). 
For a leakage-extended register \(\mathsf{Q}\), the Pauli map \(\mathcal{W}^{\mathsf{K}_\mathsf{Q}}(v) : \operatorname{End}(\mathcal{K}_{\mathsf{Q}}) \to \operatorname{End}(\mathcal{K}_{\mathsf{Q}})\) is a standard Pauli map on the computational subspace, while \(\mathcal{W}^{\mathsf{Q}}(v) : \operatorname{End}(\mathcal{H}_{\mathsf{Q}}) \to \operatorname{End}(\mathcal{H}_{\mathsf{Q}})\) is a leakage-extended Pauli map.

Consider a Clifford unitary \(C\) on the computational subspace \(\mathcal{K}_{\mathsf{Q}}\). 
Denote by \(\mathcal{C}\) the associated superoperator, \(\mathcal{C}(Y) = C Y C^\dagger\) for any operator \(Y \in \operatorname{End}(\mathcal{K}_{\mathsf{Q}})\). 
Let \(\Phi_\mathcal{K}:\operatorname{End}(\mathcal{K}_\mathsf{Q}) \to \operatorname{End}(\mathcal{K}_\mathsf{Q}) \) be a superoperator, whose ideal (noise-free) realization is \(\mathcal{C}\). 
The map \(\Phi_\mathcal{K}\) can be twirled into a more convenient form via the Pauli twirl with respect to \(C\), denoted \(\mathcal{T}^{\mathsf{K}_{\mathsf{Q}}}_C\) (see Definition~\ref{defn:clifford-twisted-Pauli-twirl}).

Here we develop an analogous theory of Pauli twirling on leakage-extended qubit registers.
Denote by \(\Delta\) the leakage-pattern dephasing channel, which removes coherence between distinct leakage subspaces (see Definition~\ref{defn:leakage-pattern-dephasing}).  
Fix a superoperator \(\Phi:\operatorname{End}(\mathcal{H}_{\mathsf{Q}}) \to\operatorname{End}(\mathcal{H}_{\mathsf{Q}})\), a Clifford unitary \(C\) defined on \(\mathcal{K}_{\mathsf{Q}}\) that is intended to be the nominal action of \(\Phi\) in the computational subspace,  and an associated symplectic automorphism \(F_C:V_{\mathsf{K}_{\mathsf{Q}}}\to V_{\mathsf{K}_{\mathsf{Q}}}\) (see Definition~\ref{defn:clifford-unitary}). 
We define the leakage-dephased Pauli twirl (see Definition~\ref{defn:leakage-dephased-Pauli-twirl}) of \(\Phi\)  according to
\begin{equation}
    \mathcal{T}^{\mathsf{Q}}_C(\Phi)
    =
    \Delta
    \circ
    \left[
        \frac{1}{|V_{\mathsf{K}_{\mathsf{Q}}}|}
        \sum_{v\in V_{\mathsf{K}_{\mathsf{Q}}}}
        \mathcal{W}^{\mathsf{Q}}(F_C v)
        \circ
        \Phi
        \circ
        \mathcal{W}^{\mathsf{Q}}(v)
    \right]
    \circ
    \Delta.
    \label{eq:maintext-leakage-dephased-Pauli-twirl}
\end{equation}
Theorem~\ref{prop:leakage-transition-Pauli-map-decomposition} provides a convenient decomposition of Eq.~\eqref{eq:maintext-leakage-dephased-Pauli-twirl}. To state Theorem~\ref{prop:leakage-transition-Pauli-map-decomposition}, we introduce the following maps for each leakage pattern \(R\subseteq\mathsf{Q}\).
The leakage-subspace projection map \(\mathcal{P}_R\) restricts an operator on \(\mathcal{H}_{\mathsf{Q}}^{\vps}\) to \(\mathcal{H}_{\mathsf{Q}}^{R}\), and the leakage-subspace extension map \(\mathcal{E}_R\) embeds an operator on \(\mathcal{H}_{\mathsf{Q}}^{R}\) as an operator on \(\mathcal{H}_{\mathsf{Q}}^{\vps}\) (see Definitions~\ref{defn:leakage-subspace-projection-map} and \ref{defn:leakage-subspace-extension-map}). 
The filling map \(\mathcal{F}_R\) replaces the leaked systems with maximally mixed computational qubits (i.e., for \(i \in R\), system \(i\) is placed in the state \(\id/2 \in \operatorname{End}(\mathfrak{c}_i)\)), and the leakage map \(\mathcal{L}_R\) places the systems labeled \(i \in R\) in their leakage states (see Definitions~\ref{defn:filling-map} and~\ref{defn:leakage-map}). 
For \(S,T\subseteq\mathsf{Q}\), define the map \(\mathcal{D}_{T, S}^{\Phi}: \operatorname{End}(\mathcal{K}_{\mathsf{Q}}) \to \operatorname{End}(\mathcal{K}_{\mathsf{Q}})\) (see Definition~\ref{defn:leakage-transition-Pauli-map-family}) via
\begin{equation}
    \mathcal{D}_{T, S}^{\Phi}
    =
    \mathcal{T}^{\mathsf{K}_{\mathsf{Q}}}_C
    \!\left(
        \mathcal{F}_T
        \circ
        \mathcal{P}_T
        \circ
        \Phi
        \circ
        \mathcal{E}_S
        \circ
        \mathcal{L}_S
    \right)
    \circ
    \mathcal{C}^{-1} .
    \label{eq:maintext-leakage-transition-Pauli-map}
\end{equation}
Theorem~\ref{prop:leakage-transition-Pauli-map-decomposition} establishes that the leakage-dephased Pauli twirl can be written
\begin{equation}
    \mathcal{T}^{\mathsf{Q}}_C(\Phi)
    =
    \sum_{S,T\subseteq\mathsf{Q}}
    \mathcal{E}_T
    \circ
    \mathcal{L}_T
    \circ
    \mathcal{D}_{T, S}^{\Phi}
    \circ
    \mathcal{C}
    \circ
    \mathcal{F}_S
    \circ
    \mathcal{P}_S.
    \label{eq:maintext-leakage-transition-Pauli-map-decomposition}
\end{equation}
This expression admits the following interpretation. First, the sum decomposes the twirled map into branches indexed by the incoming and outgoing leakage patterns, \(S\) and \(T\), respectively.
For a given \(S\to T\) branch, the projection map \(\mathcal{P}_S\) selects the incoming leakage subspace \(S\), and then \(\mathcal{F}_S\) replaces all leaked systems labeled \(i \in S\) with maximally mixed computational qubits.
Then, the ideal Clifford \(C\) acts on the computational subspace, followed by the Pauli map \(\mathcal{D}_{T, S}^{\Phi}\). 
Finally, \(\mathcal{L}_T\) produces the outgoing leakage pattern and \(\mathcal{E}_T\) embeds the result into \(\mathcal{H}_{\mathsf{Q}}\).

When \(\Phi\) is completely positive and trace preserving (CPTP), these Pauli maps \(\mathcal{D}_{T, S}^{\Phi}\) are subnormalized, and the sum of their weights (see Definition \ref{defn:pauli-map-weight}) over the outgoing leakage patterns \(T\) is unity for each incoming pattern \(S\) (see Proposition~\ref{prop:leakage-transition-Pauli-map-normalization}). 
Thus, Eq.~\eqref{eq:maintext-leakage-transition-Pauli-map-decomposition} describes the leakage-pattern transitions induced by the twirled channel \(\mathcal{T}^{\mathsf{Q}}_C (\Phi)\) in terms of a classical Markov process, where the weight \(p(T|S)=\operatorname{wt}(\mathcal{D}_{T, S}^{\Phi})\) gives the probability of the transition \(S\mapsto T\).
For each nontrivial transition, the corresponding action on the computational subspace consists of the ideal Clifford \(C\) followed by the Pauli channel obtained by normalizing \(\mathcal{D}_{T, S}^{\Phi}\) (see Definition~\ref{defn:conditional-leakage-transition-maps}).

\subsection{Model of a QEC cycle}
\label{sec:qec-cycle-model}

Here we introduce the mathematical model of a QEC cycle that is used in the remainder of this paper.  
Diagram~\ref{circ:qec-cycle-model} gives its circuit representation. 
This abstract formulation is not the usual way of presenting a QEC cycle, but it provides a convenient framework for our analysis while accommodating a very broad class of QEC-cycle constructions. 
The results in this paper apply to any QEC cycle that admits this abstract description.
\begin{equation}
    \begin{quantikz}[row sep=0.38cm, column sep=0.32cm]
        \lstick{$\mathsf{D}: \rho_{\mathrm{in}}$~}
            & \gate[3]{\mathcal{N}}
            & \qw
            & \qw
            & \qw
            & \gate{\mathcal{W}^{\mathsf{D}}(D(\theta))}
            & \gate{\mathcal{R}_{K(\theta)}}
            & \qw \rstick{$~\rho_{\mathrm{out}}$}
            \\
        \lstick{$\mathsf{A}:\ket{0}_{\mathsf{A}}$}
            & \ghost{\mathcal{N}}
            & \qw
            & \qw
            & \qw
            & \qw
            & \qw
            & \qw \rstick{\text{Discard}}
            \\
        \lstick{$\mathsf{O}:\ket{0}_{\mathsf{O}}$}
            & \ghost{\mathcal{N}}
            & \meter{$\theta$}
            & \gate{\includegraphics[width=0.4cm]{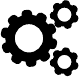}}
            & \gate[cwires={1}]{D,K}
            & \cwbend{-2}
            & \cwbend{-2}
            & \cw \rstick{\text{Discard}}
    \end{quantikz}
    \label{circ:qec-cycle-model}
\end{equation}

Consider a fixed stabilizer code on \(n\) physical qubits that encodes \(k\) logical qubits.   
We model a QEC cycle for this code using the registers \(\mathsf{D}\), \(\mathsf{A}\), and \(\mathsf{O}\) depicted in Diagram~\ref{circ:qec-cycle-model}, which are all leakage-extended qubit registers (see Definition~\ref{defn:register-of-leakage-extended-qubits}). 
The data register \(\mathsf{D}\) contains the physical qubits of the code and persists from one QEC cycle to the next.  
The ancilla register \(\mathsf{A}\) contains auxiliary 
qubits used for fault-tolerant syndrome extraction and leakage detection, and is reset between QEC cycles.  
The outcome register \(\mathsf{O}\) represents the classical outcomes produced by those processes, and is also reset between QEC cycles. 
Although the information contained in \(\mathsf{O}\) is classical, we model \(\mathsf{O}\) as a leakage-extended qubit register to model all errors associated with the QEC cycle, including leakage, as a channel on the joint \(\mathsf{D}\), \(\mathsf{A}\), and \(\mathsf{O}\) system. 
The subsequent measurement of \(\mathsf{O}\) is taken to be perfect. 
We denote by \(\mathsf{T}\) the joint leakage-extended qubit register comprising \(\mathsf{D}\), \(\mathsf{A}\), and \(\mathsf{O}\) (see Definition~\ref{defn:joint-leakage-extended-qubit-register}).

The mathematical model of a QEC cycle consists of a channel \(\Phi:\operatorname{End}(\mathcal{H}_{\mathsf{D}}) \to\operatorname{End}(\mathcal{H}_{\mathsf{D}})\) acting on \(\mathsf{D}\) alone.
The channel \(\Phi\) is derived from the extended model shown in Diagram~\ref{circ:qec-cycle-model}, read from left to right. 
The incoming state of \(\mathsf{D}\) is \(\rho_{\mathrm{in}}\).
We use an initialization map \(\mathcal{I}_{\mathsf{A}\mathsf{O}}\) to formalize the introduction of the registers \(\mathsf{A}\) and \(\mathsf{O}\), prepared in their all-zero states~\eqref{eq:ancilla-outcome-initialization}.
The first step of the QEC cycle is fault-tolerant syndrome extraction and leakage detection.
In the absence of noise or leakage, this is captured by the Clifford unitary \(C_{\mathrm{synd}}\), which records syndrome information into the outcome register, possibly using the ancilla qubits in \(\mathsf{A}\).
This construction appears in Ref.~\cite{kwiatkowski2025approximate}.
In the presence of noise and leakage, we generalize the Clifford channel \(\mathcal{C}_{\mathrm{synd}}\) to the channel \(\mathcal{N}\), which jointly acts on the leakage-extended registers \(\mathsf{D}\), \(\mathsf{A}\), and \(\mathsf{O}\); in the noise- and leakage-free limits, \(\mathcal{N}\) reproduces \(\mathcal{C}_{\mathrm{synd}}\) on the computational subspace.

Each system in \(\mathsf{O}\) is then measured perfectly, resulting in one of three outcome labels \(0\), \(1\), and \(\mathrm L\) (where
\(\mathrm L\) labels the leaked state).
The full outcome record is denoted \(\theta\in\Theta\), where \(\Theta\coloneqq \{0,1,\mathrm L\}^{\mathsf{O}}\).
The measurement and the discard of \(\mathsf{A}\) and \(\mathsf{O}\) are formalized by outcome maps \(\{\mathcal{M}_\theta\}_{\theta\in\Theta}\)~\eqref{eq:outcome-branch}, with \(\theta\) retained only as a classical outcome label. 
We use a generalized notion of the QEC cycle's decoder, which uses the classical outcome record \(\theta\) to determine both the logical correction given by the syndrome record and the subset of data qubits to return to the computational subspace via classical side processing.
The Pauli-correction function \(D:\Theta\to V_{\mathsf{K}_{\mathsf{D}}}\)~\eqref{eq:Pauli-correction-function} determines the logical correction, captured by the Pauli channel \(\mathcal{W}^{\mathsf{D}}(D(\theta))\) on \(\mathsf{D}\) (see Definition~\ref{defn:leakage-extended-Pauli-family}).
The return function \(K:\Theta\to 2^{\mathsf{D}}\)~\eqref{eq:return-function} determines the subset of the systems in the data register that should be returned to the computational subspace as maximally mixed states. 
This is captured by the leakage-extended operation \(\mathcal{R}_{K(\theta)}\) on \(\mathsf{D}\), where \(K(\theta)\) is the subset of qubits to be returned to the computational subspace.

Thus, the mathematical model of the  complete-cycle channel is given by
\begin{equation}
    \Phi
    =
    \sum_{\theta\in\Theta}
    \mathcal{R}_{K(\theta)}
    \circ
    \mathcal{W}^{\mathsf{D}}(D(\theta))
    \circ
    \mathcal{M}_\theta
    \circ
    \mathcal{N}
    \circ
    \mathcal{I}_{\mathsf{A}\mathsf{O}}, 
    \label{eq:complete-cycle}
\end{equation}
which acts on \(\mathsf{D}\) alone and is derived in detail in Appendix~\ref{sec:setup}.

To define logical error processes, consider an abstract \(k\)-qubit logical register \(\mathsf{L}\) and an abstract \((n-k)\)-qubit syndrome register
\(\mathsf{S}\) (see Definition~\ref{defn:qubit-register}).  
Let \(\CFrame\) be a Clifford isomorphism (see Definition~\ref{defn:Clifford-isomorphism}) that provides the logical--syndrome decomposition of the computational subspace of the data register~\cite{kwiatkowski2025approximate}:
\begin{equation}
    \CFrame:
    \mathcal{K}_{{\mathsf{D}}}
    \to 
    \mathcal{K}_{\mathsf{L}}\otimes\mathcal{K}_{\mathsf{S}}.
    \label{eq:maintext-Clifford-frame}
\end{equation}
In the Clifford frame, the nominal Clifford unitary \(C_{\mathrm{synd}}\) that implements fault-tolerant syndrome extraction and leakage detection in the computational subspace is replaced by the Clifford \(C_{\mathrm{fr}}\in \operatorname{End} (\mathcal{K}_{\mathsf{L}} \otimes \mathcal{K}_{\mathsf{S}} \otimes \mathcal{K}_{\mathsf{A}} \otimes \mathcal{K}_{\mathsf{O}} ) \).  
For every \(s\in\mathbb{Z}_2^{\mathsf{S}}\) and \(\ket{\psi}_{\mathsf{L}}\in\mathcal{K}_{\mathsf{L}}\), it satisfies
\begin{align}
    &C_{\mathrm{fr}}
    \left(
        \ket{\psi}_{\mathsf{L}}
        \otimes
        \ket{s}_{\mathsf{S}}
        \otimes
        \ket{0}_{\mathsf{A}}
        \otimes
        \ket{0}_{\mathsf{O}}
    \right)
    =
    \ket{\psi}_{\mathsf{L}}
    \otimes
    \ket{s}_{\mathsf{S}}
    \otimes
    \ket{\alpha(s)}_{\mathsf{A}}
    \otimes
    \ket{E(s)}_{\mathsf{O}}, 
    \label{eq:maintext-nominal-readout}
\end{align}
where \(E:\mathbb{Z}_2^{\mathsf{S}}\to\mathbb{Z}_2^{\mathsf{O}}\) is an injective encoding of the syndrome record~\eqref{eq:syndrome-record-encoding} that is determined by the ideal Clifford $C_{\rm synd}$ in the original frame~\eqref{eq:syndrome-readout-frame-conjugation}, while \(\ket{\alpha(s)}_{\mathsf{A}}\) is a state of \(\mathsf{A}\) that may depend on \(s\)~\eqref{eq:nominal-readout-vector-action}; its detailed form is unimportant. 
The interpretation of \(C_{\mathrm{synd}}\) is that it gives the ideal (noiseless) action of \(\mathcal{N}\) in the joint computational subspace of the registers \(\mathsf{D}\), \(\mathsf{A}\), and \(\mathsf{O}\).  
The channel \(\mathcal{N}\) is otherwise arbitrary, so \(C_{\mathrm{synd}}\) matters only as the reference Clifford for leakage-dephased Pauli twirling.

The channel \(\mathcal{N}\) can be simplified by leakage-dephased Pauli twirling, as in Theorem~\ref{prop:leakage-transition-Pauli-map-decomposition}.  Its twirled form is denoted 
\begin{equation}
    \label{eq:twirled-readout-map}
    \mathcal{N}^{\mathrm{tw}}
    =
    \mathcal{T}^{\mathsf{T}}_{C_{\mathrm{synd}}}(\mathcal{N}) \,
\end{equation}
and replacing \(\mathcal{N}\) by \(\mathcal{N}^{\mathrm{tw}}\) in Eq.~\eqref{eq:complete-cycle} gives the twirled complete-cycle
channel \(\Phi^{\mathrm{tw}}\) defined in Eq.~\eqref{eq:twirled complete-cycle}.  
In the remainder of this paper, we assume that the complete-cycle channel \(\Phi\) has the structure exhibited by \(\Phi^{\mathrm{tw}}\) in Eq.~\eqref{eq:twirled complete-cycle-constituent-Pauli-map-expansion}, as formalized in Eq.~\eqref{eq:complete-cycle-projector-Markov-Pauli-form}.  
This structure can arise from physically implementing leakage-dephased Pauli twirling, as in Theorem~\ref{prop:leakage-transition-Pauli-map-decomposition}, from physical guarantees or approximations, or from applying leakage-dephased Pauli twirling separately to the constituent processes of the QEC cycle.
The subsequent results in this paper depend only on the resulting structure in Eq.~\eqref{eq:complete-cycle-projector-Markov-Pauli-form}, not on
how it is obtained.

\subsection{Logical Pauli maps of a QEC-cycle}
\label{sec:qec-cycle-logical-Pauli-maps}

Our second main result is that the complete QEC-cycle channel \(\Phi\) decomposes into logical Pauli maps conditioned on joint transitions of the syndrome and data-leakage pattern.  
This is formalized in detail in Theorem~\ref{thm:single-cycle} and described at a high level in this subsection.

First, for \(s_{\mathrm{in}},s_{\mathrm{out}} \in\mathbb{Z}_2^{\mathsf{S}}\), define the syndrome transition map
\(
    \mathcal{S}_{s_{\mathrm{out}}, s_{\mathrm{in}}}:
    \operatorname{End}(\mathcal{K}_{\mathsf{S}})
    \to
    \operatorname{End}(\mathcal{K}_{\mathsf{S}})
\)
by
\(
    \mathcal{S}_{s_{\mathrm{out}}, s_{\mathrm{in}}}(Y)
    \coloneqq
    \MatEl{s_{\mathrm{in}}}{Y}{s_{\mathrm{in}}}_{\mathsf{S}}
    \ZProj{s_{\mathrm{out}}}_{\mathsf{S}}
\)
for \(Y\in\operatorname{End}(\mathcal{K}_{\mathsf{S}})\)~\eqref{eq:syndrome-transition-map}.
Theorem~\ref{thm:single-cycle} states that there is a family of subnormalized Pauli maps
\(
    \mathcal{J}^{R_{\mathrm{out}}, R_{\mathrm{in}}}
    _{s_{\mathrm{out}},s_{\mathrm{in}}}:
    \operatorname{End}(\mathcal{K}_{\mathsf{L}})
    \to
    \operatorname{End}(\mathcal{K}_{\mathsf{L}})
\),
indexed by \(R_{\mathrm{in}},R_{\mathrm{out}}\subseteq\mathsf{D}\) and \(s_{\mathrm{in}},s_{\mathrm{out}} \in\mathbb{Z}_2^{\mathsf{S}}\), such that
\begin{equation}
    \begin{aligned}
    \Phi
    = \hspace{-2.5mm}
    \sum_{\substack{R_{\mathrm{in}},R_{\mathrm{out}}\subseteq\mathsf{D} \\ s_{\mathrm{in}},s_{\mathrm{out}} \in\mathbb{Z}_2^{\mathsf{S}}}}
    \hspace{-1mm}
        \mathcal{E}_{R_{\mathrm{out}}}^{\mathsf{D}}
        \circ
        \mathcal{L}_{R_{\mathrm{out}}}^{\mathsf{D}}
        \circ
        \mathcal{C}_{\mathrm F}^{-1}
    \circ
        \mathcal{J}^{R_{\mathrm{out}}, R_{\mathrm{in}}}
        _{s_{\mathrm{out}},s_{\mathrm{in}}}
        \otimes
        \mathcal{S}_{s_{\mathrm{out}}, s_{\mathrm{in}}}
    \circ
        \mathcal{C}_{\mathrm F}^{\vpd}
        \circ
        \mathcal{F}_{R_{\mathrm{in}}}^{\mathsf{D}}
        \circ
        \mathcal{P}_{R_{\mathrm{in}}}^{\mathsf{D}}, 
    \end{aligned}
    \label{eq:maintext-physical-single-cycle-decomposition}
\end{equation}
where \( \mathcal{C}_{\mathrm F}^{\vpd}(X) = \CFrame X \CFrame^{\dagger}\) moves any operator \(X \in \operatorname{End}(\mathcal{K}_{\mathsf{D}})\)  between the computational data register and the logical--syndrome frame.  
A superscript \(\mathsf{D}\), e.g., indicates the specialization of the corresponding leakage-sector map to the data register \(\mathsf{D}\).
Reading Eq.~\eqref{eq:maintext-physical-single-cycle-decomposition} from right to left, 
the projection map \(\mathcal{P}_{R_{\mathrm{in}}}^{\mathsf{D}}\) selects the incoming data-leakage subspace, 
the filling map \(\mathcal{F}_{R_{\mathrm{in}}}^{\mathsf{D}}\) replaces the leaked systems labeled \(i \in R_{\mathrm{in}}\) with maximally mixed computational qubits,
and \(\mathcal{C}_{\mathrm F}^{\,}\) expresses the resulting operator in the logical--syndrome frame.  
Next, each term \( \mathcal{J} \otimes \mathcal{S}\) with fixed \(s_{\rm in}\), \(s_{\rm out}\) applies the syndrome transition \(s_{\mathrm{in}}\to s_{\mathrm{out}}\) (via \(\mathcal{S}\)), together with the logical Pauli map \(\mathcal{J}^{R_{\mathrm{out}}, R_{\mathrm{in}}}
_{s_{\mathrm{out}},s_{\mathrm{in}}}\).  
Finally, \(\mathcal{C}_{\mathrm F}^{-1}\) maps operators back into the computational subspace \(\mathcal{K}_{\mathsf{D}}\), 
the leakage map \(\mathcal{L}_{R_{\mathrm{out}}}^{\mathsf{D}}\) leaks into the subspace associated with the outgoing data-leakage pattern, 
and the extension map \(\mathcal{E}_{R_{\mathrm{out}}}^{\mathsf{D}}\) embeds that leakage subspace into the full space of the data register.

Here, the maps \(\mathcal{J}^{R_{\mathrm{out}}, R_{\mathrm{in}}} _{s_{\mathrm{out}},s_{\mathrm{in}}}\) depend jointly on the incoming
and outgoing syndrome and data-leakage pattern.  
When the logical register is considered in isolation, the syndrome and data-leakage pattern together constitute a hidden memory that persists between cycles and determines which logical Pauli map acts.  
This dependence is the source of the emergent non-Markovianity of the isolated logical dynamics discussed in Ref.~\cite{ziyad2025emergent}. 
To formalize the interpretation of the syndrome and data-leakage pattern as the hidden memory of a QEC cycle, we define the hidden-memory state space \(\mathsf{M} \coloneqq 2^{\mathsf{D}}\times\mathbb{Z}_2^{\mathsf{S}}\) (see Definition~\ref{def:hidden-memory-index}).  
An element \(m=(R,s)\in\mathsf{M}\) specifies a data-leakage pattern \(R\subseteq\mathsf{D}\) and a syndrome \(s\in\mathbb{Z}_2^{\mathsf{S}}\).  
The model of a QEC cycle can then be expressed in terms of logical-Pauli transfer matrices, whose rows and columns are indexed by outgoing and incoming hidden-memory states in \(\mathsf{M}\), respectively.  
These transfer matrices are used to analyze memory experiments, as described in detail in the next subsection.

\subsection{Application to memory experiments}
\label{sec:memory-experiments-main}

In this paper, we define a memory experiment to consist of an input state \(\rho\) on \(\mathcal{H}_{\mathsf{D}}\), followed by \(k\geq1\) QEC cycles, followed by a POVM measurement with a distinguished POVM element \(E_{\mathrm{succ}}\) that indicates the outcome ``success.''  Typically, success is understood to mean
successful recovery of the logical information encoded in \(\rho\).
Modeling each cycle by the complete-cycle channel \(\Phi\) in
Eq.~\eqref{eq:complete-cycle}, the success probability is
\begin{equation}
    p_{\mathrm{succ}}(k) \coloneqq \trace \! \left[ E_{\mathrm{succ}} \Phi^k(\rho) \right].
    \label{eq:maintext-memory-experiment-success-probability}
\end{equation}
To analyze \(p_{\mathrm{succ}}(k)\), for each logical Pauli label
\(\ell\in V_{\mathsf{L}}\) we first define a logical-Pauli transfer matrix
\(T_\ell\in\mathbb{R}^{\mathsf{M}\times\mathsf{M}}\)
(see Definition~\ref{def:memory-logical-Pauli-transfer-matrices}).  Each
\(T_\ell\) is determined by the \(P_\ell\) eigenvalues of the logical
Pauli maps
\(\mathcal{J}^{R_{\mathrm{out}}, R_{\mathrm{in}}}
_{s_{\mathrm{out}},s_{\mathrm{in}}}\).
The family of transfer matrices
\(\{T_\ell\}_{\ell\in V_{\mathsf{L}}}\) contains the information needed
to calculate \(\Phi^k(\rho)\) for any input state \(\rho\) and
\(k\geq1\), as formalized in
Corollary~\ref{cor:memory-repeated-cycle-logical-Pauli-dynamics}.
Then, for each input state \(\rho\) we define a state-preparation vector
\(\boldsymbol r_\ell(\rho)\in\mathbb{C}^{\mathsf{M}}\)
(see Definition~\ref{def:memory-logical-Pauli-block-coefficient-vectors}).
Finally, for each POVM element \(E\), we define a measurement vector
\(\boldsymbol e_\ell(E)\in\mathbb{C}^{\mathsf{M}}\)
(see Definition~\ref{def:memory-logical-Pauli-measurement-vectors}).
Thus the components of both vectors are indexed by \(\mathsf{M}\).

Our third main result expresses \(p_{\mathrm{succ}}(k)\) in terms of
\(\boldsymbol r_\ell(\rho)\),
\(\boldsymbol e_\ell(E_{\mathrm{succ}})\), and \(T_\ell\) for
\(\ell\in V_{\mathsf{L}}\).  This is formalized in
Theorem~\ref{thm:memory-logical-Pauli-transfer-formula}, which states
that
\begin{equation}
    p_{\mathrm{succ}}(k)
    =
    d_{\mathsf{L}}
    \sum_{\ell\in V_{\mathsf{L}}}
    \boldsymbol e_\ell(E_{\mathrm{succ}})^{\mathsf{T}}
    T_\ell^k
    \boldsymbol r_\ell(\rho).
    \label{eq:maintext-memory-experiment-transfer-formula}
\end{equation}

Here \(d_{\mathsf{L}}=\dim(\mathcal{K}_{\mathsf{L}})\), and
\(\ell\in V_{\mathsf{L}}\) labels the logical Pauli \(P_\ell\).
Matrix multiplication sums over intermediate hidden-memory states, so
\(T_\ell^k\) propagates the \(P_\ell\) coefficient through \(k\)
cycles.  Consequently,
\(
    d_{\mathsf{L}}
    \boldsymbol e_\ell(E_{\mathrm{succ}})^{\mathsf{T}}
    T_\ell^k
    \boldsymbol r_\ell(\rho)
\)
is the contribution of \(P_\ell\) to the success probability, and
the sum over \(\ell\) in
Eq.~\eqref{eq:maintext-memory-experiment-transfer-formula}
reconstructs the full success probability.

In favorable settings, the individual summands in
Eq.~\eqref{eq:maintext-memory-experiment-transfer-formula} may be
isolated either using physically motivated assumptions about state
preparation and measurement or by adapting the character-weighting
techniques of Ref.~\cite{helsen2019new}; if so, the summands may be analyzed separately.

\subsection{Discussion}
\label{sec:discussion-main}

We envision several additional applications of these results.  First,
Theorem~\ref{prop:leakage-transition-Pauli-map-decomposition} is
deliberately formulated as an abstract structural result for maps on
leakage-extended qubit registers.  Because it does not depend on the
details of a QEC cycle, we expect it to be useful more broadly in the
study of leakage and other multilevel quantum processes.  Second,
Theorem~\ref{thm:single-cycle} resolves the logical noise of a QEC
cycle into contributions conditioned on the incoming and outgoing
syndrome and data-leakage pattern.  For any QEC scheme and decoder
described by our model, this decomposition makes explicit how each
transition of these variables contributes to the resulting logical
Pauli map.  We hope that this perspective will inform the analysis and
design of leakage-aware decoders.  Third,
Theorem~\ref{thm:memory-logical-Pauli-transfer-formula} extends the
program of Ref.~\cite{kwiatkowski2025approximate}, which developed
approximate logical-error models for QEC memory experiments.  The
present results show that the data-leakage pattern plays a role closely
analogous to that of the syndrome: both are persistent hidden-memory
variables that mediate the effective logical dynamics.  In
Ref.~\cite{kwiatkowski2025approximate}, physically motivated assumptions
on the distribution of the syndrome made it possible to establish the
existence of accurate approximate Markovian models in a broad class of
settings.  An important direction for future work is to determine
whether analogous assumptions can be justified for the joint
syndrome-leakage process and, if so, whether they lead to similarly
accurate reduced-memory models.

Several extensions of the QEC-cycle model are particularly important.
Although the final Pauli correction and return operation may depend on
the complete outcome record of a cycle, our model does not permit an
outcome obtained partway through the cycle to determine subsequent
operations within that same cycle.  Such adaptive decisions arise, for
example, in some constructions of flagged fault tolerance.  It appears
plausible that our results can be generalized to these settings by
representing the different adaptive branches as distinct QEC-cycle
maps, or by composing several such maps into a larger effective cycle,
but we leave a formal treatment to future work.  We also do not
consider windowed decoding, in which decoding decisions depend on
outcome records collected over several QEC cycles.  Extending the
hidden-memory framework to windowed decoders, perhaps by augmenting the
memory state with the decoder's internal state or by treating an entire
decoding window as a composite cycle, is an especially important
direction.  Finally, we assume that both the decoder-selected Pauli
correction and the return map are implemented perfectly.  Allowing
these operations to be noisy is unlikely to change the essential
structure of the results, provided that their errors admit an
appropriate leakage-resolved Pauli description, but establishing this
extension rigorously is also left for future work.

\section{Context of this work and LLM usage}
This project began in July 2026 as an extension of our previous work in Ref.~\cite{kwiatkowski2025approximate}.
Our original goal was to generalize our previous results to noisy QEC with qubits that also experience leakage.
It was clear that in general, allowing arbitrary error processes on leaked levels would make it quite difficult to model QEC and prove useful structural results.
However, based on our prior research experience on randomized benchmarking, Pauli twirling, and other randomized methods, it seemed quite plausible that Pauli twirling, once appropriately adapted for leakage, could provide useful simplifications.
We made an initial investigation into this question using the LLMs that were available to us at the time, predominantly GPT 5 models.
Based on that investigation, we decided it was worthwhile to start a full research project, eventually culminating in this paper.

Throughout the project, we continuously made decisions about the scope, assumptions, organization, and technical strategy of the work.
We made these decisions using extensive LLM assistance to suggest ideas, try out different ways to organize results and concepts, fill in details, and find related work. 
We found that the direct outputs of LLMs were almost always unsuitable for inclusion in the final paper, largely because the LLMs by default would use confusing and ambiguous terminology for specific mathematical objects, and would routinely use vague language and heuristic ideas instead of rigorous proofs.
As a result, we found that the best way to efficiently make progress was for us to choose how to break our results down into tractable pieces, essentially by specifying every definition, lemma, and theorem.
We chose the organization in such a way the individual lemmas were simple enough that we felt like the proofs output by LLMs were trustworthy, and usable after some (sometimes substantial) editorial effort and manual checking.
As such, we have done our best to make sure that the proofs are easily readable and easily checkable.

As part of this project, we also used LLMs to construct formal proofs of our results in LEAN4. 
We did not manually check any of the formal proofs, and consequently we do not view them as a true part of this publication.
We instead view them as yet another part of our due diligence to guard against errors and inaccurate statements.
The proofs are included with the arxiv submission and we encourage any interested parties to verify them.

\begin{acknowledgments}
      The authors thank Aliza Siddiqui, Shawn Geller, Scott Glancy, and Emanuel Knill for helpful comments and discussions.
  AJF acknowledges support from the Professional Research Experience Program (PREP) operated jointly by NIST and the University of Colorado. This work includes contributions of the National Institute of Standards and Technology, which are not subject to U.S. copyright. 
\end{acknowledgments}

\bibliography{ref.bib} 

@misc{kwiatkowski2025approximate,
  author        = {Kwiatkowski, Alex and Friedman, Aaron J. and Geller, Shawn and Ziyad, Jalan A. and Glancy, Scott and Knill, Emanuel},
  title         = {Constructing an approximate logical {M}arkovian model of consecutive {QEC} cycles of a stabilizer code},
  year          = {2025},
  eprint        = {2509.16887},
  archivePrefix = {arXiv},
  doi           = {10.48550/arXiv.2509.16887}
}

@misc{ziyad2025emergent,
  author        = {Ziyad, Jalan A. and Blume-Kohout, Robin and Rudinger, Kenneth},
  title         = {Emergent Non-{M}arkovianity in Logical Qubit Dynamics},
  year          = {2025},
  eprint        = {2512.08893},
  archivePrefix = {arXiv},
  doi           = {10.48550/arXiv.2512.08893}
}

@article{aliferis2007faulttolerant,
  author        = {Aliferis, Panos and Terhal, Barbara M.},
  title         = {Fault-tolerant quantum computation for local leakage faults},
  journal       = {Quantum Information and Computation},
  volume        = {7},
  number        = {1--2},
  pages         = {139--156},
  year          = {2007},
  doi           = {10.26421/QIC7.1-2-9}
}

@article{fowler2013coping,
  author        = {Fowler, Austin G.},
  title         = {Coping with qubit leakage in topological codes},
  journal       = {Physical Review A},
  volume        = {88},
  number        = {4},
  pages         = {042308},
  year          = {2013},
  doi           = {10.1103/PhysRevA.88.042308},
  eprint        = {1308.6642},
  archivePrefix = {arXiv}
}

@article{ghosh2013understanding,
  author        = {Ghosh, Joydip and Fowler, Austin G. and Martinis, John M. and Geller, Michael R.},
  title         = {Understanding the effects of leakage in superconducting quantum-error-detection circuits},
  journal       = {Physical Review A},
  volume        = {88},
  number        = {6},
  pages         = {062329},
  year          = {2013},
  doi           = {10.1103/PhysRevA.88.062329},
  eprint        = {1306.0925},
  archivePrefix = {arXiv}
}

@article{ghosh2015leakageresilient,
  author        = {Ghosh, Joydip and Fowler, Austin G.},
  title         = {Leakage-resilient approach to fault-tolerant quantum computing with superconducting elements},
  journal       = {Physical Review A},
  volume        = {91},
  number        = {2},
  pages         = {020302},
  year          = {2015},
  doi           = {10.1103/PhysRevA.91.020302},
  eprint        = {1406.2404},
  archivePrefix = {arXiv}
}

@article{suchara2015leakage,
  author        = {Suchara, Martin and Cross, Andrew W. and Gambetta, Jay M.},
  title         = {Leakage suppression in the toric code},
  journal       = {Quantum Information and Computation},
  volume        = {15},
  number        = {11--12},
  pages         = {997--1016},
  year          = {2015},
  doi           = {10.26421/QIC15.11-12-8},
  eprint        = {1410.8562},
  archivePrefix = {arXiv}
}

@article{battistel2021hardware,
  author        = {Battistel, Francesco and Varbanov, Boris M. and Terhal, Barbara M.},
  title         = {Hardware-Efficient Leakage-Reduction Scheme for Quantum Error Correction with Superconducting Transmon Qubits},
  journal       = {PRX Quantum},
  volume        = {2},
  number        = {3},
  pages         = {030314},
  year          = {2021},
  doi           = {10.1103/PRXQuantum.2.030314},
  eprint        = {2102.08336},
  archivePrefix = {arXiv}
}

@article{mcewen2021removing,
  author        = {McEwen, M. and Kafri, D. and Chen, Z. and Atalaya, J. and Satzinger, K. J. and Quintana, C. and Klimov, P. V. and Sank, D. and Gidney, C. and Fowler, A. G. and Arute, F. and Arya, K. and Buckley, B. and Burkett, B. and Bushnell, N. and Chiaro, B. and Collins, R. and Demura, S. and Dunsworth, A. and Erickson, C. and Foxen, B. and Giustina, M. and Huang, T. and Hong, S. and Jeffrey, E. and Kim, S. and Kechedzhi, K. and Kostritsa, F. and Laptev, P. and Megrant, A. and Mi, X. and Mutus, J. and Naaman, O. and Neeley, M. and Neill, C. and Niu, M. and Paler, A. and Redd, N. and Roushan, P. and White, T. C. and Yao, J. and Yeh, P. and Zalcman, A. and Chen, Yu and Smelyanskiy, V. N. and Martinis, John M. and Neven, H. and Kelly, J. and Korotkov, A. N. and Petukhov, A. G. and Barends, R.},
  title         = {Removing leakage-induced correlated errors in superconducting quantum error correction},
  journal       = {Nature Communications},
  volume        = {12},
  number        = {1},
  pages         = {1761},
  year          = {2021},
  doi           = {10.1038/s41467-021-21982-y},
  eprint        = {2102.06131},
  archivePrefix = {arXiv}
}

@article{bultink2020protecting,
  author        = {Bultink, C. C. and O'Brien, T. E. and Vollmer, R. and Muthusubramanian, N. and Beekman, M. W. and Rol, M. A. and Fu, X. and Tarasinski, B. and Ostroukh, V. and Varbanov, B. and Bruno, A. and DiCarlo, L.},
  title         = {Protecting quantum entanglement from leakage and qubit errors via repetitive parity measurements},
  journal       = {Science Advances},
  volume        = {6},
  number        = {12},
  pages         = {eaay3050},
  year          = {2020},
  doi           = {10.1126/sciadv.aay3050},
  eprint        = {1905.12731},
  archivePrefix = {arXiv}
}

@article{varbanov2020leakage,
  author        = {Varbanov, Boris M. and Battistel, Francesco and Tarasinski, Brian M. and Ostroukh, Viacheslav P. and O'Brien, Thomas E. and DiCarlo, Leonardo and Terhal, Barbara M.},
  title         = {Leakage detection for a transmon-based surface code},
  journal       = {npj Quantum Information},
  volume        = {6},
  number        = {1},
  pages         = {102},
  year          = {2020},
  doi           = {10.1038/s41534-020-00330-w},
  eprint        = {2002.07119},
  archivePrefix = {arXiv}
}

@article{ziad2025local,
  author        = {Ziad, Abbas B. and Zalawadiya, Ankit and Topal, Canberk and Camps, Joan and Geh{\'e}r, Gy{\"o}rgy P. and Stafford, Matthew P. and Turner, Mark L.},
  title         = {Local clustering decoder as a fast and adaptive hardware decoder for the surface code},
  journal       = {Nature Communications},
  volume        = {16},
  number        = {1},
  pages         = {11048},
  year          = {2025},
  doi           = {10.1038/s41467-025-66773-x},
  eprint        = {2411.10343},
  archivePrefix = {arXiv}
}

@article{marshall2025incoherent,
  author        = {Marshall, Jeffrey and Kafri, Dvir},
  title         = {Incoherent approximation of leakage in quantum error correction},
  journal       = {Physical Review Applied},
  volume        = {23},
  number        = {5},
  pages         = {054025},
  year          = {2025},
  doi           = {10.1103/PhysRevApplied.23.054025},
  eprint        = {2312.10277},
  archivePrefix = {arXiv}
}

@article{manabe2025efficient,
  author        = {Manabe, Hidetaka and Suzuki, Yasunari and Darmawan, Andrew S.},
  title         = {Efficient simulation of leakage errors in quantum error correcting codes using tensor network methods},
  journal       = {New Journal of Physics},
  volume        = {27},
  number        = {11},
  pages         = {114512},
  year          = {2025},
  doi           = {10.1088/1367-2630/ae1529},
  eprint        = {2308.08186},
  archivePrefix = {arXiv}
}

@misc{vendrell2026plaquette,
  author        = {Conchello Vendrell, Raul and D{\'i}az L{\'o}pez, Carlos and Dhand, Ish and Kapoor, Kshitij and Laureti, Davide and Massaro, Marcello and Nayak, Pranjal and Ogloblin, Ivan and Plenio, Martin B. and Prasanna Kumar, Shreya and Santandrea, Matteo and Seshadri, Varun and Sz{\'a}va, Antal and Vincent, Trevor and Weber, Raphael},
  title         = {Plaquette: A hardware-aware design platform for fault-tolerant quantum computers},
  year          = {2026},
  eprint        = {2607.08767},
  archivePrefix = {arXiv},
  doi           = {10.48550/arXiv.2607.08767}
}

@article{wallman2016noise,
  author        = {Wallman, Joel J. and Emerson, Joseph},
  title         = {Noise tailoring for scalable quantum computation via randomized compiling},
  journal       = {Physical Review A},
  volume        = {94},
  number        = {5},
  pages         = {052325},
  year          = {2016},
  doi           = {10.1103/PhysRevA.94.052325},
  eprint        = {1512.01098},
  archivePrefix = {arXiv}
}

@article{geller2013efficient,
  author        = {Geller, Michael R. and Zhou, Zhongyuan},
  title         = {Efficient error models for fault-tolerant architectures and the {P}auli twirling approximation},
  journal       = {Physical Review A},
  volume        = {88},
  number        = {1},
  pages         = {012314},
  year          = {2013},
  doi           = {10.1103/PhysRevA.88.012314},
  eprint        = {1305.2021},
  archivePrefix = {arXiv}
}

@article{helsen2019new,
  author  = {Helsen, Jonas and Xue, Xiao and Vandersypen, Lieven M. K. and Wehner, Stephanie},
  title   = {A New Class of Efficient Randomized Benchmarking Protocols},
  journal = {npj Quantum Information},
  volume  = {5},
  number  = {1},
  pages   = {71},
  year    = {2019},
  doi     = {10.1038/s41534-019-0182-7}
}

@article{google2021exponential,
  author  = {{Google Quantum AI}},
  title   = {Exponential suppression of bit or phase errors with cyclic error correction},
  journal = {Nature},
  volume  = {595},
  pages   = {383--387},
  year    = {2021},
  doi     = {10.1038/s41586-021-03588-y}
}

@article{google2023suppressing,
  author        = {{Google Quantum AI}},
  title         = {Suppressing quantum errors by scaling a surface code logical qubit},
  journal       = {Nature},
  volume        = {614},
  pages         = {676--681},
  year          = {2023},
  doi           = {10.1038/s41586-022-05434-1},
  eprint        = {2207.06431},
  archivePrefix = {arXiv}
}

@article{google2025belowthreshold,
  author  = {{Google Quantum AI and Collaborators}},
  title   = {Quantum error correction below the surface code threshold},
  journal = {Nature},
  volume  = {638},
  pages   = {920--926},
  year    = {2025},
  doi     = {10.1038/s41586-024-08449-y}
}

@article{xin2026improved,
  author        = {Xin, Yuejie and van der Meer, Sean L. M. and Serra-Peralta, Marc and Vroomans, Tim H. F. and Finkel, Matvey and Veen, Hendrik M. and Beekman, Marc W. and DiCarlo, Leonardo},
  title         = {Improved error correction with leakage reduction units built into qubit measurement in a superconducting quantum processor},
  journal       = {PRX Quantum},
  year          = {2026},
  doi           = {10.1103/y66v-988s},
  eprint        = {2511.17460},
  archivePrefix = {arXiv}
}

@misc{berthusen2026experimental,
  author        = {Berthusen, Noah and Lavasani, Ali and Benhemou, Asmae and Allman, M. S. and Dreiling, Joan and Estey, Brian and Foltz, Cameron and Jacobs, Trent and Mills, Michael and Park, Annie Jihyun and Reed, Adam P. and Hayes, David and Metodi, Tzvetan S. and Potter, Andrew C.},
  title         = {Experimental validation of a compact fault-tolerant architecture for trapped ions},
  year          = {2026},
  eprint        = {2609.03194},
  archivePrefix = {arXiv},
  doi           = {10.48550/arXiv.2609.03194}
}

@article{xu2018integrity,
  author        = {Xu, Xiaosi and de Beaudrap, Niel and O'Gorman, Joe and Benjamin, Simon C.},
  title         = {An integrity measure to benchmark quantum error correcting memories},
  journal       = {New Journal of Physics},
  volume        = {20},
  number        = {2},
  pages         = {023009},
  year          = {2018},
  doi           = {10.1088/1367-2630/aaa372},
  eprint        = {1707.09951},
  archivePrefix = {arXiv}
}

@article{pal2022relaxation,
  author        = {Pal, Amit Kumar and Schindler, Philipp and Erhard, Alexander and Rivas, {\'A}ngel and Martin-Delgado, Miguel-Angel and Blatt, Rainer and Monz, Thomas and M{\"u}ller, Markus},
  title         = {Relaxation times do not capture logical qubit dynamics},
  journal       = {Quantum},
  volume        = {6},
  pages         = {632},
  year          = {2022},
  doi           = {10.22331/q-2022-01-24-632},
  eprint        = {2012.07911},
  archivePrefix = {arXiv}
}

@article{gidney2022stability,
  author        = {Gidney, Craig},
  title         = {Stability Experiments: The Overlooked Dual of Memory Experiments},
  journal       = {Quantum},
  volume        = {6},
  pages         = {786},
  year          = {2022},
  doi           = {10.22331/q-2022-08-24-786},
  eprint        = {2204.13834},
  archivePrefix = {arXiv}
}

@article{paetznick2026improved,
  author        = {Paetznick, Adam and Reichardt, Ben W. and da Silva, Marcus P. and Ryan-Anderson, Ciaran and others},
  title         = {Improved quantum processor logical error rates via correction and detection},
  journal       = {Nature},
  volume        = {654},
  number        = {8118},
  pages         = {349--355},
  year          = {2026},
  doi           = {10.1038/s41586-026-10628-y},
  eprint        = {2404.02280},
  archivePrefix = {arXiv}
}

@article{caune2026realtime,
  author        = {Caune, Laura and Skoric, Luka and Blunt, Nick S. and others},
  title         = {Demonstrating real-time and low-latency quantum error correction with superconducting qubits},
  journal       = {Nature Communications},
  volume        = {17},
  pages         = {7383},
  year          = {2026},
  doi           = {10.1038/s41467-026-73331-6},
  eprint        = {2410.05202},
  archivePrefix = {arXiv}
}

@misc{atomcomputing2026toric,
  author        = {{Atom Computing and Collaborators}},
  title         = {Quantum error correction with the toric code},
  year          = {2026},
  eprint        = {2606.04079},
  archivePrefix = {arXiv},
  doi           = {10.48550/arXiv.2606.04079}
}

@article{iyer2022diagnostics,
  author        = {Iyer, Pavithran and Jain, Aditya and Bartlett, Stephen D. and Emerson, Joseph},
  title         = {Efficient diagnostics for quantum error correction},
  journal       = {Physical Review Research},
  volume        = {4},
  number        = {4},
  pages         = {043218},
  year          = {2022},
  doi           = {10.1103/PhysRevResearch.4.043218},
  eprint        = {2108.10830},
  archivePrefix = {arXiv}
}

@article{fazio2026characterizing,
  author        = {Fazio, Nicholas and Freund, Robert and Sannamoth, Debankan and Steiner, Alex and Marciniak, Christian D. and Rispler, Manuel and Harper, Robin and Monz, Thomas and Emerson, Joseph and Bartlett, Stephen D.},
  title         = {Characterizing Physical and Logical Errors in a Transversal {CNOT} Gate via Cycle Error Reconstruction},
  journal       = {Physical Review X},
  volume        = {16},
  number        = {1},
  pages         = {011030},
  year          = {2026},
  doi           = {10.1103/qfwc-584t},
  eprint        = {2504.11980},
  archivePrefix = {arXiv}
}

@misc{combes2017logical,
  author        = {Combes, Joshua and Granade, Christopher and Ferrie, Christopher and Flammia, Steven T.},
  title         = {Logical Randomized Benchmarking},
  year          = {2017},
  eprint        = {1702.03688},
  archivePrefix = {arXiv},
  doi           = {10.48550/arXiv.1702.03688}
}

@article{wagner2023learning,
  author        = {Wagner, Thomas and Kampermann, Hermann and Bru{\ss}, Dagmar and Kliesch, Martin},
  title         = {Learning Logical {P}auli Noise in Quantum Error Correction},
  journal       = {Physical Review Letters},
  volume        = {130},
  number        = {20},
  pages         = {200601},
  year          = {2023},
  doi           = {10.1103/PhysRevLett.130.200601},
  eprint        = {2209.09267},
  archivePrefix = {arXiv}
}

@article{harper2023correlated,
  author        = {Harper, Robin and Flammia, Steven T.},
  title         = {Learning Correlated Noise in a 39-Qubit Quantum Processor},
  journal       = {PRX Quantum},
  volume        = {4},
  number        = {4},
  pages         = {040311},
  year          = {2023},
  doi           = {10.1103/PRXQuantum.4.040311},
  eprint        = {2303.00780},
  archivePrefix = {arXiv}
}

@article{miao2023overcoming,
  author  = {Miao, Kevin C. and McEwen, Matt and Atalaya, Juan and others},
  title   = {Overcoming leakage in quantum error correction},
  journal = {Nature Physics},
  volume  = {19},
  pages   = {1780--1786},
  year    = {2023},
  doi     = {10.1038/s41567-023-02226-w}
}

@article{postler2024steane,
  author  = {Postler, Lukas and Butt, Friederike and Pogorelov, Ivan
             and Marciniak, Christian D. and Heu{\ss}en, Sascha
             and Blatt, Rainer and Schindler, Philipp and Rispler, Manuel
             and M{\"u}ller, Markus and Monz, Thomas},
  title   = {Demonstration of Fault-Tolerant Steane Quantum Error Correction},
  journal = {PRX Quantum},
  volume  = {5},
  number  = {3},
  pages   = {030326},
  year    = {2024},
  doi     = {10.1103/PRXQuantum.5.030326},
  eprint  = {2312.09745},
  archivePrefix = {arXiv}
}

@article{lacroix2025scaling,
  author  = {Lacroix, N. and Bourassa, A. and Heras, F. J. H. and others},
  title   = {Scaling and logic in the colour code on a superconducting quantum processor},
  journal = {Nature},
  volume  = {645},
  pages   = {614--619},
  year    = {2025},
  doi     = {10.1038/s41586-025-09061-4}
}

@article{dur2005standard,
  author        = {D{\"u}r, Wolfgang and Hein, Markus and Cirac, J. Ignacio and Briegel, Hans J.},
  title         = {Standard forms of noisy quantum operations via depolarization},
  journal       = {Physical Review A},
  volume        = {72},
  number        = {5},
  pages         = {052326},
  year          = {2005},
  doi           = {10.1103/PhysRevA.72.052326}
}

@article{dankert2009exact,
  author        = {Dankert, Christoph and Cleve, Richard and Emerson, Joseph and Livine, Etera},
  title         = {Exact and approximate unitary 2-designs and their application to fidelity estimation},
  journal       = {Physical Review A},
  volume        = {80},
  number        = {1},
  pages         = {012304},
  year          = {2009},
  doi           = {10.1103/PhysRevA.80.012304}
}

@misc{knill2004threshold,
  author        = {Knill, Emanuel},
  title         = {Fault-tolerant postselected quantum computation: {T}hreshold analysis},
  howpublished  = {arXiv},
  year          = {2004},
  url           = {https://arxiv.org/abs/quant-ph/0404104}
}

@article{dehaene2003clifford,
  author        = {Dehaene, Jeroen and De Moor, Bart},
  title         = {{Clifford} group, stabilizer states, and linear and quadratic operations over {GF}(2)},
  journal       = {Physical Review A},
  volume        = {68},
  number        = {4},
  pages         = {042318},
  year          = {2003},
  doi           = {10.1103/PhysRevA.68.042318}
}

@article{cai2019constructing,
  author        = {Cai, Zhenyu and Benjamin, Simon C.},
  title         = {Constructing smaller {P}auli twirling sets for arbitrary error channels},
  journal       = {Scientific Reports},
  volume        = {9},
  number        = {1},
  pages         = {11281},
  year          = {2019},
  doi           = {10.1038/s41598-019-46722-7},
  eprint        = {1807.04973},
  archivePrefix = {arXiv}
}

@article{erhard2019characterizing,
  author        = {Erhard, Alexander and Wallman, Joel J. and Postler, Lukas and Meth, Michael and Stricker, Roman and Martinez, Esteban A. and Schindler, Philipp and Monz, Thomas and Emerson, Joseph and Blatt, Rainer},
  title         = {Characterizing large-scale quantum computers via cycle benchmarking},
  journal       = {Nature Communications},
  volume        = {10},
  number        = {1},
  pages         = {5347},
  year          = {2019},
  doi           = {10.1038/s41467-019-13068-7},
  eprint        = {1902.08543},
  archivePrefix = {arXiv}
}

@inproceedings{flammia2022averaged,
  author        = {Flammia, Steven T.},
  title         = {Averaged circuit eigenvalue sampling},
  booktitle     = {17th Conference on the Theory of Quantum Computation, Communication and Cryptography ({TQC} 2022)},
  editor        = {Le Gall, Fran{\c{c}}ois and Morimae, Tomoyuki},
  series        = {Leibniz International Proceedings in Informatics ({LIPIcs})},
  volume        = {232},
  pages         = {4:1--4:10},
  publisher     = {Schloss Dagstuhl--Leibniz-Zentrum f{\"u}r Informatik},
  address       = {Dagstuhl, Germany},
  year          = {2022},
  doi           = {10.4230/LIPIcs.TQC.2022.4},
  eprint        = {2108.05803},
  archivePrefix = {arXiv}
}

@article{epstein2014investigating,
  author        = {Epstein, Jeffrey M. and Cross, Andrew W. and Magesan, Easwar and Gambetta, Jay M.},
  title         = {Investigating the limits of randomized benchmarking protocols},
  journal       = {Physical Review A},
  volume        = {89},
  number        = {6},
  pages         = {062321},
  year          = {2014},
  doi           = {10.1103/PhysRevA.89.062321},
  eprint        = {1308.2928},
  archivePrefix = {arXiv}
}

@article{chasseur2015complete,
  author        = {Chasseur, Tobias and Wilhelm, Frank K.},
  title         = {Complete randomized benchmarking protocol accounting for leakage errors},
  journal       = {Physical Review A},
  volume        = {92},
  number        = {4},
  pages         = {042333},
  year          = {2015},
  doi           = {10.1103/PhysRevA.92.042333},
  eprint        = {1505.00580},
  archivePrefix = {arXiv}
}

@article{wallman2016robust,
  author        = {Wallman, Joel J. and Barnhill, Marie and Emerson, Joseph},
  title         = {Robust characterization of leakage errors},
  journal       = {New Journal of Physics},
  volume        = {18},
  number        = {4},
  pages         = {043021},
  year          = {2016},
  doi           = {10.1088/1367-2630/18/4/043021},
  eprint        = {1412.4126},
  archivePrefix = {arXiv}
}

@article{wu2024leakage,
  author        = {Wu, Bujiao and Wang, Xiaoyang and Yuan, Xiao and Huang, Cupjin and Chen, Jianxin},
  title         = {Leakage benchmarking for universal gate sets},
  journal       = {Entropy},
  volume        = {26},
  number        = {1},
  pages         = {71},
  year          = {2024},
  doi           = {10.3390/e26010071},
  eprint        = {2304.07884},
  archivePrefix = {arXiv}
}

\appendix

\renewcommand{\thesubsection}{\thesection.\arabic{subsection}}
\renewcommand{\theequation}{\thesection.\arabic{equation}}




\section{Standard Pauli twirling}
\label{sec:twirling}

The Pauli-twirling construction and its basic properties reviewed in this section are standard in the community~\cite{dur2005standard,dankert2009exact}.
The use of Pauli randomization to simplify noise in quantum circuits is discussed in Refs.~\cite{knill2004threshold,wallman2016noise}.
We include the full derivations in order to establish notation and provide a self-contained treatment.

\subsection{Qubit registers and operator spaces}
\label{subsec:qubit-registers-and-operator-spaces}

\begin{defn}[Qubit Hilbert space]
\label{defn:qubit-Hilbert-space}
A \emph{qubit Hilbert space} is a Hilbert space that is isomorphic to \(\mathbb{C}^2\) and is equipped with a fixed orthonormal basis, called its \emph{computational basis}.
\end{defn}

\begin{defn}[Qubit register]
\label{defn:qubit-register}
A \emph{qubit register} \(\mathsf{N}\) consists of a finite label set (which we also denote \(\mathsf{N}\)) and, for each \(i\in\mathsf{N}\), a qubit Hilbert space \(\mathfrak{q}_i\) whose computational basis is denoted by \(\{\ket{0_i},\ket{1_i}\}\).  
Define the Hilbert space \(\mathcal{K}_{\mathsf{N}}\) of \(\mathsf{N}\) by
\begin{equation}
    \mathcal{K}_{\mathsf{N}}
    =
    \bigotimes_{i\in\mathsf{N}}\mathfrak{q}_i.
    \label{eq:qubit-register-Hilbert-space}
\end{equation}
The subscript of \(\mathcal{K}\) specifies the register's label set.
\end{defn}

Fix a qubit register and associated label set \(\mathsf{N}\).
Define \(\mathcal{A}\) to be \(\operatorname{End}(\mathcal{K}_{\mathsf{N}})\) equipped with the Hilbert--Schmidt inner product
\begin{equation}
    \langle A,B\rangle_{\mathcal{A}}
    =
    \trace(A^\dagger B),
    \label{eq:app-HS-inner-product}
\end{equation}
for all \(A,B\in\mathcal{A}\). 
Define \(\mathcal{S}\) to be \(\operatorname{End}(\mathcal{A})\) equipped with the Hilbert--Schmidt inner product
\begin{equation}
    \langle\Phi,\Psi\rangle_{\mathcal{S}}
    =
    \trace_{\mathcal{A}}(\Phi^\dagger\circ\Psi),
    \label{eq:app-superoperator-inner-product}
\end{equation}
for all \(\Phi,\Psi\in\mathcal{S}\), where \(\Phi^\dagger\) denotes the adjoint with respect to the Hilbert--Schmidt inner product on \(\mathcal{A}\).

\subsection{Pauli-label space and Pauli operators}
\label{subsec:Pauli-operators}

Define \(\mathbb{Z}_2=\{0,1\}\), with arithmetic modulo \(2\).

\begin{defn}[Pauli-label space]
\label{defn:Pauli-label-space}
Define \(\mathbb{Z}_2^{\mathsf{N}}\) to be the \(\mathbb{Z}_2\)-valued vector space of functions \(x:\mathsf{N} \to \mathbb{Z}_2\), with pointwise
operations.  
Define the \emph{Pauli-label space} \(V_{\mathsf{N}}\) of \(\mathsf{N}\) by
\begin{equation}
    V_{\mathsf{N}}
    =
    \mathbb{Z}_2^{\mathsf{N}}
    \oplus
    \mathbb{Z}_2^{\mathsf{N}}.
    \label{eq:app-Pauli-label-space}
\end{equation}
\end{defn}

For \(x\in\mathbb{Z}_2^{\mathsf{N}}\) and \(i\in\mathsf{N}\), write \(x_i=x(i)\).
For \(u,v\in\mathbb{Z}_2^{\mathsf{N}}\), define
\(u\mathbin{\cdot}v\in\mathbb{Z}\) by
\begin{equation}
    u\mathbin{\cdot}v
    =
    \sum_{i\in\mathsf{N}}u_iv_i,
    \label{eq:app-binary-dot-product}
\end{equation}
where the elements of \(\mathbb{Z}_2\) are represented by the integers \(0\) and \(1\) in the sum.

The space \(V_{\mathsf{N}}\) 
is a finite abelian group under componentwise addition.  
For \(a=(x,z)\) and \(b=(x',z')\) in \(V_{\mathsf{N}}\), one has \(a+b=(x+x',z+z')\).  
Define the standard symplectic form \([\cdot,\cdot]\colon V_{\mathsf{N}}\times V_{\mathsf{N}} \to\mathbb{Z}_2\) by
\begin{equation}
    [a,b]
    =
    x\mathbin{\cdot}z'
    +
    z\mathbin{\cdot}x'
    \pmod 2.
    \label{eq:app-symplectic-form}
\end{equation}
The form is nondegenerate: for every \(a\in V_{\mathsf{N}}\setminus\{0\}\), there exists \(v\in V_{\mathsf{N}}\) such that \([v,a]=1\).  
This follows because for \(a=(x,z)\neq0\), one can choose \(j\in\mathsf{N}\) such that either \(z_j=1\) or \(x_j=1\). 
Define \(e_j\in\mathbb{Z}_2^{\mathsf{N}}\) by \((e_j)_i=\delta_{i,j}\).
If \(z_j=1\), then \([(e_j,0),a]=1\); if \(x_j=1\), then \([(0,e_j),a]=1\).
For each \(i\in\mathsf{N}\), define \(X_i,Z_i\in\operatorname{End}(\mathfrak{q}_i)\) by
\begin{equation}
    X_i
    \coloneqq
    \ket{0_i}\!\bra{1_i}
    +
    \ket{1_i}\!\bra{0_i},
    \qquad
    Z_i
    \coloneqq
    \ket{0_i}\!\bra{0_i}
    -
    \ket{1_i}\!\bra{1_i}.
    \label{eq:computational-system-Pauli-generators}
\end{equation}
For \(i\in\mathsf{N}\) and \(a_i=(x_i,z_i)\in\mathbb{Z}_2^2\), define the single-qubit Pauli \(w_i(a_i)\in\operatorname{End}(\mathfrak{q}_i)\) by
\begin{equation}
    w_i(a_i)
    =
    \mathrm{i}^{x_iz_i}X_i^{x_i}Z_i^{z_i}.
    \label{eq:computational-site-Pauli}
\end{equation}
For \(a=(x,z)\in V_{\mathsf{N}}\), write \(a_i=(x_i,z_i)\) for every \(i\in\mathsf{N}\). 

\begin{defn}
\label{defn:Pauli-operator-computational}
A \emph{Pauli operator on \(\mathsf{N}\)} associated with the Pauli label \(a\) is given by
\begin{equation}
    \label{eq:app-Pauli-definition}
    W^{\mathsf{N}}(a)
    =
    \mathrm{i}^{x\mathbin{\cdot}z}
    \bigotimes_{i\in\mathsf{N}}X_i^{x_i}Z_i^{z_i}
    =
    \bigotimes_{i\in\mathsf{N}}w_i(a_i),
\end{equation}
where \(\mathsf{N}\) is a computational (qubit) register.
\end{defn}

When it is unambiguous, we suppress the superscript \(\mathsf{N}\) and write \(W(a)=W^{\mathsf{N}}(a)\).

\begin{lem}
\label{lem:app-Pauli-product}
For all \(a=(x,z)\) and \(b=(x',z')\) in \(V_{\mathsf{N}}\),
\begin{equation}
    W(a)W(b)
    =
    (-1)^{z\mathbin{\cdot}x'}
    \mathrm{i}^{
        x\mathbin{\cdot}z
        +
        x'\mathbin{\cdot}z'
        -
        (x+x')\mathbin{\cdot}(z+z')
    }
    W(a+b).
    \label{eq:app-Pauli-product}
\end{equation}
\end{lem}

\begin{proof}
For each \(i\in\mathsf{N}\), write \(a_i=(x_i,z_i)\) and \(b_i=(x_i',z_i')\).  
The identities \(X_i^2=Z_i^2=\id_{\mathfrak{q}_i}\) and \(Z_iX_i=-X_iZ_i\) give
\begin{align}
    w_i(a_i)w_i(b_i)
    &=
    \mathrm{i}^{x_iz_i+x_i'z_i'}
    (-1)^{z_ix_i'}
    X_i^{x_i+x_i'}Z_i^{z_i+z_i'}
    \\
    &=
    (-1)^{z_ix_i'}
    \mathrm{i}^{
        x_iz_i
        +
        x_i'z_i'
        -
        (x_i+x_i')(z_i+z_i')
    }
    w_i(a_i+b_i).
    \label{eq:single-qubit-Pauli-product-calculation}
\end{align}
Taking tensor products over \(i\in\mathsf{N}\) and using Eq.~\eqref{eq:app-Pauli-definition} gives
\begin{align}
    W(a)W(b)
    &=
    \bigotimes_{i\in\mathsf{N}}w_i(a_i)w_i(b_i)
    \\
    &=
    (-1)^{z\mathbin{\cdot}x'}
    \mathrm{i}^{
        x\mathbin{\cdot}z
        +
        x'\mathbin{\cdot}z'
        -
        (x+x')\mathbin{\cdot}(z+z')
    }
    \bigotimes_{i\in\mathsf{N}}w_i(a_i+b_i)
    \\
    &=
    (-1)^{z\mathbin{\cdot}x'}
    \mathrm{i}^{
        x\mathbin{\cdot}z
        +
        x'\mathbin{\cdot}z'
        -
        (x+x')\mathbin{\cdot}(z+z')
    }
    W(a+b).
    \label{eq:Pauli-product-calculation}
\end{align}
Then, Eq.~\eqref{eq:Pauli-product-calculation} proves Eq.~\eqref{eq:app-Pauli-product}.
\end{proof}

Define \(N=|\mathsf{N}|\)  to be the number of qubits in \(\mathsf{N}\) and define \(d=\dim(\mathcal{K}_{\mathsf{N}})=2^N\), so that \(|V_{\mathsf{N}}|=4^N=d^2=\dim\mathcal{A}\).

\begin{lem}
\label{lem:app-Pauli-identities}
For all \(a,b\in V_{\mathsf{N}}\), the Pauli operators satisfy
\begin{align}
    W(a)^\dagger
    &=
    W(a),
    \label{eq:app-Pauli-Hermiticity}
    \\
    W(a)^2
    &=
    \id_{\mathcal{K}_{\mathsf{N}}},
    \label{eq:app-Pauli-involution}
    \\
    W(b)^\dagger W(a)W(b)
    &=
    (-1)^{[b,a]}W(a),
    \label{eq:app-Pauli-conjugation}
    \\
    \trace\!\left(W(a)W(b)\right)
    &=
    d\,\delta_{a,b}.
    \label{eq:app-Pauli-orthogonality}
\end{align}
Consequently, each \(W(a)\) is unitary and Hermitian, and the family \(\{d^{-1/2}W(a)\}_{a\in V_{\mathsf{N}}}\) is an orthonormal basis for \(\mathcal{A}\).
\end{lem}

\begin{proof}
Eqs.~\eqref{eq:computational-system-Pauli-generators}, \eqref{eq:computational-site-Pauli}, and \eqref{eq:app-Pauli-definition} give Eq.~\eqref{eq:app-Pauli-Hermiticity} and \(\trace(W(c))=d\,\delta_{c,0}\) for every \(c\in V_{\mathsf{N}}\).
Setting \(b=a\) in Eq.~\eqref{eq:app-Pauli-product} gives Eq.~\eqref{eq:app-Pauli-involution}, since \((-1)^{z\mathbin{\cdot}x}\mathrm{i}^{2x\mathbin{\cdot}z}=1\).

The exponent of \(\mathrm{i}\) in Eq.~\eqref{eq:app-Pauli-product} is unchanged when \(a\) and \(b\) are interchanged.  
Comparing the two instances of Eq.~\eqref{eq:app-Pauli-product} and using Eq.~\eqref{eq:app-symplectic-form} gives
\(W(b)W(a)=(-1)^{[b,a]}W(a)W(b)\). 
Combining this equality with Eqs.~\eqref{eq:app-Pauli-Hermiticity} and \eqref{eq:app-Pauli-involution} gives Eq.~\eqref{eq:app-Pauli-conjugation}.

If \(a=b\), Eq.~\eqref{eq:app-Pauli-involution} gives \(\trace(W(a)W(b))=d\).  
If \(a\neq b\), then \(a+b\neq0\), and Eq.~\eqref{eq:app-Pauli-product} together with \(\trace(W(a+b))=0\) gives \(\trace(W(a)W(b))=0\). 
Hence, Eq.~\eqref{eq:app-Pauli-orthogonality} holds.

Eqs.~\eqref{eq:app-Pauli-Hermiticity} and \eqref{eq:app-Pauli-involution} show that each \(W(a)\) is unitary.
Eqs.~\eqref{eq:app-Pauli-Hermiticity} and \eqref{eq:app-Pauli-orthogonality} give
\begin{equation*}
    \langle W(a),W(b)\rangle_{\mathcal{A}}
    =
    d\,\delta_{a,b},
\end{equation*}
for all \(a,b\in V_{\mathsf{N}}\). 
Since \(|V_{\mathsf{N}}|=\dim\mathcal{A}\), 
 \(\{d^{-1/2}W(a)\}_{a\in V_{\mathsf{N}}}\) is an orthonormal basis for \(\mathcal{A}\).
\end{proof}

Accordingly, we simply write \(W(a)\) in lieu of \( W(a)^\dagger \) or \(W(a)^{-1}\) going forward. 

\subsection{Pauli representations on operator spaces}
\label{subsec:Pauli-conjugation}

For a finite-dimensional Hilbert space \(\mathcal{X}\) and \(U\in\operatorname{End}(\mathcal{X})\), define \(\operatorname{Ad}_U:\operatorname{End}(\mathcal{X}) \to\operatorname{End}(\mathcal{X})\) by \(\operatorname{Ad}_U(A)=UAU^\dagger\).
Next, define the map
\(\mathcal{W}^{\mathsf{N}}:V_{\mathsf{N}}\to\mathcal{S}\) by
\begin{equation}
    \mathcal{W}^{\mathsf{N}}(v)(\cdot)
    =
    \operatorname{Ad}_{W^{\mathsf{N}}(v)}(\cdot)
    =
    W^{\mathsf{N}}(v)(\cdot)W^{\mathsf{N}}(v),
    \label{eq:app-Pauli-conjugation-action}
\end{equation}
for all \(v\in V_{\mathsf{N}}\), where the dot denotes an arbitrary operator in \(\mathcal{A}\).

When the register is unambiguous, we suppress its superscript
and write \(\mathcal{W}\) for \(\mathcal{W}^{\mathsf{N}}\).

\begin{lem}
\label{lem:app-Pauli-representation}
The map \(\mathcal{W}\) is a unitary representation of the additive group \(V_{\mathsf{N}}\) on \(\mathcal{A}\).
\end{lem}

\begin{proof}
Eq.~\eqref{eq:app-Pauli-product} shows that \(W(u)W(v)\) is a unit-modulus scalar multiple of \(W(u+v)\).  
Therefore, Eqs.~\eqref{eq:app-Pauli-conjugation-action}, \eqref{eq:app-adjoint-action}, and \eqref{eq:app-adjoint-action-composition} give
\begin{align*}
    \mathcal{W}(u)\circ\mathcal{W}(v)
    &=
    \operatorname{Ad}_{W(u)W(v)}
    =
    \operatorname{Ad}_{W(u+v)}
    =
    \mathcal{W}(u+v), 
\end{align*}
for all \( u,v\in V_{\mathsf{N}}\). 
Also, \(W(0)=\id_{\mathcal{K}_{\mathsf{N}}}\), so \(\mathcal{W}(0)=\operatorname{id}_{\mathcal{A}}\).
Thus, \(\mathcal{W}\) is a representation.
Furthermore, Eqs.~\eqref{eq:app-Pauli-Hermiticity} and \eqref{eq:app-Pauli-involution} show that each \(W(v)\) is unitary, while  Eq.~\eqref{eq:app-Pauli-conjugation-action} and Lemma~\ref{lem:app-unitary-adjoint-action} show that each \(\mathcal{W}(v)\) is unitary on \(\mathcal{A}\).
\end{proof}

Define the map \(\Gamma:V_{\mathsf{N}}\to\operatorname{End}(\mathcal{S})\) by
\begin{equation}
    \Gamma(v)(\Phi)
    =
    \mathcal{W}(v) 
    \circ\Phi\circ
    \mathcal{W}(v),
    \label{eq:app-induced-superoperator-representation}
\end{equation}
for all \( v\in V_{\mathsf{N}},\ \Phi\in\mathcal{S}\). 

\begin{lem}
\label{lem:app-induced-superoperator-representation}
The map \(\Gamma\) is a unitary representation of the additive group \(V_{\mathsf{N}}\) on \(\mathcal{S}\).
\end{lem}

\begin{proof}
By Lemma~\ref{lem:app-Pauli-representation}, \(\mathcal{W}(v)^\dagger\) is unitary on \(\mathcal{A}\) for every \(v\in V_{\mathsf{N}}\).  
Therefore, Eqs.~\eqref{eq:app-adjoint-action} and \eqref{eq:app-induced-superoperator-representation}, together with Lemma~\ref{lem:app-unitary-adjoint-action}, show that \(\Gamma(v)\) is unitary on \(\mathcal{S}\) for every \(v\in V_{\mathsf{N}}\).
Lemma~\ref{lem:app-Pauli-representation} and the commutativity of \(V_{\mathsf{N}}\) give \(\mathcal{W}(v)\circ\mathcal{W}(u)=\mathcal{W}(v+u)=\mathcal{W}(u+v)\).
Therefore, for \(u,v\in V_{\mathsf{N}}\) and \(\Phi\in\mathcal{S}\), Eq.~\eqref{eq:app-induced-superoperator-representation} gives
\begin{align*}
    \left(\Gamma(u)\circ\Gamma(v)\right)(\Phi)
    &=
    \mathcal{W}(u) 
    \circ\mathcal{W}(v) 
    \circ\Phi\circ
    \mathcal{W}(v)
    \circ\mathcal{W}(u)
    \\
    &=
    \mathcal{W}(u+v) 
    \circ\Phi\circ
    \mathcal{W}(u+v)
    \\
    &=
    \Gamma(u+v)(\Phi).
\end{align*}
Finally, Eq.~\eqref{eq:app-induced-superoperator-representation} and \(\mathcal{W}(0)=\operatorname{id}_{\mathcal{A}}\) give \(\Gamma(0)=\operatorname{id}_{\mathcal{S}}\).
\end{proof}

\subsection{Standard Pauli twirl and Pauli maps}
\label{subsec:standard-Pauli-twirl}

\begin{lem}
\label{lem:finite-group-averaging-projection}
Let \(\mathcal{X}\) be a finite-dimensional complex Hilbert space, let \(G\) be a finite group, and let \(\rho:G\to\operatorname{U}(\mathcal{X})\) be a unitary representation. 
Define the operator \(P_G\in\operatorname{End}(\mathcal{X})\) by
\begin{equation}
    P_G
    =
    \frac{1}{|G|}\sum_{g\in G}\rho(g),
    \label{eq:finite-group-averaging-operator}
\end{equation}
and define the invariant subspace \(\mathcal{X}^G\subseteq\mathcal{X}\) by
\begin{equation}
    \mathcal{X}^G
    =
    \left\{
        x\in\mathcal{X}:
        \rho(g)x=x
        \text{ for every }g\in G
    \right\}.
    \label{eq:finite-group-invariant-subspace}
\end{equation}
Then \(P_G\) is the orthogonal projection onto \(\mathcal{X}^G\).
\end{lem}

\begin{proof}
Left multiplication permutes the elements of \(G\), so Eq.~\eqref{eq:finite-group-averaging-operator} gives \(\rho(g)P_G=P_G\) for every \(g\in G\).  
Thus, \(\operatorname{ran}(P_G)\subseteq\mathcal{X}^G\).  
Conversely, Eq.~\eqref{eq:finite-group-averaging-operator} gives \(P_Gx=x\) for every \(x\in\mathcal{X}^G\).  
Finally, unitarity of \(\rho\) and the fact that inversion permutes the elements of \(G\) give
\begin{equation*}
    P_G^\dagger
    =
    \frac{1}{|G|}\sum_{g\in G}\rho(g^{-1})
    =
    P_G.
\end{equation*}
Hence, \(P_G\) is the orthogonal projection onto \(\mathcal{X}^G\).
\end{proof}

\begin{defn}[Standard Pauli twirl]
\label{defn:standard-Pauli-twirl}
Define the \emph{standard Pauli twirl} \(\mathcal{T}^{\mathsf{N}}:\mathcal{S}\to\mathcal{S}\) by:
\begin{equation}
    \mathcal{T}^{\mathsf{N}}(\Phi)
    =
    \frac{1}{|V_{\mathsf{N}}|}
    \sum_{v\in V_{\mathsf{N}}}
    \Gamma(v)(\Phi),
    \label{eq:standard-twirl-def}
\end{equation}
for all \(\Phi\in\mathcal{S}\). 
\end{defn}

When the register is unambiguous, we suppress its superscript and write \(\mathcal{T}\) for \(\mathcal{T}^{\mathsf{N}}\).

Define the invariant subspace \(\mathcal{S}^{V_{\mathsf{N}}}\subseteq\mathcal{S}\) of \(\Gamma\) by
\begin{equation}
    \mathcal{S}^{V_{\mathsf{N}}}
    =
    \left\{
        \Phi\in\mathcal{S}:
        \Gamma(v)(\Phi)=\Phi
        \text{ for every }v\in V_{\mathsf{N}}
    \right\}.
    \label{eq:app-invariant-superoperator-definition}
\end{equation}

\begin{lem}
\label{lem:app-invariant-Pauli-superoperators}
The standard Pauli twirl \(\mathcal{T}\) is the orthogonal projection onto \(\mathcal{S}^{V_{\mathsf{N}}}\).  
Moreover, \(\mathcal{S}^{V_{\mathsf{N}}}\) can be expressed as
\begin{equation}
    \mathcal{S}^{V_{\mathsf{N}}}
    =
    \operatorname{span}_{\mathbb{C}}
    \left\{
        \mathcal{W}(e) : e\in V_{\mathsf{N}}
    \right\},
    \label{eq:app-invariant-superoperator-space}
\end{equation}
and the family \(\{ 
\mathcal{W}(e) \}_{e\in V_{\mathsf{N}}}\) is an orthogonal basis for \(\mathcal{S}^{V_{\mathsf{N}}}\) satisfying
\begin{equation}
    \left\langle
        \mathcal{W}(e), \mathcal{W}(f)
    \right\rangle_{\mathcal{S}}
    =
    d^2\delta_{e,f},
    \label{eq:app-Pauli-superoperator-orthogonality}
\end{equation}
for all \(e,f\in V_{\mathsf{N}}\). 
\end{lem}

\begin{proof}
Apply Lemma~\ref{lem:finite-group-averaging-projection} to the unitary representation \(\Gamma\) from Lemma~\ref{lem:app-induced-superoperator-representation}, with \(G=V_{\mathsf{N}}\), \(\mathcal{X}=\mathcal{S}\), and \(\rho=\Gamma\).  
By Eqs.~\eqref{eq:standard-twirl-def} and \eqref{eq:app-invariant-superoperator-definition}, the resulting averaging operator and invariant subspace are \(\mathcal{T}\) and \(\mathcal{S}^{V_{\mathsf{N}}}\), respectively.  
Thus, \(\mathcal{T}\) is the orthogonal projection onto \(\mathcal{S}^{V_{\mathsf{N}}}\), and \(\dim(\mathcal{S}^{V_{\mathsf{N}}})=\trace_{\mathcal{S}}(\mathcal{T})\).

For every \(v\in V_{\mathsf{N}}\), applying Eqs.~\eqref{eq:app-induced-superoperator-representation} and \eqref{eq:app-elementary-superoperator-definition} with \(\mathcal{X}=\mathcal{A}\) identifies \(\Gamma(v)\) with \(\mathcal{L}_{\mathcal{W}(v) 
,\mathcal{W}(v)}\).
Therefore, Eq.~\eqref{eq:app-elementary-superoperator-trace} gives
\begin{align*}
    \trace_{\mathcal{S}}(\Gamma(v))
    &=
    \trace_{\mathcal{A}}
    \!\left(\mathcal{W}(v) 
    \right)
    \trace_{\mathcal{A}}
    \!\left(\mathcal{W}(v)\right)
    =
    \left|
        \trace_{\mathcal{A}}(\mathcal{W}(v))
    \right|^2.
\end{align*}
Similarly, Eqs.~\eqref{eq:app-Pauli-conjugation-action} and \eqref{eq:app-elementary-superoperator-definition}, applied with \(\mathcal{X}=\mathcal{K}_{\mathsf{N}}\), identify \(\mathcal{W}(v)\) with \(\mathcal{L}_{W(v),W(v) 
}\).  
Hence, Eq.~\eqref{eq:app-elementary-superoperator-trace} gives
\begin{equation*}
    \trace_{\mathcal{A}}(\mathcal{W}(v))
    =
    \trace_{\mathcal{K}_{\mathsf{N}}}(W(v))
    \trace_{\mathcal{K}_{\mathsf{N}}}(W(v) 
    )
    =
    \left|
        \trace_{\mathcal{K}_{\mathsf{N}}}(W(v))
    \right|^2.
\end{equation*}
Consequently,
\begin{equation*}
    \trace_{\mathcal{S}}(\Gamma(v))
    =
    \left|
        \trace_{\mathcal{K}_{\mathsf{N}}}(W(v))
    \right|^4.
\end{equation*}
Since \(W(0)=\id_{\mathcal{K}_{\mathsf{N}}}\), applying Eq.~\eqref{eq:app-Pauli-orthogonality} with the second label zero gives \(\trace_{\mathcal{K}_{\mathsf{N}}}(W(v)) =d\,\delta_{v,0}\) for every \(v\in V_{\mathsf{N}}\).
Taking \(\trace_{\mathcal{S}}\) in Eq.~\eqref{eq:standard-twirl-def} and using \(\trace_{\mathcal{S}}(\Gamma(v)) =|\trace_{\mathcal{K}_{\mathsf{N}}}(W(v))|^4\), \(\trace_{\mathcal{K}_{\mathsf{N}}}(W(v)) =d\,\delta_{v,0}\), and \(|V_{\mathsf{N}}|=d^2\) gives
\begin{align}
    \dim(\mathcal{S}^{V_{\mathsf{N}}})
    =
    \frac{1}{|V_{\mathsf{N}}|}
    \sum_{v\in V_{\mathsf{N}}}
    \trace_{\mathcal{S}}(\Gamma(v))
    =
    \frac{d^4}{|V_{\mathsf{N}}|}
    =
    |V_{\mathsf{N}}|.
    \label{eq:app-invariant-superoperator-dimension-bound}
\end{align}
For \(e,v\in V_{\mathsf{N}}\), Eqs.~\eqref{eq:app-Pauli-conjugation-action} and \eqref{eq:app-induced-superoperator-representation}, together with Lemma~\ref{lem:app-Pauli-representation} and the commutativity of \(V_{\mathsf{N}}\), give
\begin{align*}
    \Gamma(v)\bigl( 
    \mathcal{W}(e) \bigr)
    =
    \mathcal{W}(v) 
    \circ\mathcal{W}(e)
    \circ\mathcal{W}(v)
    =
    \mathcal{W}(v) 
    \circ\mathcal{W}(v)
    \circ\mathcal{W}(e)
    =
    \mathcal{W}(e)
    .
\end{align*}
Thus, \( 
\mathcal{W}(e) \in\mathcal{S}^{V_{\mathsf{N}}}\) for every \(e\in V_{\mathsf{N}}\).
Then, Eqs.~\eqref{eq:app-adjoint-action-inner-product}, \eqref{eq:app-Pauli-Hermiticity}, and \eqref{eq:app-Pauli-orthogonality}, applied with
\(\mathcal{X}=\mathcal{K}_{\mathsf{N}}\), give
\begin{align*}
    \left\langle
        \mathcal{W}(e),\mathcal{W}(f)
    \right\rangle_{\mathcal{S}}
    &=
    \left|
        \trace_{\mathcal{K}_{\mathsf{N}}}
        \!\left(W(e) 
        W(f)\right)
    \right|^2
    =
    d^2\delta_{e,f}.
\end{align*}
Finally, Eq.~\eqref{eq:app-invariant-superoperator-dimension-bound} shows that \(\mathcal{S}^{V_{\mathsf{N}}}\) has dimension
\(|V_{\mathsf{N}}|\).  
Thus, \(\{
\mathcal{W}(e)\}_{e\in V_{\mathsf{N}}}\) is an orthogonal basis for \(\mathcal{S}^{V_{\mathsf{N}}}\).
\end{proof}

\begin{defn}[Pauli map]
\label{defn:pauli-map}
A \emph{Pauli map} on \(\mathcal{A}\) is a map
\(\mathcal{D}\in\mathcal{S}\) for which there exists a family
\(\{q_e\}_{e\in V_{\mathsf{N}}}\subseteq\mathbb{C}\) such that
\begin{equation}
    \mathcal{D}
    =
    \sum_{e\in V_{\mathsf{N}}}
    q_e\,
    \mathcal{W}(e).
    \label{eq:pauli-map-def}
\end{equation}
\end{defn}

\begin{lem}
\label{lem:pauli-map-structure}
Let \(\mathcal{D}\) be a Pauli map.  
There is a unique family \(\{q_e\}_{e\in V_{\mathsf{N}}}\subseteq\mathbb{C}\) satisfying Eq.~\eqref{eq:pauli-map-def}.  
Fo every \(e\in V_{\mathsf{N}}\), 
\begin{equation}
    q_e
    =
    \frac{1}{d^2}
    \left\langle
\mathcal{W}(e),
        \mathcal{D}
    \right\rangle_{\mathcal{S}}.
    \label{eq:pauli-map-coefficient-extraction}
\end{equation}
\end{lem}

\begin{proof}
Let \(\{r_e\}_{e\in V_{\mathsf{N}}}\subseteq\mathbb{C}\) satisfy \(\mathcal{D}=\sum_{e\in V_{\mathsf{N}}}r_e 
\mathcal{W}(e)\).
For every \(f\in V_{\mathsf{N}}\), Eq.~\eqref{eq:app-Pauli-superoperator-orthogonality} gives
\begin{align*}
    \frac{1}{d^2}
    \left\langle
        \mathcal{W}(f),
        \mathcal{D}
    \right\rangle_{\mathcal{S}}
    &=
    \frac{1}{d^2}
    \sum_{e\in V_{\mathsf{N}}}
    r_e
    \left\langle
    \mathcal{W}(f), \mathcal{W}(e)
    \right\rangle_{\mathcal{S}}
    =
    \sum_{e\in V_{\mathsf{N}}}r_e\delta_{f,e}
    =
    r_f.
\end{align*}
Thus, every family satisfying Eq.~\eqref{eq:pauli-map-def} satisfies Eq.~\eqref{eq:pauli-map-coefficient-extraction}, and the family is unique.
\end{proof}

\begin{prop}
\label{prop:standard-twirl-Pauli-expansion}
Let \(\mathcal{E}\in\mathcal{S}\).  For \(e\in V_{\mathsf{N}}\), define
\(q_e\in\mathbb{C}\) by
\begin{equation}
    q_e
    =
    \frac{1}{d^2}
    \left\langle
    \mathcal{W}(e),
        \mathcal{E}
    \right\rangle_{\mathcal{S}}.
    \label{eq:standard-twirl-general-coefficient}
\end{equation}
Then \(\mathcal{T}(\mathcal{E})\) 
is a Pauli map satisfying
\begin{equation}
    \mathcal{T}(\mathcal{E})
    =
    \sum_{e\in V_{\mathsf{N}}}
    q_e\,
\mathcal{W}(e).
    \label{eq:standard-twirl-form}
\end{equation}
\end{prop}

\begin{proof}
By Lemma~\ref{lem:app-invariant-Pauli-superoperators}, the family \( \{d^{-1}
\mathcal{W}(e)\}_{e\in V_{\mathsf{N}}} \) is an orthonormal basis for \(\mathcal{S}^{V_{\mathsf{N}}}\) and \(\mathcal{T}\) is the orthogonal projection onto \(\mathcal{S}^{V_{\mathsf{N}}}\).  
Therefore, for every \(\mathcal{E}\in\mathcal{S}\), the orthogonal-projection formula gives
\begin{align*}
    \mathcal{T}(\mathcal{E})
    &=
    \sum_{e\in V_{\mathsf{N}}}
    \left\langle
        d^{-1}
\mathcal{W}(e),
        \mathcal{E}
    \right\rangle_{\mathcal{S}}
    d^{-1}
\mathcal{W}(e)
    =
    \sum_{e\in V_{\mathsf{N}}}
    \frac{1}{d^2}
    \left\langle
\mathcal{W}(e),
        \mathcal{E}
    \right\rangle_{\mathcal{S}}
\mathcal{W}(e).
\end{align*}
\end{proof}

\subsection{Pauli channels and subnormalized Pauli maps}
\label{subsec:Pauli-map-normalization-properties}

\begin{defn}[Weight of a Pauli map]
\label{defn:pauli-map-weight}
For a Pauli map \(\mathcal{D}\), define its \emph{weight} by
\begin{equation}
    \operatorname{wt}(\mathcal{D})
    =
    \frac{1}{d}
    \trace\!\left(
        \mathcal{D}(\id_{\mathcal{K}_{\mathsf{N}}})
    \right).
    \label{eq:pauli-map-weight}
\end{equation}
\end{defn}

\begin{lem}
\label{lem:pauli-map-weight-and-trace-scaling}
Let \(\mathcal{D}\) be a Pauli map, and let \(\{q_e\}_{e\in V_{\mathsf{N}}}\subseteq\mathbb{C}\) be its unique coefficient family from Lemma~\ref{lem:pauli-map-structure}.  
Then
\begin{equation}
    \operatorname{wt}(\mathcal{D})
    =
    \sum_{e\in V_{\mathsf{N}}}q_e.
    \label{eq:pauli-map-weight-sum}
\end{equation}
Consequently, for all \(A\in\mathcal{A}\), 
\begin{equation}
    \trace(\mathcal{D}(A))
    =
    \operatorname{wt}(\mathcal{D})\trace(A).
    \label{eq:pauli-map-trace-scaling}
\end{equation}
\end{lem}

\begin{proof}
Eqs.~\eqref{eq:app-adjoint-action}, \eqref{eq:app-Pauli-Hermiticity}, and \eqref{eq:app-Pauli-involution} give \(
\mathcal{W}(e)(\id_{\mathcal{K}_{\mathsf{N}}}) =\id_{\mathcal{K}_{\mathsf{N}}}\) for every \(e\in V_{\mathsf{N}}\).
Substituting Eq.~\eqref{eq:pauli-map-def} into Eq.~\eqref{eq:pauli-map-weight} therefore gives Eq.~\eqref{eq:pauli-map-weight-sum}.  
Finally, unitary invariance of the trace, Eq.~\eqref{eq:pauli-map-def}, and Eq.~\eqref{eq:pauli-map-weight-sum} give Eq.~\eqref{eq:pauli-map-trace-scaling}.
\end{proof}

\begin{defn}[Trace-nonincreasing map]
\label{defn:trace-nonincreasing-map}
A positive linear map \(\mathcal{E}\in\mathcal{S}\) is \emph{trace nonincreasing} if
\begin{equation}
    \trace(\mathcal{E}(X))
    \leq
    \trace(X)~~
    \text{for every}~~X\succeq0.
    \label{eq:trace-nonincreasing-definition}
\end{equation}
\end{defn}

\begin{defn}[Pauli channels and subnormalized Pauli maps]
\label{defn:pauli-map-normalization}
A Pauli map is a \emph{Pauli channel} if it is completely positive and trace preserving (CPTP).  A Pauli map is a \emph{subnormalized Pauli map} if it is completely positive and trace nonincreasing.
\end{defn}

\begin{lem}
\label{lem:pauli-map-positivity-and-normalization}
Let \(\mathcal{D}\) be a Pauli map with coefficient family \(\{q_e\}_{e\in V_{\mathsf{N}}}\).  
Then \(\mathcal{D}\) is a Pauli channel if and only if
\begin{equation}
    q_e\in[0,1] ~~\mathrm{for~all}~e\in V_{\mathsf{N}}~~\mathrm{and}~~
    \sum_{e\in V_{\mathsf{N}}} q_e=1,
    \label{eq:pauli-channel-coefficients}
\end{equation}
and \(\mathcal{D}\) is a subnormalized Pauli map if and only if
\begin{equation}
    q_e\in[0,1] ~\mathrm{for~all}~e\in V_{\mathsf{N}} ~~\mathrm{and}~~
    \sum_{e\in V_{\mathsf{N}}}q_e\leq1.
    \label{eq:subnormalized-pauli-map-coefficients}
\end{equation}
\end{lem}

\begin{proof}
Suppose first that \(q_e\geq0\) for every \(e\in V_{\mathsf{N}}\).
Then Eq.~\eqref{eq:pauli-map-def} has the Kraus form
\begin{equation}
    \mathcal{D}(A)
    =
    \sum_{e\in V_{\mathsf{N}}}
    \bigl(\sqrt{q_e}W(e)\bigr)
    A
    \bigl(\sqrt{q_e}W(e)\bigr)^\dagger,
    \label{eq:pauli-map-kraus-form}
\end{equation}
for all \(A\in\mathcal{A}\), so \(\mathcal{D}\) is completely positive.

Conversely, suppose that \(\mathcal{D}\) is completely positive.  
By the finite-dimensional Kraus representation theorem, there exist \(r\in\mathbb{N}\) and a family \(\{A_\mu\}_{\mu=1}^r\subseteq\mathcal{A}\) such that
\begin{equation}
    \mathcal{D}(X)
    =
    \sum_{\mu=1}^r A_\mu^{\vpd} X A_\mu^\dagger,
    \label{eq:app-Kraus-representation}
\end{equation}
for all \(X\in\mathcal{A}\). 
Next, substituting Eq.~\eqref{eq:app-Kraus-representation} into Eq.~\eqref{eq:pauli-map-coefficient-extraction} and using Lemma~\ref{lem:app-unitary-adjoint-action} and Eqs.~\eqref{eq:app-superoperator-inner-product} and \eqref{eq:app-elementary-superoperator-trace} gives
\begin{align}
    q_e
    &=
    \frac{1}{d^2}
    \sum_{\mu=1}^r
    \trace_{\mathcal{A}}
    \!\left[
        X\longmapsto
        W(e)^\dagger A_\mu^{\vpd} X A_\mu^\dagger W(e)
    \right]
    \nonumber\\
    &=
    \frac{1}{d^2}
    \sum_{\mu=1}^r
    \trace\!\left(W(e)^\dagger A_\mu^{\vpd} \right)
    \trace\!\left(A_\mu^\dagger W(e)\right)
    \nonumber\\
    &=
    \frac{1}{d^2}
    \sum_{\mu=1}^r
    \left|
        \trace\!\left(W(e)^\dagger A_\mu^{\vpd} \right)
    \right|^2
    \geq
    0.
    \label{eq:standard-twirl-projection-coefficient}
\end{align}
Thus, \(\mathcal{D}\) is completely positive if and only if \(q_e\geq0\) for every \(e\in V_{\mathsf{N}}\).

Next, Eqs.~\eqref{eq:pauli-map-weight-sum} and \eqref{eq:pauli-map-trace-scaling} show that \(\mathcal{D}\) is trace preserving if and only if \(\sum_{e\in V_{\mathsf{N}}}q_e=1\).  
If \(\mathcal{D}\) is completely positive, the same equations and Definition~\ref{defn:trace-nonincreasing-map} show that \(\mathcal{D}\) is trace nonincreasing if and only if \(\sum_{e\in V_{\mathsf{N}}}q_e\leq1\). 
In either case, nonnegativity and the bound on the sum imply that \(q_e\leq1\) for every \(e\in V_{\mathsf{N}}\). 
The lemma follows from Definition~\ref{defn:pauli-map-normalization}.
\end{proof}

\begin{prop}
\label{prop:standard-twirl-positivity}
The standard Pauli twirl \(\mathcal{T}\) 
preserves the classes of completely positive, trace-preserving, and trace-nonincreasing maps.
\end{prop}

\begin{proof}
For \(\mathcal{E}\in\mathcal{S}\) and \(v\in V_{\mathsf{N}}\), Eqs.~\eqref{eq:app-Pauli-conjugation-action}, \eqref{eq:app-induced-superoperator-representation}, and \eqref{eq:app-adjoint-action-adjoint} give
\begin{equation*}
    \Gamma(v)(\mathcal{E})
    =
    \mathcal{W}(v)
    \circ
    \mathcal{E}
    \circ
\mathcal{W}(v).
\end{equation*}
Because each adjoint action in this expression is conjugation by a
unitary, the transformation
\(\mathcal{E}\longmapsto\Gamma(v)(\mathcal{E})\) preserves complete
positivity, trace preservation, and the class of trace-nonincreasing
maps.  The average in Eq.~\eqref{eq:standard-twirl-def} therefore
preserves these classes.
\end{proof}

\section{Pauli twirling with respect to a Clifford}
\label{sec:Clifford-twisted-Pauli-twirl}

The material reviewed in this section is standard in the community~\cite{dehaene2003clifford,knill2004threshold,wallman2016noise}.
We include the full derivations in order to establish notation and provide a self-contained treatment.

\begin{defn}[Clifford unitary]
\label{defn:clifford-unitary}
A unitary \(C\) on \(\mathcal{K}_{\mathsf{N}}\) is a \emph{Clifford unitary} if there exist a symplectic automorphism \(F_C:V_{\mathsf{N}}\to V_{\mathsf{N}}\) and a map \(\eta_C:V_{\mathsf{N}}\to\{+1,-1\}\) such that
\begin{equation}
    \mathcal{C} (W(a)) = CW(a)C^\dagger
    =
    \eta_C(a)W(F_Ca),
    \label{eq:app-Clifford-action}
\end{equation}
for all labels \(a\in V_{\mathsf{N}}\), using the convention of Eq.~\eqref{eq:adjoint-unitary-convention} for \(\mathcal{C}\). 
\end{defn}

Fix a Clifford unitary \(C\), a symplectic automorphism \(F_C:V_{\mathsf{N}}\to V_{\mathsf{N}}\), and a map \(\eta_C:V_{\mathsf{N}}\to\{+1,-1\}\) satisfying Eq.~\eqref{eq:app-Clifford-action}.  
The maps \(F_C\) and \(\eta_C\) are unique. 
To see this, suppose that a symplectic automorphism \(\widetilde F_C:V_{\mathsf{N}}\to V_{\mathsf{N}}\) and a map \(\widetilde\eta_C:V_{\mathsf{N}}\to\{+1,-1\}\) also satisfy Eq.~\eqref{eq:app-Clifford-action}. 
Then, for every \(a\in V_{\mathsf{N}}\), the two instances of that equation give \(\eta_C(a)W(F_Ca)=\widetilde\eta_C(a)W(\widetilde F_Ca)\).
Equation~\eqref{eq:app-Pauli-orthogonality} then gives \(F_Ca=\widetilde F_Ca\), and the same equality leads to \(\eta_C(a)=\widetilde\eta_C(a)\).

\begin{defn}[Pauli twirl with respect to \(C\)]
\label{defn:clifford-twisted-Pauli-twirl}
Define the \emph{Pauli twirl with respect to \(C\)}
\(\mathcal{T}^{\mathsf{N}}_C:\mathcal{S}\to\mathcal{S}\) by
\begin{equation}
    \mathcal{T}^{\mathsf{N}}_C(\Phi)
    =
    \frac{1}{|V_{\mathsf{N}}|}
    \sum_{v\in V_{\mathsf{N}}}
    \mathcal{W}(F_C v)
    \circ
    \Phi
    \circ
\mathcal{W}(v),
    \label{eq:clifford-twirl-def}
\end{equation}
for all \(\Phi\in\mathcal{S}\). 
\end{defn}

When the register is unambiguous, we similarly write \(\mathcal{T}_C\) for \(\mathcal{T}^{\mathsf{N}}_C\).

For a unitary operator \( U \in \operatorname{End}(\mathcal{H})\), define the associated maps
\begin{equation}
    \label{eq:adjoint-unitary-convention}
    \mathcal{U} (Y) \coloneqq U Y U^\dagger  \qquad \text{and} \qquad \mathcal{U}^{-1} (Y) \coloneqq U^{\dagger} Y U , 
\end{equation}
for all \( Y \in \operatorname{End}(\mathcal{H})\). 
The convention is that the standard font denotes the unitary and the caligraphic font denotes its adjoint action, as described above. 

Define the map \(R_C:\mathcal{S}\to\mathcal{S}\) by
\begin{equation}
    R_C(\Phi)
    =
    \Phi\circ \mathcal{C}^\dagger
    \label{eq:clifford-right-composition}
\end{equation}
for all \(\Phi\in\mathcal{S}\) and any unitary \(C\) with associated adjoint map \(\mathcal{C}^{\dagger}\).

\begin{lem}
\label{lem:clifford-twirl-standard-reduction}
The map \(R_C\) is unitary on \(\mathcal{S}\).  
Its inverse is given by
\begin{equation}
    R_C^{-1}(\Phi)
    =
    \Phi\circ \mathcal{C},
    \label{eq:clifford-right-composition-inverse}
\end{equation}
for all \( \Phi\in\mathcal{S}\), and
\begin{equation}
    \mathcal{T}_C
    =
    R_C^{-1}\circ\mathcal{T}\circ R_C.
    \label{eq:twisted-standard-conjugacy}
\end{equation}
\end{lem}

\begin{proof}
For \(\Phi,\Psi\in\mathcal{S}\), Eqs.~\eqref{eq:app-superoperator-inner-product}, \eqref{eq:app-adjoint-action-composition}, and
\eqref{eq:app-adjoint-action-adjoint}, together with cyclicity of \(\trace_{\mathcal{A}}\), give
\begin{align}
    \left\langle R_C(\Phi),R_C(\Psi)\right\rangle_{\mathcal{S}}
    &=
    \trace_{\mathcal{A}}
    \!\left(
        \mathcal{C} \circ \Phi^\dagger \circ \Psi \circ \mathcal{C}^{-1}
    \right) 
    = \trace_{\mathcal{A}}(\Phi^\dagger\circ\Psi)
    = \langle\Phi,\Psi\rangle_{\mathcal{S}}.
    \label{eq:clifford-right-composition-unitarity}
\end{align}
Thus, \(R_C\) is unitary.
Next, Eq.~\eqref{eq:app-adjoint-action-composition} gives \(\mathcal{C} \circ \mathcal{C}^{-1} = \mathcal{C}^{-1} \circ \mathcal{C} =\operatorname{id}_{\mathcal{A}}\), which proves Eq.~\eqref{eq:clifford-right-composition-inverse}.
Rearranging Eq.~\eqref{eq:app-Clifford-action} gives \(C^\dagger W(F_Cv)C=\eta_C(v)W(v)\). 
Since \(\operatorname{Ad}_{\eta_C(v)W(v)}=
\mathcal{W}(v)\), Eq.~\eqref{eq:app-adjoint-action-composition} gives
\begin{equation}
   \mathcal{C}^{-1} \circ\mathcal{W}(F_Cv) \circ \mathcal{C} = \mathcal{W}(v),
    \label{eq:Clifford-adjoint-intertwining}
\end{equation}
for all \(v\in V_{\mathsf{N}}\). 
For \(\Phi\in\mathcal{S}\), Eqs.~\eqref{eq:standard-twirl-def}, \eqref{eq:app-Pauli-conjugation-action}, \eqref{eq:app-induced-superoperator-representation}, \eqref{eq:app-adjoint-action-adjoint}, \eqref{eq:clifford-twirl-def}, and
\eqref{eq:Clifford-adjoint-intertwining}, together with the change of variables \(u=F_Cv\) (which is valid because \(F_C\) is an automorphism
of \(V_{\mathsf{N}}\) by Definition~\ref{defn:clifford-unitary}), give
\begin{align}
    \left(R_C^{-1}\circ\mathcal{T}\circ R_C\right)(\Phi)
    &=
    \frac{1}{|V_{\mathsf{N}}|}
    \sum_{u\in V_{\mathsf{N}}}
    \mathcal{W}(u) 
    \circ\Phi
    \circ \mathcal{C}^{-1}
    \circ \mathcal{W}(u)
    \circ \mathcal{C}
    \nonumber\\
    &=
    \frac{1}{|V_{\mathsf{N}}|}
    \sum_{v\in V_{\mathsf{N}}}
    \mathcal{W}(F_Cv) \circ\Phi
    \circ \mathcal{W}(v)
    =
    \mathcal{T}_C(\Phi),
    \label{eq:twist-to-standard}
\end{align}
which holds for every \(\Phi\in\mathcal{S}\), and thus proves Eq.~\eqref{eq:twisted-standard-conjugacy}.
\end{proof}

For \(\mathcal{E}\in\mathcal{S}\), define the map \(\mathcal{D}_{\mathcal{E}}\in\mathcal{S}\) by \(\mathcal{D}_{\mathcal{E}} \coloneqq \mathcal{T}(R_C(\mathcal{E}))\).
Proposition~\ref{prop:standard-twirl-Pauli-expansion} shows that \(\mathcal{D}_{\mathcal{E}}\) is a Pauli map.  
Denote by \(\{q_e[\mathcal{E}]\}_{e\in V_{\mathsf{N}}}\subseteq\mathbb{C}\) its unique coefficient family from Lemma~\ref{lem:pauli-map-structure}.

\begin{prop}
\label{prop:clifford-twirl}
For every \(\mathcal{E}\in\mathcal{S}\),
\begin{equation}
    \mathcal{T}_C(\mathcal{E})
    =
    \mathcal{D}_{\mathcal{E}}\circ \mathcal{C}
    =
    \sum_{e\in V_{\mathsf{N}}}
    q_e[\mathcal{E}]\,
\mathcal{W}(e)\circ \mathcal{C}.
    \label{eq:clifford-twirl-form}
\end{equation}
Moreover,
\begin{equation}
    q_e[\mathcal{E}]
    =
    \frac{1}{d^2}
    \left\langle
\mathcal{W}(e),
        R_C(\mathcal{E})
    \right\rangle_{\mathcal{S}}
    =
    \frac{1}{d^2}
    \left\langle
\mathcal{W}(e)\circ \mathcal{C},
        \mathcal{E}
    \right\rangle_{\mathcal{S}},
    \label{eq:clifford-twirl-coefficient}
\end{equation}
for all \(e\in V_{\mathsf{N}}\). 
\end{prop}

\begin{proof}
Eqs.~\eqref{eq:twisted-standard-conjugacy} and
\eqref{eq:clifford-right-composition-inverse}, together with the
definition \(\mathcal{D}_{\mathcal{E}}=\mathcal{T}(R_C(\mathcal{E}))\),
give
\(\mathcal{T}_C(\mathcal{E})
=\mathcal{D}_{\mathcal{E}}\circ\operatorname{Ad}_C\).
By Lemma~\ref{lem:pauli-map-structure},
\(\mathcal{D}_{\mathcal{E}}
=\sum_{e\in V_{\mathsf{N}}}q_e[\mathcal{E}]
\mathcal{W}(e)\).
Substituting this equality gives Eq.~\eqref{eq:clifford-twirl-form}.
Applying Proposition~\ref{prop:standard-twirl-Pauli-expansion} to \(R_C(\mathcal{E})\) and using the uniqueness in Lemma~\ref{lem:pauli-map-structure} gives the first equality in Eq.~\eqref{eq:clifford-twirl-coefficient}.  
Unitarity of \(R_C\) (from Lemma~\ref{lem:clifford-twirl-standard-reduction}) and the inverse
formula in Eq.~\eqref{eq:clifford-right-composition-inverse} give
\begin{equation}
    \left\langle
\mathcal{W}(e),
        R_C(\mathcal{E})
    \right\rangle_{\mathcal{S}}
    =
    \left\langle
        R_C^{-1}\!\left(
\mathcal{W}(e)\right),
        \mathcal{E}
    \right\rangle_{\mathcal{S}}
    =
    \left\langle
\mathcal{W}(e)\circ \mathcal{C},
        \mathcal{E}
    \right\rangle_{\mathcal{S}},
    \label{eq:clifford-twirl-coefficient-transfer}
\end{equation}
which proves the second equality in Eq.~\eqref{eq:clifford-twirl-coefficient}.
\end{proof}

\begin{cor}
\label{cor:clifford-twirl-properties}
Let \(\mathcal{E}\in\mathcal{S}\) be completely positive and trace nonincreasing.  
Then \(\mathcal{D}_{\mathcal{E}}\) is a subnormalized Pauli map.
\end{cor}

\begin{proof}
By Eq.~\eqref{eq:clifford-right-composition}, \(R_C(\mathcal{E})=\mathcal{E}\circ\mathcal{C}^{-1}\).
The map \(\mathcal{C}^{-1}\) is CPTP, so \(R_C(\mathcal{E})\) is completely positive and trace nonincreasing. 
Proposition~\ref{prop:standard-twirl-positivity} then shows that \(\mathcal{D}_{\mathcal{E}}=\mathcal{T}(R_C(\mathcal{E}))\) is completely positive and trace nonincreasing, while Proposition~\ref{prop:standard-twirl-Pauli-expansion} shows that it is a Pauli map.  
The conclusion follows from Definition~\ref{defn:pauli-map-normalization}.
\end{proof}

\section{Leakage-dephased Pauli Twirling}
\label{sec:total-Hilbert-space-and-leakage-subspaces}

\begin{defn}[Leakage-extended qubit space]
\label{defn:leakage-extended-qubit}
A \emph{leakage-extended qubit space} is a Hilbert space \(\mathfrak{h}\) equipped with an orthogonal direct-sum decomposition
\begin{equation}
    \mathfrak{h}
    =
    \mathfrak{c}\oplus\mathfrak{l}.
    \label{eq:leakage-extended-qubit-decomposition}
\end{equation}
The \emph{computational subspace} \(\mathfrak{c}\) is a single-qubit Hilbert space, as in Definition~\ref{defn:qubit-Hilbert-space}.  
The \emph{leakage subspace} \(\mathfrak{l}\) is isomorphic to \(\mathbb{C}\).
\end{defn}

\begin{defn}[Leakage-extended qubit register]
\label{defn:register-of-leakage-extended-qubits}
A \emph{leakage-extended qubit register} \(\mathsf{Q}\) consists of a finite nonempty set of labels, also denoted \(\mathsf{Q}\), and, for each \(i\in\mathsf{Q}\), a leakage-extended qubit space \(\mathfrak{h}_i\).
For each \(i\in\mathsf{Q}\), denote the computational and leakage subspaces of \(\mathfrak{h}_i\) by \(\mathfrak{c}_i\) and \(\mathfrak{l}_i\), respectively, and denote the computational basis for \(\mathfrak{c}_i\) by \(\{\ket{0_i},\ket{1_i}\}\).  
Define the \emph{total Hilbert space} \(\mathcal{H}_{\mathsf{Q}}\) of \(\mathsf{Q}\) by
\begin{equation}
    \mathcal{H}_{\mathsf{Q}}
    =
    \bigotimes_{i\in\mathsf{Q}}\mathfrak{h}_i.
    \label{eq:leakage-extended-qubit-register-Hilbert-space}
\end{equation}
\end{defn}

Fix a leakage-extended qubit register and associated label set \(\mathsf{Q}\). 

\subsection{Leakage-subspace decomposition}

All tensor products in this subsection are indexed by the indicated labels. 
We use the canonical unitary identifications obtained by regrouping tensor products with the same labeled systems (tensor factors).

\begin{defn}[Leakage pattern]
\label{defn:leakage-pattern}
A \emph{leakage pattern} of \(\mathsf{Q}\) is a subset of the label set \(S\subseteq\mathsf{Q}\).  
Write \(\overline S=\mathsf{Q}\setminus S\), and define
\begin{equation}
    \mathfrak{l}_S
    =
    \bigotimes_{i\in S}\mathfrak{l}_i,
    \qquad
    \mathfrak{c}_{\overline S}
    =
    \bigotimes_{i\in\overline S}\mathfrak{c}_i.
    \label{eq:leakage-pattern-factors}
\end{equation}
\end{defn}

For \(i\in\mathsf{Q}\), the condition \(i\in S\) indicates that the system labeled by \(i\) is leaked, whereas \(i\in\overline S\) indicates
that it is computational.

\begin{defn}[Leakage subspace]
\label{defn:leakage-subspace}
For a leakage pattern \(S\subseteq\mathsf{Q}\), define the corresponding \emph{leakage subspace} of \(\mathcal{H}_{\mathsf{Q}}\) 
by
\begin{equation}
    \mathcal{H}_{\mathsf{Q}}^{S}
    =
    \mathfrak{l}_S\otimes\mathfrak{c}_{\overline S}
    \subseteq
    \mathcal{H}_{\mathsf{Q}}.
    \label{eq:leakage-subspace}
\end{equation}
\end{defn}

For each \(i\in\mathsf{Q}\), define the orthogonal projectors on \(\mathfrak{h}_i\) by
\begin{equation}
    P_i^{\mathrm c}
    =
    \id_{\mathfrak{c}_i}\oplus 0_{\mathfrak{l}_i},
    \qquad
    P_i^{\ell}
    =
    0_{\mathfrak{c}_i}\oplus\id_{\mathfrak{l}_i}.
    \label{eq:local-computational-leakage-projectors}
\end{equation}
We use the same symbols for their extensions to \(\mathcal{H}_{\mathsf{Q}}\), which act as the corresponding projector on the system labeled by \(i\) and as the identity on every other system.
They satisfy
\begin{equation}
    P_i^{\mathrm c}+P_i^{\ell}
    =
    \id_{\mathcal{H}_{\mathsf{Q}}},
    \qquad
    P_i^{\mathrm c}P_i^{\ell}
    =
    P_i^{\ell}P_i^{\mathrm c}
    =
    0.
    \label{eq:local-projector-relations}
\end{equation}

\begin{defn}[Leakage-pattern projector]
\label{defn:leakage-pattern-projector}
For a leakage pattern \(S\subseteq\mathsf{Q}\), define the \emph{leakage-pattern projector} \(\Pi_S:\mathcal{H}_{\mathsf{Q}}\to\mathcal{H}_{\mathsf{Q}}\) by
\begin{equation}
    \Pi_S
    =
    \left(
        \prod_{i\in S}P_i^{\ell}
    \right)
    \left(
        \prod_{i\in\overline S}P_i^{\mathrm c}
    \right).
    \label{eq:leakage-pattern-projector}
\end{equation}
An empty product defaults to \(\id_{\mathcal{H}_{\mathsf{Q}}}\).
\end{defn}

By Eq.~\eqref{eq:local-projector-relations}, the family \(\{\Pi_S\}_{S\subseteq\mathsf{Q}}\) is pairwise orthogonal and satisfies
\begin{equation}
    \sum_{S\subseteq\mathsf{Q}}\Pi_S
    =
    \prod_{i\in\mathsf{Q}}(P_i^{\mathrm c}+P_i^{\ell})
    =
    \id_{\mathcal{H}_{\mathsf{Q}}}.
    \label{eq:leakage-pattern-resolution}
\end{equation}
Combining Eqs.~\eqref{eq:leakage-subspace} and \eqref{eq:leakage-pattern-projector} gives \(\operatorname{image}(\Pi_S)=\mathcal{H}_{\mathsf{Q}}^{S}\) for every \(S\subseteq\mathsf{Q}\).  
Consequently, the resolution of the identity in Eq.~\eqref{eq:leakage-pattern-resolution} gives
\begin{equation}
    \mathcal{H}_{\mathsf{Q}}
    =
    \bigoplus_{S\subseteq\mathsf{Q}}\mathcal{H}_{\mathsf{Q}}^{S}.
    \label{eq:leakage-subspace-direct-sum}
\end{equation}

\subsection{Leakage-pattern dephasing}

\begin{defn}[Leakage-pattern dephasing]
\label{defn:leakage-pattern-dephasing}
Define the \emph{leakage-pattern dephasing channel} \(\Delta:\operatorname{End}(\mathcal{H}_{\mathsf{Q}})\to \operatorname{End}(\mathcal{H}_{\mathsf{Q}})\) by
\begin{equation}
    \Delta(X)
    =
    \sum_{S\subseteq\mathsf{Q}}\Pi_SX\Pi_S,
    \label{eq:Delta}
\end{equation}
for all \( X\in\operatorname{End}(\mathcal{H}_{\mathsf{Q}})\). 
\end{defn}

Since each \(\Pi_S\) is an orthogonal projector, the family \(\{\Pi_S\}_{S\subseteq\mathsf{Q}}\) is a Kraus family for \(\Delta\). 
Hence, \(\Delta\) is completely positive.  
Since \(\Pi_S^\dagger\Pi_S^{\vpd}=\Pi_S^{\vpd}\) for every \(S\subseteq\mathsf{Q}\), Eq.~\eqref{eq:leakage-pattern-resolution} gives \(\sum_{S \subseteq \mathsf{Q}} \Pi_S^\dagger \Pi_S^{\vpd} =\id_{\mathcal{H}_{\mathsf{Q}}}\), so \(\Delta\) is also trace preserving. 
Thus, \( \Delta \) is CPTP. 
Pairwise orthogonality of the leakage-pattern projectors gives
\begin{align}
    (\Delta\circ\Delta)(X)
    &=
    \sum_{S,T\subseteq\mathsf{Q}}
    \Pi_S\Pi_TX\Pi_T\Pi_S
    \nonumber\\
    &=
    \sum_{S\subseteq\mathsf{Q}}\Pi_SX\Pi_S
    =
    \Delta(X),
    \label{eq:Delta-idempotence}
\end{align}
for all \(X\in\operatorname{End}(\mathcal{H}_{\mathsf{Q}})\). 

\begin{lem}
\label{lem:random-sign-dephasing}
For each \(R\subseteq\mathsf{Q}\), define the operator \(U_R:\mathcal{H}_{\mathsf{Q}}\to\mathcal{H}_{\mathsf{Q}}\) by
\begin{equation}
    U_R
    =
    \prod_{i\in R}(P_i^{\mathrm c}-P_i^{\ell}),
    \label{eq:random-sign-unitary}
\end{equation}
where \(U_{\varnothing}=\id_{\mathcal{H}_{\mathsf{Q}}}\).  
Every \(U_R\) is a Hermitian unitary, and
\begin{equation}
    \Delta
    =
    2^{-|\mathsf{Q}|}
    \sum_{R\subseteq\mathsf{Q}}
    \operatorname{Ad}_{U_R} =     
    2^{-|\mathsf{Q}|} \sum_{R\subseteq\mathsf{Q}} \mathcal{U}_R.
    \label{eq:random-sign-dephasing}
\end{equation}
\end{lem}

\begin{proof}
Each \(P_i^{\mathrm c}-P_i^{\ell}\) is Hermitian, and Eq.~\eqref{eq:local-projector-relations} gives \((P_i^{\mathrm c}-P_i^{\ell})^2=\id_{\mathcal{H}_{\mathsf{Q}}}\).
Since operators supported on distinctly labeled systems commute, every \(U_R\) is a Hermitian unitary.

For \(i\in\mathsf{Q}\) and \(S\subseteq\mathsf{Q}\), Eqs.~\eqref{eq:local-projector-relations} and \eqref{eq:leakage-pattern-projector} give \((P_i^{\mathrm c}-P_i^{\ell})\Pi_S=-\Pi_S\) when \(i\in S\), and \((P_i^{\mathrm c}-P_i^{\ell})\Pi_S=\Pi_S\) when \(i\notin S\).
Thus, for \(R,S\subseteq\mathsf{Q}\), \(U_R\Pi_S=(-1)^{|R\cap S|}\Pi_S\).  
Multiplying Eq.~\eqref{eq:leakage-pattern-resolution} on the left by \(U_R\) gives \(U_R=\sum_{S\subseteq\mathsf{Q}}(-1)^{|R\cap S|}\Pi_S\).

If \(S=T\), then \((-1)^{|R\cap S|+|R\cap T|}=1\) for every \(R\subseteq\mathsf{Q}\).  
If \(S\neq T\), choose \(i\in S\mathbin{\triangle}T\), where \(\triangle\) denotes the symmetric difference.  
The involution \(R\mapsto R\mathbin{\triangle}\{i\}\) pairs the summands with opposite signs, since \(i\) belongs to exactly one of \(S\) and \(T\).
Consequently,
\begin{equation}
    2^{-|\mathsf{Q}|}
    \sum_{R\subseteq\mathsf{Q}}
    (-1)^{|R\cap S|+|R\cap T|}
    =
    \begin{cases}
        1,
        & S=T,\\
        0,
        & S\neq T
    \end{cases} \, ,
    \label{eq:random-sign-character-sum}
\end{equation}
for all \(S,T\subseteq\mathsf{Q}\).
For \(X\in\operatorname{End}(\mathcal{H}_{\mathsf{Q}})\), substituting \(U_R=\sum_{S\subseteq\mathsf{Q}}(-1)^{|R\cap S|}\Pi_S\) and using Eq.~\eqref{eq:random-sign-character-sum} gives
\begin{align}
    2^{-|\mathsf{Q}|}
    \sum_{R\subseteq\mathsf{Q}}
    \mathcal{U}_R(X)
    &=
    \sum_{S,T\subseteq\mathsf{Q}}
    \left[
        2^{-|\mathsf{Q}|}
        \sum_{R\subseteq\mathsf{Q}}
        (-1)^{|R\cap S|+|R\cap T|}
    \right]
    \Pi_SX\Pi_T
    \nonumber\\
    &=
    \sum_{S\subseteq\mathsf{Q}}
    \Pi_SX\Pi_S
    =
    \Delta(X).
    \label{eq:random-sign-average-calculation}
\end{align}
\end{proof}

\subsection{Leakage-extended Paulis and the leakage-dephased Pauli twirl}

\begin{defn}[Computational qubit register]
\label{defn:computational-qubit-register}
Define the \emph{computational qubit register associated with \(\mathsf{Q}\)}, denoted \(\mathsf{K}_{\mathsf{Q}}\), to be the qubit register (see Definition~\ref{defn:qubit-register}) with label set \(\mathsf{Q}\) and family of qubit Hilbert spaces \(\{\mathfrak{c}_i\}_{i\in\mathsf{Q}}\).  Denote its Hilbert space by \(\mathcal{K}_{\mathsf{Q}} =\bigotimes_{i\in\mathsf{Q}}\mathfrak{c}_i\).
\end{defn}

Let \(V_{\mathsf{K}_{\mathsf{Q}}}\) be the Pauli-label space of \(\mathsf{K}_{\mathsf{Q}}\), as in Definition~\ref{defn:Pauli-label-space}.
For \(a=(x,z)\in V_{\mathsf{K}_{\mathsf{Q}}}\), write \(a_i=(x_i,z_i)\) for every \(i\in\mathsf{Q}\), and let \(w_i^{\mathsf{K}_{\mathsf{Q}}}(a_i)\) and \(W^{\mathsf{K}_{\mathsf{Q}}}(a)\) be the corresponding Pauli operators from
Eqs.~\eqref{eq:computational-system-Pauli-generators}, \eqref{eq:computational-site-Pauli}, and \eqref{eq:app-Pauli-definition}, specialized to \(\mathsf{K}_{\mathsf{Q}}\) (i.e., defined only on the computational register \(\mathsf{K}_{\mathsf{Q}})\) and associated Hilbert space \( \mathcal{K}_{\mathsf{Q}}\). 

\begin{defn}[Leakage-extended Pauli family]
\label{defn:leakage-extended-Pauli-family}
For every \(a\in V_{\mathsf{K}_{\mathsf{Q}}}\), define the \emph{leakage-extended Pauli} \(W^{\mathsf{Q}}(a)\in \operatorname{End}(\mathcal{H}_{\mathsf{Q}})\) according to
\begin{equation}
    W^{\mathsf{Q}}(a)
    =
    \bigotimes_{i\in\mathsf{Q}}
    \left(
        w_i^{\mathsf{K}_{\mathsf{Q}}}(a_i)
        \oplus
        \id_{\mathfrak{l}_i}
    \right).
    \label{eq:leakage-extended-Pauli}
\end{equation}
We call \(\{W^{\mathsf{Q}}(a)\}_{a\in V_{\mathsf{K}_{\mathsf{Q}}}}\) the \emph{leakage-extended Pauli family associated with \(\mathsf{Q}\)}. 
The computational-only Pauli family is distinguished by the superscript \(\mathsf{K}_{\mathsf{Q}}\) instead of \(\mathsf{Q}\). 
\end{defn}

The inclusions \(\mathfrak{c}_i\hookrightarrow\mathfrak{h}_i\) induce a canonical inclusion \(\mathcal{K}_{\mathsf{Q}}\hookrightarrow
\mathcal{H}_{\mathsf{Q}}\), corresponding extension by 0.   
We identify \(\mathcal{K}_{\mathsf{Q}}\) with its image under this inclusion. 
For every \(a\in V_{\mathsf{K}_{\mathsf{Q}}}\), applying Lemma~\ref{lem:app-Pauli-identities} to each single-qubit system in Eq.~\eqref{eq:leakage-extended-Pauli} shows that \(W^{\mathsf{Q}}(a)\) is a Hermitian unitary and that its restriction to \(\mathcal{K}_{\mathsf{Q}}\) is \(W^{\mathsf{K}_{\mathsf{Q}}}(a)\).

Fix a Clifford unitary \(C\) on \(\mathcal{K}_{\mathsf{Q}}\) and a symplectic automorphism \(F_C:V_{\mathsf{K}_{\mathsf{Q}}}\to V_{\mathsf{K}_{\mathsf{Q}}}\) associated with \(C\), as in Definition~\ref{defn:clifford-unitary}.

\begin{defn}[Leakage-dephased Pauli twirl]
\label{defn:leakage-dephased-Pauli-twirl}
Define the \emph{leakage-dephased Pauli twirl} \(\mathcal{T}^{\mathsf{Q}}_C: \operatorname{End}\!\left(\operatorname{End}(\mathcal{H}_{\mathsf{Q}})\right) \to \operatorname{End} \! \left( \operatorname{End}(\mathcal{H}_{\mathsf{Q}})\right)\) by
\begin{equation}
    \mathcal{T}^{\mathsf{Q}}_C(\Phi)
    =
    \Delta
    \circ
    \left[
        \frac{1}{|V_{\mathsf{K}_{\mathsf{Q}}}|}
        \sum_{v\in V_{\mathsf{K}_{\mathsf{Q}}}}
        \mathcal{W}^{\mathsf{Q}}(F_C v)
        \circ
        \Phi
        \circ
        \mathcal{W}^{\mathsf{Q}}(v)
    \right]
    \circ
    \Delta,
    \label{eq:leakage-dephased-Pauli-twirl}
\end{equation}
for all \(\Phi \in \operatorname{End} \! \left(\operatorname{End}(\mathcal{H}_{\mathsf{Q}})\right)\), where \( \mathcal{W}^{\mathsf{Q}}(v) = \operatorname{Ad}_{W^{\mathsf{Q}}(v)}\). 
\end{defn}

If \(\Phi\) is CPTP, then every map in the sum in Eq.~\eqref{eq:leakage-dephased-Pauli-twirl} is also CPTP.  
Because the sum contains \(|V_{\mathsf{K}_{\mathsf{Q}}}|\) terms and is multiplied by \(1/|V_{\mathsf{K}_{\mathsf{Q}}}|\), the bracketed map is
CPTP. 
Since \(\Delta\) is CPTP, it follows that \(\mathcal{T}^{\mathsf{Q}}_C(\Phi)\) is also CPTP whenever \(\Phi\) is CPTP.

\subsection{Filling and leakage maps}

For a leakage pattern \(S\subseteq\mathsf{Q}\), define \(\mathfrak{c}_S=\bigotimes_{i\in S}\mathfrak{c}_i\).  
Under the canonical tensor-product identifications, \(\mathcal{K}_{\mathsf{Q}} =\mathfrak{c}_S\otimes\mathfrak{c}_{\overline S}\).

\begin{defn}[Filling map]
\label{defn:filling-map}
For \(S\subseteq\mathsf{Q}\), define the \emph{filling map} \(\mathcal{F}_S: \operatorname{End}  (\mathcal{H}_{\mathsf{Q}}^{S}) \to \operatorname{End}  (\mathcal{K}_{\mathsf{Q}})\) by
\begin{equation}
    \mathcal{F}_S(X)
    =
    \frac{1}{2^{|S|}} \id_{\mathfrak{c}_S}
    \otimes
    \trace_{\mathfrak{l}_S}(X),
    \label{eq:fill-map}
\end{equation}
for all \(X\in\operatorname{End}(\mathcal{H}_{\mathsf{Q}}^{S})\), where \(\trace_{\mathfrak{l}_S}\) denotes the partial trace over \(\mathfrak{l}_S\) with respect to the factorization in Eq.~\eqref{eq:leakage-subspace}.
\end{defn}

\begin{defn}[Leakage map]
\label{defn:leakage-map}
For \(S\subseteq\mathsf{Q}\), define the \emph{leakage map} \(\mathcal{L}_S: \operatorname{End} (\mathcal{K}_{\mathsf{Q}}) \to \operatorname{End}  (\mathcal{H}_{\mathsf{Q}}^{S})\) by
\begin{equation}
    \mathcal{L}_S(Y)
    =
    \id_{\mathfrak{l}_S}
    \otimes
    \trace_{\mathfrak{c}_S}(Y),
    \label{eq:leak-map}
\end{equation}
for all \(Y \in \operatorname{End}(\mathcal{K}_{\mathsf{Q}})\), where \(\trace_{\mathfrak{c}_S}\) denotes the partial trace over \(\mathfrak{c}_S\) with respect to the factorization \(\mathcal{K}_{\mathsf{Q}} =\mathfrak{c}_S\otimes\mathfrak{c}_{\overline S}\).
\end{defn}

\begin{lem}
\label{lem:filling-leakage-map-properties}
For every \(S\subseteq\mathsf{Q}\), the maps \(\mathcal{F}_S\) and \(\mathcal{L}_S\) are CPTP and
\begin{equation}
    \mathcal{L}_S\circ\mathcal{F}_S
    =
    \operatorname{id}_{\operatorname{End}(\mathcal{H}_{\mathsf{Q}}^{S})}.
    \label{eq:leak-fill-identity}
\end{equation}
\end{lem}

\begin{proof}
For every \(S\subseteq\mathsf{Q}\), \(\id_{\mathfrak{c}_S}/2^{|S|} =\bigotimes_{i\in S}(\id_{\mathfrak{c}_i}/2)\) is a density operator on \(\mathfrak{c}_S\).  
Since \(\mathfrak{l}_S\) is one dimensional, \(\id_{\mathfrak{l}_S}\) is the unique density operator on
\(\mathfrak{l}_S\).  The map \(\mathcal{F}_S\) is the composition of \(\trace_{\mathfrak{l}_S}\) with tensoring by \(\id_{\mathfrak{c}_S}/2^{|S|}\), and \(\mathcal{L}_S\) is the composition of \(\trace_{\mathfrak{c}_S}\) with tensoring by \(\id_{\mathfrak{l}_S}\). 
Hence, both maps are CPTP. 
For every \(X\in\operatorname{End}(\mathcal{H}_{\mathsf{Q}}^{S})\), Eqs.~\eqref{eq:fill-map} and \eqref{eq:leak-map} give \((\mathcal{L}_S\circ\mathcal{F}_S)(X) =\id_{\mathfrak{l}_S}\otimes \trace_{\mathfrak{l}_S}(X)=X\), where the second equality uses the fact that \(\mathfrak{l}_S\) is one dimensional.
\end{proof}

\begin{defn}[Computational replacement channel]
\label{defn:computational-replacement-channel}
For \(S\subseteq\mathsf{Q}\), define the \emph{computational replacement channel} \(\mathcal{G}_S: \operatorname{End}(\mathcal{K}_{\mathsf{Q}}) \to \operatorname{End}(\mathcal{K}_{\mathsf{Q}})\) by
\begin{equation}
    \mathcal{G}_S
    =
    \mathcal{F}_S
    \circ
    \mathcal{L}_S.
    \label{eq:computational-replacement-map}
\end{equation}
\end{defn}

Combining Eqs.~\eqref{eq:fill-map} and \eqref{eq:leak-map} gives
\begin{equation}
    \mathcal{G}_S(Y)
    =
    \frac{1}{2^{|S|}} \id_{\mathfrak{c}_S}
    \otimes
    \trace_{\mathfrak{c}_S}(Y),
    \label{eq:computational-replacement-action}
\end{equation}
for all \( Y \in \operatorname{End}(\mathcal{K}_{\mathsf{Q}})\). 
Define the subgroup of Pauli labels supported on \(S\) by
\begin{equation}
    V_S
    =
    \left\{
        a\in V_{\mathsf{K}_{\mathsf{Q}}}:
        a_i=(0,0)\ \text{for every }i\in\overline S
    \right\}.
    \label{eq:system-supported-Pauli-labels}
\end{equation}
For every \(i\in\mathsf{Q}\) and \(A\in\operatorname{End}(\mathfrak{c}_i)\), we have that
\begin{equation}
    \frac{1}{2} \id_{\mathfrak{c}_i} \trace(A)
    =
    \frac14
    \sum_{a_i\in\mathbb{Z}_2^2}
    w_i^{\mathsf{K}_{\mathsf{Q}}}(a_i) \, 
    A
    \, w_i^{\mathsf{K}_{\mathsf{Q}}}(a_i) 
    .
    \label{eq:single-system-depolarizing-Pauli-average}
\end{equation}
It follows from
Eq.~\eqref{eq:single-system-depolarizing-Pauli-average} that
\begin{equation}
    \mathcal{G}_S
    =
    \frac{1}{4^{|S|}}
    \sum_{a\in V_S}
    \mathcal{W}^{\mathsf{K}_{\mathsf{Q}}} (a). 
    \label{eq:computational-replacement-Pauli-channel}
\end{equation}
Since \(|V_S|=4^{|S|}\), the coefficients in Eq.~\eqref{eq:computational-replacement-Pauli-channel} are nonnegative and sum to one.  
It follows from Definition~\ref{defn:pauli-map-normalization} that \(\mathcal{G}_S\) is a Pauli channel.
Combining Eqs.~\eqref{eq:leakage-pattern-projector} and \eqref{eq:leakage-extended-Pauli} gives
\begin{equation}
    W^{\mathsf{Q}}(a)\Pi_S
    =
    \Pi_S W^{\mathsf{Q}}(a),
    \label{eq:leakage-extended-Pauli-subspace-commutation}
\end{equation}
for all \(a\in V_{\mathsf{K}_{\mathsf{Q}}}, S\subseteq\mathsf{Q}\). 
Thus, \(\mathcal{H}_{\mathsf{Q}}^{S}\) is invariant under every \(W^{\mathsf{Q}}(a)\).

\begin{defn}[Subspace-\(S\) Pauli]
\label{defn:subspace-Pauli}
For a leakage pattern \(S\subseteq\mathsf{Q}\) and a Pauli label \(a\in V_{\mathsf{K}_{\mathsf{Q}}}\), define the \emph{subspace-\(S\) Pauli} \(W_{S}^{\mathsf{Q}}(a)\in \operatorname{End}(\mathcal{H}_{\mathsf{Q}}^{S})\) by
\begin{equation}
    W_{S}^{\mathsf{Q}}(a) = \left.W^{\mathsf{Q}}(a)\right|_{\mathcal{H}_{\mathsf{Q}}^{S}}.
    \label{eq:subspace-Pauli}
\end{equation}
The restriction is well defined because \(W^{\mathsf{Q}}(a)\) preserves the subspace \(\mathcal{H}_{\mathsf{Q}}^{S}\).
\end{defn}

Combining Eqs.~\eqref{eq:leakage-subspace} and \eqref{eq:leakage-extended-Pauli} gives
\begin{equation}
    W_{S}^{\mathsf{Q}}(a)
    =
    \id_{\mathfrak{l}_S}
    \otimes
    \bigotimes_{i\in\overline S}
    w_i^{\mathsf{K}_{\mathsf{Q}}}(a_i).
    \label{eq:subspace-Pauli-factorization}
\end{equation}
For every \(S\subseteq\mathsf{Q}\) and \(a\in V_{\mathsf{K}_{\mathsf{Q}}}\), the subspace-\(S\) Pauli \(W_{S}^{\mathsf{Q}}(a)\) is a Hermitian unitary.

\begin{lem}
\label{lem:filling-leakage-Pauli-covariance}
For every \(S\subseteq\mathsf{Q}\) and \(v\in V_{\mathsf{K}_{\mathsf{Q}}}\),
\begin{align}
    \mathcal{F}_S
    \circ
    \mathcal{W}_{S}^{\mathsf{Q}}(v)
    &=
    \mathcal{W}^{\mathsf{K}_{\mathsf{Q}}}(v)
    \circ
    \mathcal{F}_S,
    \label{eq:fill-covariance}
    \\
    \mathcal{L}_S
    \circ
    \mathcal{W}^{\mathsf{K}_{\mathsf{Q}}}(v)
    &=
    \mathcal{W}_{S}^{\mathsf{Q}}(v)
    \circ
    \mathcal{L}_S.
    \label{eq:leak-covariance}
\end{align}
\end{lem}

\begin{proof}
The equality in Eq.~\eqref{eq:fill-covariance} follows from Eqs.~\eqref{eq:fill-map}, \eqref{eq:app-Pauli-definition}, and \eqref{eq:subspace-Pauli-factorization}, since conjugation by \(\bigotimes_{i\in S}w_i^{\mathsf{K}_{\mathsf{Q}}}(v_i)\) fixes \(\id_{\mathfrak{c}_S}/2^{|S|}\).
The equality in Eq.~\eqref{eq:leak-covariance} follows from Eqs.~\eqref{eq:leak-map}, \eqref{eq:app-Pauli-definition}, and \eqref{eq:subspace-Pauli-factorization}, together with invariance of \(\trace_{\mathfrak{c}_S}\) under unitary conjugation on \(\mathfrak{c}_S\).
\end{proof}

\subsection{Leakage-subspace inclusion, projection, and extension maps}

For each \(S\subseteq\mathsf{Q}\), denote by \(\iota_S:\mathcal{H}_{\mathsf{Q}}^{S}\hookrightarrow
\mathcal{H}_{\mathsf{Q}}\) the canonical inclusion. 
Its adjoint \(\iota_S^\dagger:\mathcal{H}_{\mathsf{Q}}\to \mathcal{H}_{\mathsf{Q}}^{S}\) is the orthogonal projection onto \(\mathcal{H}_{\mathsf{Q}}^{S}\), with codomain restricted to \(\mathcal{H}_{\mathsf{Q}}^{S}\). 
Hence,
\begin{equation}
    \iota_S^\dagger \iota_S^{\vpd}
    =
    \id_{\mathcal{H}_{\mathsf{Q}}^{S}}
    \qquad \mathrm{and} \qquad
    \iota^{\vpd}_S\iota_S^\dagger
    =
    \Pi_S.
    \label{eq:leakage-subspace-inclusion-identities}
\end{equation}

\begin{defn}[Leakage-subspace projection map]
\label{defn:leakage-subspace-projection-map}
For \(S\subseteq\mathsf{Q}\), define the \emph{leakage-subspace projection map} \(\mathcal{P}_S: \operatorname{End}(\mathcal{H}_{\mathsf{Q}}) \to \operatorname{End}(\mathcal{H}_{\mathsf{Q}}^{S})\) by
\begin{equation}
    \mathcal{P}_S(X)
    =
    \iota_S^\dagger X\iota^{\vpd}_S,
    \label{eq:leakage-subspace-projection-map}
\end{equation}
for all \( X\in\operatorname{End}(\mathcal{H}_{\mathsf{Q}}) \). 
\end{defn}

\begin{defn}[Leakage-subspace extension map]
\label{defn:leakage-subspace-extension-map}
For \(S\subseteq\mathsf{Q}\), define the \emph{leakage-subspace extension map} \(\mathcal{E}_S: \operatorname{End}(\mathcal{H}_{\mathsf{Q}}^{S}) \to \operatorname{End}(\mathcal{H}_{\mathsf{Q}})\) by
\begin{equation}
    \mathcal{E}_S(Y)
    =
    \iota_S^{\vpd} Y \iota_S^\dagger,
    \label{eq:leakage-subspace-extension-map}
\end{equation}
for all \(Y\in\operatorname{End}(\mathcal{H}_{\mathsf{Q}}^{S})\). 
\end{defn}

The operator \(\mathcal{E}_S(Y)\) is supported on \(\mathcal{H}_{\mathsf{Q}}^{S}\), and its restriction to \(\mathcal{H}_{\mathsf{Q}}^{S}\) is \(Y\).
Both \(\mathcal{P}_S\) and \(\mathcal{E}_S\) are completely positive, since each has a single Kraus operator.  
For \(Y\in\operatorname{End}(\mathcal{H}_{\mathsf{Q}}^{S})\) and \(X\in\operatorname{End}(\mathcal{H}_{\mathsf{Q}})\) with \(X\succeq0\), Eq.~\eqref{eq:leakage-subspace-inclusion-identities} gives
\begin{equation}
    \trace(\mathcal{E}_S(Y))
    =
    \trace(Y)
    \qquad \mathrm{and} \qquad
    \trace(\mathcal{P}_S(X))
    =
    \trace(\Pi_SX)
    \leq
    \trace(X).
    \label{eq:leakage-subspace-map-traces}
\end{equation}
Thus, \(\mathcal{E}_S\) is trace preserving and \(\mathcal{P}_S\) is trace nonincreasing.
Equation~\eqref{eq:leakage-subspace-inclusion-identities} also gives
\begin{align}
    \mathcal{P}_S\circ\mathcal{E}_S
    &=
    \operatorname{id}_{\operatorname{End}(\mathcal{H}_{\mathsf{Q}}^{S})},
    \label{eq:projection-extension-identity}
    \\
    (\mathcal{E}_S\circ\mathcal{P}_S)(X)
    &=
    \Pi_SX\Pi_S.
    \label{eq:extension-projection-compression}
\end{align}
Next, combining Eqs.~\eqref{eq:Delta} and \eqref{eq:extension-projection-compression} gives
\begin{equation}
    \Delta
    =
    \sum_{S\subseteq\mathsf{Q}}
    \mathcal{E}_S\circ\mathcal{P}_S,
    \label{eq:Delta-leakage-subspace-maps}
\end{equation}
and, for \(S,T\subseteq\mathsf{Q}\), pairwise orthogonality of the leakage subspaces gives
\begin{equation}
    \mathcal{P}_T\circ\mathcal{E}_S
    =
    \begin{cases}
        \operatorname{id}_{\operatorname{End}(\mathcal{H}_{\mathsf{Q}}^{S})}
        & T=S\\
        0_{\operatorname{End}(\mathcal{H}_{\mathsf{Q}}^{S})
        \to\operatorname{End}(\mathcal{H}_{\mathsf{Q}}^{T})}
        & T\neq S
    \end{cases} \, .
    \label{eq:cross-subspace-orthogonality}
\end{equation}
Indeed, \(\iota_T^\dagger \iota_S^{\vpd}=0\) when \(T\neq S\), while Eq.~\eqref{eq:projection-extension-identity} covers the case \(T=S\).

\subsection{Leakage-transition branches}

\begin{defn}[Leakage-transition branches]
\label{defn:leakage-subspace-branches}
Let \(\Phi: \operatorname{End}(\mathcal{H}_{\mathsf{Q}}) \to \operatorname{End}(\mathcal{H}_{\mathsf{Q}})\) be a linear map.  
For every \(S,T\subseteq\mathsf{Q}\), define the \emph{\(S\to T\) leakage-transition branch} \(\Phi_{T, S}: \operatorname{End}(\mathcal{H}_{\mathsf{Q}}^{S}) \to \operatorname{End}(\mathcal{H}_{\mathsf{Q}}^{T})\) by
\begin{equation}
    \Phi_{T, S}
    =
    \mathcal{P}_T
    \circ
    \Phi
    \circ
    \mathcal{E}_S,
    \label{eq:leakage-subspace-branch}
\end{equation}
and define the \emph{family of leakage-transition branches associated with \(\Phi\)} to be \(\{\Phi_{T, S}\}_{S,T\subseteq\mathsf{Q}}\).
\end{defn}

\begin{lem}
\label{lem:leakage-subspace-branch-properties}
Let \(\Phi: \operatorname{End}(\mathcal{H}_{\mathsf{Q}}) \to \operatorname{End}(\mathcal{H}_{\mathsf{Q}})\) be a CPTP linear map and let \(\{\Phi_{T, S}\}_{S,T\subseteq\mathsf{Q}}\) be the family of leakage-transition branches associated with \(\Phi\), as in Definition~\ref{defn:leakage-subspace-branches}.  
Then, for every \(S,T\subseteq\mathsf{Q}\), the map \(\Phi_{T, S}\) is completely positive and trace nonincreasing.
\end{lem}

\begin{proof}
For every \(S,T\subseteq\mathsf{Q}\), combining Eq.~\eqref{eq:leakage-subspace-branch} with the fact that \(\mathcal{E}_S\) is CPTP and that \(\mathcal{P}_T\) is completely positive and trace nonincreasing gives the result.
\end{proof}

For a fixed Clifford unitary \(C\), denote by \(\mathcal{T}^{\mathsf{K}_{\mathsf{Q}}}_C\) the Pauli twirl with respect to \(C\), as in Definition~\ref{defn:clifford-twisted-Pauli-twirl}, specialized to \(\mathsf{K}_{\mathsf{Q}}\) and \(C\).

\begin{prop}
\label{prop:leakage-subspace-lifting}
Let \(\Phi: \operatorname{End}(\mathcal{H}_{\mathsf{Q}}) \to \operatorname{End}(\mathcal{H}_{\mathsf{Q}})\)
be a linear map. 
Let \(\{\Phi_{T, S}\}_{S,T\subseteq\mathsf{Q}}\) and \(\{(\mathcal{T}^{\mathsf{Q}}_C(\Phi))_{T, S}\}_{ S,T\subseteq\mathsf{Q}}\) be the families of leakage-transition branches associated with \(\Phi\) and \(\mathcal{T}^{\mathsf{Q}}_C(\Phi)\), respectively, as in Definition~\ref{defn:leakage-subspace-branches}.
Then, for every \(S,T\subseteq\mathsf{Q}\),
\begin{equation}
    (\mathcal{T}^{\mathsf{Q}}_C(\Phi))_{T, S}
    =
    \mathcal{L}_T
    \circ
    \mathcal{T}^{\mathsf{K}_{\mathsf{Q}}}_C
    \!\left(
        \mathcal{F}_T
        \circ
        \Phi_{T, S}
        \circ
        \mathcal{L}_S
    \right)
    \circ
    \mathcal{F}_S.
    \label{eq:leakage-subspace-lifting-identity}
\end{equation}
\end{prop}

\begin{proof}
Fix \(S,T\subseteq\mathsf{Q}\). 
Combining Eqs.~\eqref{eq:leakage-extended-Pauli-subspace-commutation}, \eqref{eq:subspace-Pauli}, \eqref{eq:leakage-subspace-projection-map}, and \eqref{eq:leakage-subspace-extension-map} gives
\begin{align}
    \mathcal{W}^{\mathsf{Q}}(v)
    \circ
    \mathcal{E}_S
    &=
    \mathcal{E}_S
    \circ
    \mathcal{W}_{S}^{\mathsf{Q}}(v)
    \label{eq:leakage-subspace-extension-Pauli-intertwining}
    \\
    \mathcal{P}_T
    \circ
    \mathcal{W}^{\mathsf{Q}}(F_C v)
    &=
    \mathcal{W}_{T}^{\mathsf{Q}}(F_C v)
    \circ
    \mathcal{P}_T
    \label{eq:leakage-subspace-projection-Pauli-intertwining}
\end{align}
for every \(v\in V_{\mathsf{K}_{\mathsf{Q}}}\).  
Combining Eqs.~\eqref{eq:Delta-leakage-subspace-maps} and \eqref{eq:cross-subspace-orthogonality} also gives
\begin{equation}
    \mathcal{P}_T\circ\Delta
    =
    \mathcal{P}_T,
    \qquad
    \Delta\circ\mathcal{E}_S
    =
    \mathcal{E}_S.
    \label{eq:leakage-subspace-dephasing-absorption}
\end{equation}
Expanding \(\mathcal{T}^{\mathsf{K}_{\mathsf{Q}}}_C\) by Eq.~\eqref{eq:clifford-twirl-def} and using Eqs.~\eqref{eq:fill-covariance}, \eqref{eq:leak-covariance}, and \eqref{eq:leak-fill-identity} gives
\begin{align}
    &\mathcal{L}_T
    \circ
    \mathcal{T}^{\mathsf{K}_{\mathsf{Q}}}_C
    \!\left(
        \mathcal{F}_T
        \circ
        \Phi_{T, S}
        \circ
        \mathcal{L}_S
    \right)
    \circ
    \mathcal{F}_S
    \nonumber\\
    &\quad=
    \frac{1}{|V_{\mathsf{K}_{\mathsf{Q}}}|}
    \sum_{v\in V_{\mathsf{K}_{\mathsf{Q}}}}
    \mathcal{L}_T
    \circ
    \mathcal{W}^{\mathsf{K}_{\mathsf{Q}}}(F_C v)^\dagger
    \circ
    \mathcal{F}_T
    \circ
    \Phi_{T, S}
    \circ
    \mathcal{L}_S
    \circ
    \mathcal{W}^{\mathsf{K}_{\mathsf{Q}}}(v)
    \circ
    \mathcal{F}_S
    \nonumber\\
    &\quad=
    \frac{1}{|V_{\mathsf{K}_{\mathsf{Q}}}|}
    \sum_{v\in V_{\mathsf{K}_{\mathsf{Q}}}}
    \mathcal{W}_{T}^{\mathsf{Q}}(F_C v)
    \circ
    (\mathcal{L}_T\circ\mathcal{F}_T)
    \circ
    \Phi_{T, S}
    \circ
    \mathcal{W}_{S}^{\mathsf{Q}}(v)
    \circ
    (\mathcal{L}_S\circ\mathcal{F}_S)
    \nonumber\\
    &\quad=
    \frac{1}{|V_{\mathsf{K}_{\mathsf{Q}}}|}
    \sum_{v\in V_{\mathsf{K}_{\mathsf{Q}}}}
    \mathcal{W}_{T}^{\mathsf{Q}}(F_C v)
    \circ
    \Phi_{T, S}
    \circ
    \mathcal{W}_{S}^{\mathsf{Q}}(v).
    \label{eq:leakage-subspace-lifting-covariance-reduction}
\end{align}
Substituting Eq.~\eqref{eq:leakage-subspace-branch} into Eq.~\eqref{eq:leakage-subspace-lifting-covariance-reduction} and applying Eqs.~\eqref{eq:leakage-subspace-extension-Pauli-intertwining} and \eqref{eq:leakage-subspace-projection-Pauli-intertwining} termwise gives
\begin{align}
    &\frac{1}{|V_{\mathsf{K}_{\mathsf{Q}}}|}
    \sum_{v\in V_{\mathsf{K}_{\mathsf{Q}}}}
    \mathcal{W}_{T}^{\mathsf{Q}}(F_C v)
    \circ
    \Phi_{T, S}
    \circ
    \mathcal{W}_{S}^{\mathsf{Q}}(v)
    \nonumber\\
    &\quad=
    \mathcal{P}_T
    \circ
    \left[
        \frac{1}{|V_{\mathsf{K}_{\mathsf{Q}}}|}
        \sum_{v\in V_{\mathsf{K}_{\mathsf{Q}}}}
        \mathcal{W}^{\mathsf{Q}}(F_C v)
        \circ
        \Phi
        \circ
        \mathcal{W}^{\mathsf{Q}}(v)
    \right]
    \circ
    \mathcal{E}_S
    \nonumber\\
    &\quad=
    \mathcal{P}_T
    \circ
    \Delta
    \circ
    \left[
        \frac{1}{|V_{\mathsf{K}_{\mathsf{Q}}}|}
        \sum_{v\in V_{\mathsf{K}_{\mathsf{Q}}}}
        \mathcal{W}^{\mathsf{Q}}(F_C v)
        \circ
        \Phi
        \circ
        \mathcal{W}^{\mathsf{Q}}(v)
    \right]
    \circ
    \Delta
    \circ
    \mathcal{E}_S
    \nonumber\\
    &\quad=
    \mathcal{P}_T
    \circ
    \mathcal{T}^{\mathsf{Q}}_C(\Phi)
    \circ
    \mathcal{E}_S
    =
    (\mathcal{T}^{\mathsf{Q}}_C(\Phi))_{T, S}.
    \label{eq:leakage-subspace-lifting-twirl-reduction}
\end{align}
The second equality uses Eq.~\eqref{eq:leakage-subspace-dephasing-absorption}, the third uses Eq.~\eqref{eq:leakage-dephased-Pauli-twirl}, and the last uses Eq.~\eqref{eq:leakage-subspace-branch}.  
Combining Eqs.~\eqref{eq:leakage-subspace-lifting-covariance-reduction} and \eqref{eq:leakage-subspace-lifting-twirl-reduction} proves Eq.~\eqref{eq:leakage-subspace-lifting-identity}.
\end{proof}

\subsection{Leakage-transition Pauli maps}

\begin{defn}[Leakage-transition Pauli-map family]
\label{defn:leakage-transition-Pauli-map-family}
Let \(\Phi: \operatorname{End}(\mathcal{H}_{\mathsf{Q}}) \to \operatorname{End}(\mathcal{H}_{\mathsf{Q}})\) be a linear map  and let
\(\{\Phi_{T, S}\}_{S,T\subseteq\mathsf{Q}}\) be the family of leakage-transition branches associated with \(\Phi\), as in Definition~\ref{defn:leakage-subspace-branches}.  
For every \(S,T\subseteq\mathsf{Q}\), define \(\mathcal{D}_{T, S}^{\Phi}: \operatorname{End}(\mathcal{K}_{\mathsf{Q}}) \to \operatorname{End}(\mathcal{K}_{\mathsf{Q}})\) by
\begin{equation}
    \mathcal{D}_{T, S}^{\Phi}
    =
    \mathcal{T}^{\mathsf{K}_{\mathsf{Q}}}_C
    \!\left(
        \mathcal{F}_T
        \circ
        \Phi_{T, S}
        \circ
        \mathcal{L}_S
    \right)
    \circ
    \mathcal{C}^{-1},
    \label{eq:leakage-transition-Pauli-map}
\end{equation}
and dfine the \emph{leakage-transition Pauli-map family associated with \(\Phi\)} to be \(\{\mathcal{D}_{T, S}^{\Phi}\}_{S,T\subseteq\mathsf{Q}}\).
\end{defn}

By Proposition~\ref{prop:clifford-twirl}, applied with \( \mathcal{E} = \mathcal{F}_T \circ \Phi_{T, S} \circ \mathcal{L}_S \), the map \(\mathcal{D}_{T, S}^{\Phi}\) is a Pauli map for every \(S,T\subseteq\mathsf{Q}\).  
Combining Eq.~\eqref{eq:leakage-transition-Pauli-map} with
\( 
\mathcal{C}^{-1} \circ \mathcal{C} = \operatorname{id}_{ \operatorname{End}(\mathcal{K}_{\mathsf{Q}}) } \) gives
\begin{equation}
    \mathcal{T}^{\mathsf{K}_{\mathsf{Q}}}_C
    \!\left(
        \mathcal{F}_T
        \circ
        \Phi_{T, S}
        \circ
        \mathcal{L}_S
    \right)
    =
    \mathcal{D}_{T, S}^{\Phi}
    \circ
    \mathcal{C}.
    \label{eq:leakage-transition-Pauli-map-factorization}
\end{equation}


\begin{prop}
\label{prop:leakage-transition-Pauli-map-normalization}
Let \(\Phi: \operatorname{End}(\mathcal{H}_{\mathsf{Q}}) \to \operatorname{End}(\mathcal{H}_{\mathsf{Q}})\)
be a CPTP map and let \( \{\mathcal{D}_{T, S}^{\Phi}\}_{S,T\subseteq\mathsf{Q}} \) be the leakage-transition Pauli-map family associated with \(\Phi\), as in Definition~\ref{defn:leakage-transition-Pauli-map-family}. 
Then \(\mathcal{D}_{T, S}^{\Phi}\) is a subnormalized Pauli map for every \(S,T\subseteq\mathsf{Q}\), and
\begin{equation}
    \sum_{T\subseteq\mathsf{Q}}
    \operatorname{wt}
    \!\left(
        \mathcal{D}_{T, S}^{\Phi}
    \right)
    =
    1, 
    \label{eq:leakage-transition-Pauli-map-normalization}
\end{equation}
for all \(S\subseteq\mathsf{Q}\). 
\end{prop}

\begin{proof}
Denote by \(\{\Phi_{T, S}\}_{S,T\subseteq\mathsf{Q}}\) the family of leakage-transition branches associated with \(\Phi\), as in Definition~\ref{defn:leakage-subspace-branches}. 
Fix \(S,T\subseteq\mathsf{Q}\) and set \(\mathcal{E}=\mathcal{F}_T\circ\Phi_{T, S}\circ\mathcal{L}_S\).
Lemma~\ref{lem:leakage-subspace-branch-properties} shows that \(\Phi_{T, S}\) is completely positive and trace nonincreasing. 
Lemma~\ref{lem:filling-leakage-map-properties} then shows that \(\mathcal{E}\) is completely positive and trace
nonincreasing. 
Next, Eq.~\eqref{eq:leakage-transition-Pauli-map}, Proposition~\ref{prop:clifford-twirl} (applied with this \(\mathcal{E}\)), and
\( 
\mathcal{C} \circ \mathcal{C}^{-1} = \operatorname{id}_{\operatorname{End}(\mathcal{K}_{\mathsf{Q}}) } \) give \(\mathcal{D}_{T, S}^{\Phi}=\mathcal{D}_{\mathcal{E}}\).
Corollary~\ref{cor:clifford-twirl-properties} therefore shows that \(\mathcal{D}_{T, S}^{\Phi}\) is a subnormalized Pauli map.
Since \(S,T\) were arbitrary, this holds for every \(S,T\subseteq\mathsf{Q}\).

Let \(\tau=2^{-|\mathsf{Q}|}\id_{\mathcal{K}_{\mathsf{Q}}}\), and fix \(S\subseteq\mathsf{Q}\).  
The state \(\tau\) is invariant under \(C\) and every \(W^{\mathsf{K}_{\mathsf{Q}}}(v)\). 
Definition~\ref{defn:pauli-map-weight}, linearity of the trace, Eqs.~\eqref{eq:clifford-twirl-def} and \eqref{eq:leakage-transition-Pauli-map-factorization}, and unitary invariance of the trace give the first two equalities in
\begin{align}
    \sum_{T\subseteq\mathsf{Q}}
    \operatorname{wt}
    \!\left(
        \mathcal{D}_{T, S}^{\Phi}
    \right)
    &=
    \sum_{T\subseteq\mathsf{Q}}
    \trace
    \!\left(
        \mathcal{D}_{T, S}^{\Phi}(\tau)
    \right)
    \nonumber\\
    &=
    \sum_{T\subseteq\mathsf{Q}}
    \trace
    \!\left(
        (\mathcal{F}_T
        \circ
        \Phi_{T, S}
        \circ
        \mathcal{L}_S)(\tau)
    \right)
    \nonumber\\
    &=
    \sum_{T\subseteq\mathsf{Q}}
    \trace
    \!\left(
        (\mathcal{P}_T
        \circ
        \Phi
        \circ
        \mathcal{E}_S
        \circ
        \mathcal{L}_S)(\tau)
    \right)
    \nonumber\\
    &=
    \trace
    \!\left(
        \left(
            \sum_{T\subseteq\mathsf{Q}}\Pi_T
        \right)
        \Phi
        \!\left(
            (\mathcal{E}_S\circ\mathcal{L}_S)(\tau)
        \right)
    \right)
    \nonumber\\
    &=
    \trace
    \!\left(
        \Phi
        \!\left(
            (\mathcal{E}_S\circ\mathcal{L}_S)(\tau)
        \right)
    \right)
    =
    \trace(\tau)
    =
    1.
    \label{eq:conditional-Pauli-weight-normalization-calculation}
\end{align}
The third equality uses trace preservation of \(\mathcal{F}_T\) from Lemma~\ref{lem:filling-leakage-map-properties} and Eq.~\eqref{eq:leakage-subspace-branch}.  
The fourth uses the trace identity in Eq.~\eqref{eq:leakage-subspace-map-traces}.  
The fifth uses Eq.~\eqref{eq:leakage-pattern-resolution}. 
The last uses trace preservation of \(\Phi\), \(\mathcal{E}_S\), and \(\mathcal{L}_S\).
\end{proof}

\begin{cor}
Under the 
assumptions of Proposition~\ref{prop:leakage-transition-Pauli-map-normalization}, the map \( \sum_{T\subseteq\mathsf{Q}} \mathcal{D}_{T, S}^{\Phi} \) is a Pauli channel on \(\mathcal{K}_{\mathsf{Q}}\) for every \(S\subseteq\mathsf{Q}\).
\end{cor}

\begin{proof}
Each summand is a Pauli map with nonnegative coefficients, and Eqs.~\eqref{eq:leakage-transition-Pauli-map-normalization} and \eqref{eq:pauli-map-weight-sum} imply that the coefficients of the sum add to one.
\end{proof}

\begin{thm}
\label{prop:leakage-transition-Pauli-map-decomposition}
Let \(\Phi: \operatorname{End}(\mathcal{H}_{\mathsf{Q}}) \to \operatorname{End}(\mathcal{H}_{\mathsf{Q}})\)
be a linear map and let \( \{\mathcal{D}_{T, S}^{\Phi}\}_{S,T\subseteq\mathsf{Q}} \) be the leakage-transition Pauli-map family associated with \(\Phi\), as in Definition~\ref{defn:leakage-transition-Pauli-map-family}. 
Then
\begin{equation}
    \mathcal{T}^{\mathsf{Q}}_C(\Phi)
    =
    \sum_{S,T\subseteq\mathsf{Q}}
    \mathcal{E}_T
    \circ
    \mathcal{L}_T
    \circ
    \mathcal{D}_{T, S}^{\Phi}
    \circ
    \mathcal{C}
    \circ
    \mathcal{F}_S
    \circ
    \mathcal{P}_S.
    \label{eq:leakage-transition-Pauli-map-decomposition}
\end{equation}
\end{thm}

\begin{proof}
Combining Eqs.~\eqref{eq:leakage-dephased-Pauli-twirl} and \eqref{eq:Delta-idempotence} gives \( \mathcal{T}^{\mathsf{Q}}_C(\Phi) = \Delta\circ\mathcal{T}^{\mathsf{Q}}_C(\Phi)\circ\Delta \).
Next, denote by \(\{(\mathcal{T}^{\mathsf{Q}}_C(\Phi))_{T, S}\}_{ S,T\subseteq\mathsf{Q}}\) the family of leakage-transition branches associated with \(\mathcal{T}^{\mathsf{Q}}_C(\Phi)\), as in Definition~\ref{defn:leakage-subspace-branches}.
Substituting Eq.~\eqref{eq:Delta-leakage-subspace-maps} on both sides of \(\mathcal{T}^{\mathsf{Q}}_C(\Phi)\) gives
\begin{align*}
    \mathcal{T}^{\mathsf{Q}}_C(\Phi)
    &=
    \left(
        \sum_{T\subseteq\mathsf{Q}}
        \mathcal{E}_T\circ\mathcal{P}_T
    \right)
    \circ
    \mathcal{T}^{\mathsf{Q}}_C(\Phi)
    \circ
    \left(
        \sum_{S\subseteq\mathsf{Q}}
        \mathcal{E}_S\circ\mathcal{P}_S
    \right)
    \\
    &=
    \sum_{S,T\subseteq\mathsf{Q}}
    \mathcal{E}_T
    \circ
    (\mathcal{T}^{\mathsf{Q}}_C(\Phi))_{T, S}
    \circ
    \mathcal{P}_S.
\end{align*}
Combining Eqs.~\eqref{eq:leakage-subspace-lifting-identity} and \eqref{eq:leakage-transition-Pauli-map-factorization} gives
\begin{equation}
    (\mathcal{T}^{\mathsf{Q}}_C(\Phi))_{T, S}
    =
    \mathcal{L}_T
    \circ
    \mathcal{D}_{T, S}^{\Phi}
    \circ
    \mathcal{C}
    \circ
    \mathcal{F}_S.
    \label{eq:twirled-branch-factorization}
\end{equation}
Finally, substituting Eq.~\eqref{eq:twirled-branch-factorization} termwise gives Eq.~\eqref{eq:leakage-transition-Pauli-map-decomposition}.
\end{proof}

\subsection{Leakage-pattern Markov process}
\label{subsec:Pauli-leakage-Markov-process}

Fix a CPTP map \(\Phi: \operatorname{End}(\mathcal{H}_{\mathsf{Q}}) \to \operatorname{End}(\mathcal{H}_{\mathsf{Q}})\). 
Denote by \( \{\mathcal{D}_{T, S}^{\Phi}\}_{S,T\subseteq\mathsf{Q}} \) the leakage-transition Pauli-map family associated with \(\Phi\), as in Definition~\ref{defn:leakage-transition-Pauli-map-family}.

\begin{defn}[Leakage-pattern transition weights]
\label{defn:leakage-pattern-transition-weights}
For \(S,T\subseteq\mathsf{Q}\), define the leakage-pattern transition weight \(p(T| S)\) by
\begin{equation}
    p(T| S)
    =
    \operatorname{wt}\!\left(
        \mathcal{D}_{T, S}^{\Phi}
    \right).
    \label{eq:leakage-pattern-transition-weight}
\end{equation}
\end{defn}

The conclusions of Proposition~\ref{prop:leakage-transition-Pauli-map-normalization} give \(p(T| S)\geq0\) for all \(S,T\subseteq\mathsf{Q}\) and \(\sum_{T\subseteq\mathsf{Q}}p(T| S)=1\) for every \(S\subseteq\mathsf{Q}\). 
Thus, \(p\) is a Markov transition kernel on the set of leakage patterns.

\begin{defn}[Conditional leakage-transition maps]
\label{defn:conditional-leakage-transition-maps}
For \(S,T\subseteq\mathsf{Q}\) satisfying \(p(T| S)>0\), define the map \( \mathcal{C}_{T, S}^{\Phi}: \operatorname{End}(\mathcal{K}_{\mathsf{Q}}) \to \operatorname{End}(\mathcal{K}_{\mathsf{Q}}) \) by
\begin{equation}
    \mathcal{C}_{T, S}^{\Phi}
    =
    \frac{1}{p(T| S)}
    \mathcal{D}_{T, S}^{\Phi}.
    \label{eq:conditional-leakage-transition-map}
\end{equation}
\end{defn}

\begin{cor}
\label{cor:leakage-pattern-Pauli-channel-decomposition}
For every \(S,T\subseteq\mathsf{Q}\) satisfying \(p(T| S)>0\), \(\mathcal{C}_{T, S}^{\Phi}\) is a Pauli channel and
\begin{equation}
    \mathcal{T}^{\mathsf{Q}}_C(\Phi)
    =
    \sum_{\substack{
        S,T\subseteq\mathsf{Q}\\
        p(T| S)>0
    }}
    p(T| S)
    \mathcal{E}_T
    \circ
    \mathcal{L}_T
    \circ
    \mathcal{C}_{T, S}^{\Phi}
    \circ
    \mathcal{C}
    \circ
    \mathcal{F}_S
    \circ
    \mathcal{P}_S.
    \label{eq:leakage-pattern-Pauli-channel-decomposition}
\end{equation}
\end{cor}

\begin{proof}
Fix \(S,T\subseteq\mathsf{Q}\), and let \(\{q_v\}_{v\in V_{\mathsf{K}_{\mathsf{Q}}}}\) be the unique Pauli-map coefficient family of \(\mathcal{D}_{T, S}^{\Phi}\), as in Lemma~\ref{lem:pauli-map-structure}.  
Proposition~\ref{prop:leakage-transition-Pauli-map-normalization} implies that \(\mathcal{D}_{T, S}^{\Phi}\) is a subnormalized Pauli map, so Lemma~\ref{lem:pauli-map-positivity-and-normalization} gives \(q_v\geq0\) for every \(v\in V_{\mathsf{K}_{\mathsf{Q}}}\). 
Combining Eqs.~\eqref{eq:leakage-pattern-transition-weight} and \eqref{eq:pauli-map-weight-sum} gives
\( p(T| S) = \sum_{v\in V_{\mathsf{K}_{\mathsf{Q}}}}q_v \).

Suppose that \(p(T| S)>0\). 
By Eq.~\eqref{eq:conditional-leakage-transition-map}, the Pauli-map coefficient family of \(\mathcal{C}_{T, S}^{\Phi}\) is \( \{q_v/p(T| S)\}_{v\in V_{\mathsf{K}_{\mathsf{Q}}}} \). 
These coefficients lie in \([0,1]\) and sum to one. 
Hence, \(\mathcal{C}_{T, S}^{\Phi}\) is a Pauli channel, by Lemma~\ref{lem:pauli-map-positivity-and-normalization}.

Suppose instead that \(p(T| S)=0\). 
Since every \(q_v\) is nonnegative and their sum is zero, \(q_v=0\) for every \(v\in V_{\mathsf{K}_{\mathsf{Q}}}\). 
Hence, \(\mathcal{D}_{T, S}^{\Phi}=0\).

Substituting \( \mathcal{D}_{T, S}^{\Phi} = p(T| S)\mathcal{C}_{T, S}^{\Phi} \) for the positive-weight terms and \(\mathcal{D}_{T, S}^{\Phi}=0\) for the zero-weight terms in Eq.~\eqref{eq:leakage-transition-Pauli-map-decomposition} gives Eq.~\eqref{eq:leakage-pattern-Pauli-channel-decomposition}.
\end{proof}

For a density operator \(\rho\in\operatorname{End}(\mathcal{H}_{\mathsf{Q}})\), define the scalars \(\pi_S\) and \(\pi'_T\) by
\begin{equation}
    \pi_S
    =
    \trace(\mathcal{P}_S(\rho)),
    \qquad \mathrm{and} \qquad
    \pi'_T
    =
    \trace\!\left(
        \mathcal{P}_T(\mathcal{T}^{\mathsf{Q}}_C(\Phi)(\rho))
    \right),
    \label{eq:incoming-outgoing-leakage-distributions}
\end{equation}
for all \(S,T\subseteq\mathsf{Q}\). 
The families \(\{\pi_S\}_{S\subseteq\mathsf{Q}}\) and \(\{\pi'_T\}_{T\subseteq\mathsf{Q}}\) can be understood as the probability distributions over leakage patterns of the incoming state \(\rho\) and the outgoing state \(\mathcal{T}^{\mathsf{Q}}_C(\Phi)(\rho)\), respectively.

\begin{cor}
\label{cor:leakage-pattern-Markov-update}
The families \(\{\pi_S\}_{S\subseteq\mathsf{Q}}\) and \(\{\pi'_T\}_{T\subseteq\mathsf{Q}}\) are probability distributions on the set of leakage patterns, and
\begin{equation}
    \pi'_T
    =
    \sum_{S\subseteq\mathsf{Q}}
    p(T| S)\pi_S,
    \label{eq:leakage-Markov-update}
\end{equation}
for all \(T\subseteq\mathsf{Q}\). 
\end{cor}

\begin{proof}
Since \(\rho\succeq0\), Eq.~\eqref{eq:leakage-subspace-projection-map} gives \(\pi_S\geq0\) for every \(S\subseteq\mathsf{Q}\). 
Combining Eqs.~\eqref{eq:leakage-subspace-map-traces} and \eqref{eq:leakage-pattern-resolution} gives \(\sum_{S\subseteq\mathsf{Q}}\pi_S=\trace(\rho)=1\).
Thus, \(\{\pi_S\}_{S\subseteq\mathsf{Q}}\) is a probability distribution.
Next, fix \(T\subseteq\mathsf{Q}\). 
Applying \(\mathcal{P}_T\) to Eq.~\eqref{eq:leakage-pattern-Pauli-channel-decomposition}, using Eq.~\eqref{eq:cross-subspace-orthogonality} termwise, and taking the trace gives
\begin{align}
    \pi'_T
    &=
    \sum_{\substack{
        S\subseteq\mathsf{Q}\\
        p(T| S)>0
    }}
    p(T| S)
    \trace\!\left[
        \left(
            \mathcal{L}_T
            \circ
            \mathcal{C}_{T, S}^{\Phi}
            \circ
            \mathcal{C}
            \circ
            \mathcal{F}_S
        \right)
        \bigl(\mathcal{P}_S(\rho)\bigr)
    \right]
    \nonumber\\
    &=
    \sum_{\substack{
        S\subseteq\mathsf{Q}\\
        p(T| S)>0
    }}
    p(T| S)\pi_S
    =
    \sum_{S\subseteq\mathsf{Q}}
    p(T| S)\pi_S.
    \label{eq:leakage-Markov-update-calculation}
\end{align}
The second equality uses trace preservation of \(\mathcal{F}_S\) and \(\mathcal{L}_T\) from Lemma~\ref{lem:filling-leakage-map-properties}, trace preservation of \(\mathcal{C}\), and trace preservation of \(\mathcal{C}_{T, S}^{\Phi}\) from Corollary~\ref{cor:leakage-pattern-Pauli-channel-decomposition}. 
The last equality inserts the terms satisfying \(p(T| S)=0\).

Since \(p(T| S)\geq0\), Eq.~\eqref{eq:leakage-Markov-update} gives \(\pi'_T\geq0\) for every \(T\subseteq\mathsf{Q}\). 
Summing Eq.~\eqref{eq:leakage-Markov-update} over \(T\) gives
\(
    \sum_{T\subseteq\mathsf{Q}}\pi'_T
    =
    \sum_{S\subseteq\mathsf{Q}}\pi_S
    \sum_{T\subseteq\mathsf{Q}}p(T| S)
    =
    1
\).
\end{proof}

\section{Model of a QEC cycle on leakage-extended qubits}
\label{sec:setup}

\subsection{Joint registers and Cliffords}
\label{subsec:joint-registers-and-Cliffords}

\begin{defn}[Joint register of qubit registers]
\label{defn:joint-register}
Let \(\mathsf{N}_1\) and \(\mathsf{N}_2\) be qubit registers with disjoint label sets.
For \(j=1,2\) and \(i\in\mathsf{N}_j\), denote the qubit Hilbert space at label \(i\) and its computational basis by \(\mathfrak{q}_i\) and \(\{\ket{0_i},\ket{1_i}\}\), respectively.
Define their \emph{joint register} \(\mathsf{N}\) to be the qubit register with label set
\[
\mathsf{N} = \mathsf{N}_1 \sqcup \mathsf{N}_2
\]
and family of qubit Hilbert spaces \(\{\mathfrak{q}_i\}_{i\in\mathsf{N}}\), equipped with these computational bases.
\end{defn}

\begin{defn}[Joint register of leakage-extended qubit registers]
\label{defn:joint-leakage-extended-qubit-register}
Let \(\mathsf{Q}_1\) and \(\mathsf{Q}_2\) be leakage-extended qubit registers with disjoint label sets.
For \(j=1,2\) and \(i\in\mathsf{Q}_j\), denote the leakage-extended qubit space at label \(i\) by \(\mathfrak{h}_i\), its specified decomposition by \(\mathfrak{h}_i=\mathfrak{c}_i\oplus\mathfrak{l}_i\), and the computational basis of \(\mathfrak{c}_i\) by \(\{\ket{0_i},\ket{1_i}\}\).
Define their \emph{joint register} \(\mathsf{Q}\) to be the leakage-extended qubit register with label set
\[
\mathsf{Q} = \mathsf{Q}_1 \sqcup \mathsf{Q}_2
\]
and family of leakage-extended qubit spaces \(\{\mathfrak{h}_i\}_{i\in\mathsf{Q}}\), equipped with these decompositions and computational bases.
\end{defn}
We implictly use the canonical unitary identifications obtained by regrouping tensor products with the same labeled systems (tensor factors).
For both constructions, the Hilbert space of the joint register is identified with the tensor product of the Hilbert spaces of the two constituent registers.
If \(\mathsf{N}\) is the joint register of the qubit registers \(\mathsf{N}_1\) and \(\mathsf{N}_2\), we identify its Pauli labels with pairs of constituent Pauli labels by restriction to the two label sets.
Under these identifications,
\begin{subequations}
\label{eq:joint-register-Pauli-identifications}
\begin{align}
        V_{\mathsf{N}}
        &=
        V_{\mathsf{N}_1}\times V_{\mathsf{N}_2}, \label{joint-register-Pauli-identifications-labels}
        \\
        \left[
            (a_1,a_2),
            (b_1,b_2)
        \right]
        &=
        [a_1,b_1]+[a_2,b_2], \label{joint-register-Pauli-identifications-commutator}
        \\
        W^{\mathsf{N}}\!\left((a_1,a_2)\right)
        &=
        W^{\mathsf{N}_1}(a_1)\otimes W^{\mathsf{N}_2}(a_2),
\end{align}
\end{subequations}
for \(a_j,b_j\in V_{\mathsf{N}_j}\), \(j=1,2\), with the symplectic-form sum taken in \(\mathbb{Z}_2\).

The family \(\{W^{\mathsf{N}_1}(a_1)\otimes W^{\mathsf{N}_2}(a_2)\}_{(a_1,a_2)\in V_{\mathsf{N}_1}\times V_{\mathsf{N}_2}}\) is the \emph{product Pauli family} of \(\mathsf{N}_1\) and \(\mathsf{N}_2\).
If \(\mathsf{Q}\) is the joint register of leakage-extended qubit registers \(\mathsf{Q}_1\) and \(\mathsf{Q}_2\), then \(\mathsf{K}_{\mathsf{Q}}\) is the joint register of the computational qubit registers \(\mathsf{K}_{\mathsf{Q}_1}\) and \(\mathsf{K}_{\mathsf{Q}_2}\).

Both constructions and the corresponding identifications extend immediately to any finite nonempty family of registers of the same type with pairwise disjoint label sets.
The product Pauli family likewise consists of tensor products of one Pauli operator from each constituent qubit register.
The preceding definitions and results formulated for an arbitrary qubit register or leakage-extended qubit register therefore apply to joint registers through these identifications.
Whenever a finite family of registers is treated jointly in the remainder of this discussion, its constituent label sets are assumed to be pairwise disjoint.

\begin{defn}[Clifford isomorphism]
\label{defn:Clifford-isomorphism}
Let \(\mathsf{M}\) and \(\mathsf{N}\) be qubit registers.  
A unitary isomorphism
\begin{equation}
\label{eq:Clifford-isomorphism}
    C: \mathcal{K}_{\mathsf{M}} \to  \mathcal{K}_{\mathsf{N}}
\end{equation}
is a \emph{Clifford isomorphism} if there exist a symplectic isomorphism \(F_C:V_{\mathsf{M}}\to V_{\mathsf{N}}\) and a map
\(\eta_C:V_{\mathsf{M}}\to\{+1,-1\}\) such that
\begin{equation}
    \mathcal{C} (W^{\mathsf{M}}(a) ) = C W^{\mathsf{M}}(a) C^\dagger = \eta_C(a)W^{\mathsf{N}}(F_C a),
    \label{eq:Clifford-isomorphism-action}
\end{equation}
for all \(a\in V_{\mathsf{M}}\). 
\end{defn}
When \(\mathsf{M}=\mathsf{N}\), Definition~\ref{defn:Clifford-isomorphism} agrees with Definition~\ref{defn:clifford-unitary}.  
The uniqueness argument following Definition~\ref{defn:clifford-unitary} also applies to \(F_C\) and \(\eta_C\).  
Equation~\eqref{eq:Clifford-isomorphism-action} shows that adjoints, compositions of composable Clifford isomorphisms, and tensor products of Clifford isomorphisms are Clifford.  
Their induced symplectic isomorphisms are, respectively, \(F_C^{-1}\), \(F_D\circ F_C\) for \(D\circ C\), and the product of the constituent symplectic isomorphisms.

\subsection{QEC-cycle construction}
\label{subsec:qec-cycle-construction}

Fix a stabilizer code on \(n\) physical qubits that encodes \(k\) logical qubits.  
Let \(\mathsf{D}\), \(\mathsf{A}\), and \(\mathsf{O}\) be leakage-extended qubit registers, as in Definition~\ref{defn:register-of-leakage-extended-qubits}, where \(|\mathsf{D}|=n\). 
Denote by \(\mathsf{T}\) the joint register of \( \mathsf{D}\), \(\mathsf{A}\), and \(\mathsf{O}\), and also the set of joint labels. 
Let \(\mathsf{K}_{\mathsf{D}}\), \(\mathsf{K}_{\mathsf{A}}\), \(\mathsf{K}_{\mathsf{O}}\), and \(\mathsf{K}_{\mathsf{T}}\) be the computational qubit registers associated with \(\mathsf{D}\), \(\mathsf{A}\), \(\mathsf{O}\), and \(\mathsf{T}\), respectively, as in Definition~\ref{defn:computational-qubit-register}. 
Then \(\mathsf{K}_{\mathsf{T}}\) is the joint register of \(\mathsf{K}_{\mathsf{D}}\), \(\mathsf{K}_{\mathsf{A}}\), and \(\mathsf{K}_{\mathsf{O}}\).

Let \(\mathsf{L}\) and \(\mathsf{S}\) be qubit registers (and the associated label sets) as in Definition~\ref{defn:qubit-register}, where \(|\mathsf{L}|=k\) and \(|\mathsf{S}|=n-k\). 
Choose a Clifford isomorphism
\[
    \CFrame:
    \mathcal{K}_{\mathsf{D}} 
    \to 
    \mathcal{K}_{\mathsf{L}}\otimes\mathcal{K}_{\mathsf{S}}.
\]
A QEC cycle on \(\mathsf{D}\) is modeled by a CPTP linear map \(\operatorname{End}(\mathcal{H}_{\mathsf{D}})
\to\operatorname{End}(\mathcal{H}_{\mathsf{D}})\).
Denote by \(\ket{0}_{\mathsf{A}}\in \mathcal{K}_{\mathsf{A}} \) and \(\ket{0}_{\mathsf{O}}\in\mathcal{K}_{\mathsf{O}} \) the all-zero product computational-basis vectors.  
Define the initialization map \(\mathcal{I}_{\mathsf{A}\mathsf{O}}:\operatorname{End}(\mathcal{H}_{\mathsf{D}}) \to \operatorname{End}(\mathcal{H}_{\mathsf{T}})\) as
\begin{equation}
    \mathcal{I}_{\mathsf{A}\mathsf{O}}(Z) \coloneqq Z \otimes \ZProj{0}_{\mathsf{A}} \otimes \ZProj{0}_{\mathsf{O}}, 
    \label{eq:ancilla-outcome-initialization}
\end{equation}
for all \(Z\in\operatorname{End}(\mathcal{H}_{\mathsf{D}})\). 
The tensor factors in Eq.~\eqref{eq:ancilla-outcome-initialization} are identified by the canonical labeled tensor-product identification.
The restriction of \(\mathcal{I}_{\mathsf{A}\mathsf{O}}\) to \(\operatorname{End}(\mathcal{K}_{\mathsf{D}} )\) has range in
\(\operatorname{End}(\mathcal{K}_{\mathsf{T}})\) and is denoted by the same symbol. 
Both maps are CPTP.

For \(s\in\mathbb{Z}_2^{\mathsf{S}}\), dnote by \(\ket{s}_{\mathsf{S}}\) the corresponding product computational-basis vector. 
Let \(C_{\mathrm{fr}}\) be a unitary on \(\mathcal{K}_{\mathsf{L}} \otimes\mathcal{K}_{\mathsf{S}} \otimes\mathcal{K}_{\mathsf{A}}
\otimes\mathcal{K}_{\mathsf{O}} \) that is Clifford with respect to the product Pauli family of \(\mathsf{L}\), \(\mathsf{S}\),
\(\mathsf{K}_{\mathsf{A}}\), and \(\mathsf{K}_{\mathsf{O}}\). 
Further assume that there exists an injective map
\begin{equation}
    E: \mathbb{Z}_2^{\mathsf{S}} \to \mathbb{Z}_2^{\mathsf{O}},
    \label{eq:syndrome-record-encoding}
\end{equation}
and unit vectors
\(\ket{\alpha(s)}_{\mathsf{A}} \in \mathcal{K}_{\mathsf{A}} \) (for \(s\in\mathbb{Z}_2^{\mathsf{S}}\)) such that, for every \(s\in\mathbb{Z}_2^{\mathsf{S}}\) and \(\ket{\psi}_{\mathsf{L}}\in\mathcal{K}_{\mathsf{L}}\),
\begin{align}
    &C_{\mathrm{fr}}
    \Big( \ket{\psi}_{\mathsf{L}} \otimes  \ket{s}_{\mathsf{S}} \otimes \ket{0}_{\mathsf{A}} \otimes \ket{0}_{\mathsf{O}} \Big)
    =
    \ket{\psi}_{\mathsf{L}} \otimes \ket{s}_{\mathsf{S}} \otimes \ket{\alpha(s)}_{\mathsf{A}} \otimes \ket{E(s)}_{\mathsf{O}}.
    \label{eq:nominal-readout-vector-action}
\end{align}
Define the syndrome-readout unitary \(C_{\mathrm{synd}}\) on \(\mathcal{K}_{\mathsf{T}}\) according to
\begin{equation}
    C_{\mathrm{synd}} \coloneqq 
    \left(  \CFrame^\dagger \otimes \id_{\mathcal{K}_{\mathsf{A}}  \otimes \mathcal{K}_{\mathsf{O}} }
    \right)
    C_{\mathrm{fr}}
    \left(
        \CFrame^{\vphantom{\dagger}} \otimes \id_{\mathcal{K}_{\mathsf{A}}  \otimes \mathcal{K}_{\mathsf{O}} }\right).
    \label{eq:syndrome-readout-frame-conjugation}
\end{equation}
The Clifford properties of \(C_{\mathrm{fr}}\) and \(\CFrame\) imply that \(C_{\mathrm{synd}}\) is Clifford with respect to the Pauli family of \(\mathsf{K}_{\mathsf{T}}\).

Denote by \(\mathcal{N}:\operatorname{End}(\mathcal{H}_{\mathsf{T}}) \to\operatorname{End}(\mathcal{H}_{\mathsf{T}})\) the CPTP map representing a noisy syndrome-readout operation, whose ideal (noiseless) implementation corresponds to the adjoint action of \(C_{\mathrm{synd}}\). 
The unitaries \(C_{\mathrm{synd}}\) and \(C_{\mathrm{fr}}\) are specified only on the computational spaces \(\mathcal{K}\), whereas \(\mathcal{N}\) is defined on the full space \(\mathcal{H}_{\mathsf{T}}\).  
No further restriction on the action of \(\mathcal{N}\) is imposed at this point.

Define \(\Theta\coloneqq \{0,1,\mathrm L\}^{\mathsf{O}}\).  
For each \(i\in\mathsf{O}\), define the orthogonal projectors on \(\mathfrak{h}_i=\mathfrak{c}_i\oplus\mathfrak{l}_i\) by
\begin{equation}
    P_i^0
    \coloneqq 
    \ZProj{0_i} \oplus 0_{\mathfrak{l}_i},
    \qquad
    P_i^1
    \coloneqq 
    \ZProj{1_i} \oplus 0_{\mathfrak{l}_i},
    \qquad
    P_i^{\mathrm L}
    \coloneqq 
    P_i^{\ell}.
    \label{eq:local-outcome-projectors}
\end{equation}
For \(\theta\in\Theta\), define the outcome projector
\(P_\theta^{\mathsf{O}}\in\operatorname{End}(\mathcal{H}_{\mathsf{O}})\)
by
\begin{equation}
    P_\theta^{\mathsf{O}}
    \coloneqq 
    \bigotimes_{i\in\mathsf{O}}P_i^{\theta_i}.
    \label{eq:outcome-projectors}
\end{equation}
Equations~\eqref{eq:local-computational-leakage-projectors} and
\eqref{eq:local-outcome-projectors} give
\begin{equation}
    P_\theta^{\mathsf{O}}P_{\theta'}^{\mathsf{O}}
    =
    \delta_{\theta,\theta'}P_\theta^{\mathsf{O}},
    \qquad
    \sum_{\theta\in\Theta}P_\theta^{\mathsf{O}}
    =
    \id_{\mathcal{H}_{\mathsf{O}}},.
    \label{eq:outcome-projector-resolution}
\end{equation}
for all \( \theta,\theta'\in\Theta\). 
Define the outcome-map family \(\{\mathcal{M}_\theta\}_{\theta\in\Theta}\), where  \(\mathcal{M}_\theta:\operatorname{End}(\mathcal{H}_{\mathsf{T}}) \to\operatorname{End}(\mathcal{H}_{\mathsf{D}})\) is given by
\begin{equation}
    \mathcal{M}_\theta(Z)
    \coloneqq 
    \trace_{\mathsf{A}\mathsf{O}}
    \!\left[
        \left(
            \id_{\mathcal{H}_{\mathsf{D}}\otimes\mathcal{H}_{\mathsf{A}}}
            \otimes
            P_\theta^{\mathsf{O}}
        \right)
        Z
        \left(
            \id_{\mathcal{H}_{\mathsf{D}}\otimes\mathcal{H}_{\mathsf{A}}}
            \otimes
            P_\theta^{\mathsf{O}}
        \right)
    \right],
    \label{eq:outcome-branch}
\end{equation}
for all \(Z\in\operatorname{End}(\mathcal{H}_{\mathsf{T}})\). 
Each \(\mathcal{M}_\theta\) is completely positive and trace nonincreasing, and Eq.~\eqref{eq:outcome-projector-resolution} gives
\begin{equation}
    \sum_{\theta\in\Theta}\mathcal{M}_\theta (\cdot)
    =
    \trace_{\mathsf{A}\mathsf{O}} (\cdot).
    \label{eq:outcome-branch-normalization}
\end{equation}
Let
\(\{W^{\mathsf{D}}(v)\}_{
    v\in V_{\mathsf{K}_{\mathsf{D}}}
}\)
be the leakage-extended Pauli family associated with \(\mathsf{D}\), as in Definition~\ref{defn:leakage-extended-Pauli-family}.  
Fix a Pauli-correction function 
\begin{equation}
    D: \Theta \to V_{\mathsf{K}_{\mathsf{D}}}.
    \label{eq:Pauli-correction-function}
\end{equation}
Fix a return function
\begin{equation}
    K: \Theta \to 2^{\mathsf{D}},
    \qquad K(\theta)\subseteq\mathsf{D},
    \label{eq:return-function}
\end{equation}
for all \(\theta\in\Theta\). 
For any \(J\subseteq\mathsf{D}\), define \(\mathfrak{c}_J\coloneqq \bigotimes_{i\in J}\mathfrak{c}_i\), and define the return map \(\mathcal{R}_J:\operatorname{End}(\mathcal{H}_{\mathsf{D}}) \to\operatorname{End}(\mathcal{H}_{\mathsf{D}})\) by
\begin{equation}
    \mathcal{R}_J(Z) \coloneqq  \frac{1}{2^{|J|}} \id_{\mathfrak{c}_J} \otimes \trace_J(Z), 
    \label{eq:return-map}
\end{equation}
for all \( Z\in\operatorname{End}(\mathcal{H}_{\mathsf{D}})\), where \(\id_{\mathfrak{c}_J}\) is extended by zero to all of \(\bigotimes_{i\in J}\mathfrak{h}_i\), \(\trace_J\) is the partial trace over the physical systems with labels in \(J\), and the output tensor factors are identified by the canonical labeled tensor-product identification.  
The map \(\mathcal{R}_J\) is CPTP.

For our purposes---with the stabilizer code, registers, computational bases, and Clifford frame fixed as above---the
specification of a noisy QEC cycle consists of the syndrome-readout unitary \(C_{\mathrm{synd}}\), the syndrome-readout map \(\mathcal{N}\), the Pauli-correction function \(D\), and the return function \(K\).

Define the associated complete-cycle map \(\Phi:\operatorname{End}(\mathcal{H}_{\mathsf{D}}) \to\operatorname{End}(\mathcal{H}_{\mathsf{D}})\) by
\begin{equation*}
    \Phi \coloneqq  \sum_{\theta\in\Theta} \mathcal{R}_{K(\theta)} \circ 
    \mathcal{W}^{\mathsf{D}}(D(\theta)) \circ \mathcal{M}_\theta \circ \mathcal{N} \circ \mathcal{I}_{\mathsf{A}\mathsf{O}}.
    \tag{\ref{eq:complete-cycle}}
\end{equation*}
The map \(\Phi\) is completely positive.  
For every \(Z\in\operatorname{End}(\mathcal{H}_{\mathsf{D}})\), Eq.~\eqref{eq:outcome-branch-normalization} and trace preservation of
the remaining constituent maps give
\begin{align}
    \trace\!\left(\Phi(Z)\right)
    &= \sum_{\theta\in\Theta}
    \trace\!\left(
        \mathcal{M}_\theta
        \bigl(
            \mathcal{N}\bigl(\mathcal{I}_{\mathsf{A}\mathsf{O}}(Z)\bigr)
        \bigr)
    \right) =
    \trace(Z).
    \label{eq:complete-cycle-trace-preservation}
\end{align}
Thus \(\Phi\) is a completely positive trace-preserving map on
\(\operatorname{End}(\mathcal{H}_{\mathsf{D}})\).

\subsection{Leakage-subspace structure of the twirled complete cycle}
\label{subsec:twirled complete-cycle-leakage-subspace-structure}

Let \(V_{\mathsf{K}_{\mathsf{T}}}\) be the Pauli-label space of \(\mathsf{K}_{\mathsf{T}}\) as in Definition~\ref{defn:Pauli-label-space}.  
Let \(F_{\mathrm{synd}}:V_{\mathsf{K}_{\mathsf{T}}} \to V_{\mathsf{K}_{\mathsf{T}}}\) be the symplectic automorphism associated with \(C_{\mathrm{synd}}\), as in Definition~\ref{defn:clifford-unitary}. 
Denote by \(\mathcal{T}^{\mathsf{T}}_{C_{\mathrm{synd}}}\) the leakage-dephased Pauli twirl, as in Definition~\ref{defn:leakage-dephased-Pauli-twirl}, specialized to \(\mathsf{T}\), \(C_{\mathrm{synd}}\), and \(F_{\mathrm{synd}}\).

Define the twirled syndrome-readout map \(\mathcal{N}^{\mathrm{tw}}: \operatorname{End}(\mathcal{H}_{\mathsf{T}}) \to \operatorname{End}(\mathcal{H}_{\mathsf{T}})\) by
\begin{equation*}
    \mathcal{N}^{\mathrm{tw}}
    \coloneqq 
    \mathcal{T}^{\mathsf{T}}_{C_{\mathrm{synd}}}(\mathcal{N}).
    \tag{\ref{eq:twirled-readout-map}}
\end{equation*}
Beacuse \(\mathcal{N}\) is CPTP, Eq.~\eqref{eq:leakage-dephased-Pauli-twirl} then implies that \(\mathcal{N}^{\mathrm{tw}}\) is also CPTP. 

Next, define the twirled complete-cycle map \(\Phi^{\mathrm{tw}}: \operatorname{End}(\mathcal{H}_{\mathsf{D}}) \to \operatorname{End}(\mathcal{H}_{\mathsf{D}})\) by
\begin{equation}
    \Phi^{\mathrm{tw}}
    \coloneqq 
    \sum_{\theta\in\Theta}
    \mathcal{R}_{K(\theta)}
    \circ
    \mathcal{W}^{\mathsf{D}}(D(\theta))
    \circ
    \mathcal{M}_\theta
    \circ
    \mathcal{N}^{\mathrm{tw}}
    \circ
    \mathcal{I}_{\mathsf{A}\mathsf{O}}.
    \label{eq:twirled complete-cycle}
\end{equation}
The map \(\Phi^{\mathrm{tw}}\) is completely positive.
It follows from Eq.~\eqref{eq:outcome-branch-normalization}---together with trace preservation of \(\mathcal{I}_{\mathsf{A}\mathsf{O}}\), \(\mathcal{N}^{\mathrm{tw}}\), \(\trace_{\mathsf{A}\mathsf{O}}\), \(\mathcal{W}^{\mathsf{D}}(D(\theta))\), and \(\mathcal{R}_{K(\theta)}\) for every \(\theta\in\Theta\)---that \(\Phi^{\mathrm{tw}}\) is trace preserving.

For each \(S\subseteq\mathsf{T}\), let \(\mathcal{P}_S^{\mathsf{T}}\), \(\mathcal{E}_S^{\mathsf{T}}\), \(\mathcal{F}_S^{\mathsf{T}}\), and \(\mathcal{L}_S^{\mathsf{T}}\) denote, respectively, the leakage-subspace projection map, leakage-subspace extension map, filling map, and leakage map (as in
Definitions~\ref{defn:leakage-subspace-projection-map}, \ref{defn:leakage-subspace-extension-map}, \ref{defn:filling-map}, and \ref{defn:leakage-map}, respectively), specialized to \(\mathsf{T}\).
Denote by \(\mathcal{T}^{\mathsf{K}_{\mathsf{T}}}_{C_{\mathrm{synd}}}\) the Pauli twirl with respect to \(C_{\mathrm{synd}}\) from Definition~\ref{defn:clifford-twisted-Pauli-twirl}, specialized to
\(\mathsf{K}_{\mathsf{T}}\), \(C_{\mathrm{synd}}\), and \(F_{\mathrm{synd}}\).

Using the Pauli twirl with respect to \(C_{\mathrm{synd}}\) (whose noisy implementation is given by \(\mathcal{N}\)), we define, for every \(U,T\subseteq\mathsf{T}\), the superoperator
\begin{equation}
    \mathcal{D}_{T, U}^{\mathcal{N}}
    \coloneqq 
    \mathcal{T}^{\mathsf{K}_{\mathsf{T}}}_{C_{\mathrm{synd}}}
    \!\left(
        \mathcal{F}_T^{\mathsf{T}}
        \circ
        \mathcal{P}_T^{\mathsf{T}}
        \circ
        \mathcal{N}
        \circ
        \mathcal{E}_U^{\mathsf{T}}
        \circ
        \mathcal{L}_U^{\mathsf{T}}
    \right)
    \circ
    \mathcal{C}_{\mathrm{synd}}^{-1} . 
    \label{eq:twirled-readout-leakage-transition-Pauli-map}
\end{equation}
The family \(\{\mathcal{D}_{T, U}^{\mathcal{N}}\}_{U,T\subseteq\mathsf{T}}\) is the leakage-transition Pauli-map family associated with \(\mathcal{N}\), as in Definition~\ref{defn:leakage-transition-Pauli-map-family}, specialized to \(\mathsf{T}\), \(C_{\mathrm{synd}}\), and \(F_{\mathrm{synd}}\).
Proposition~\ref{prop:leakage-transition-Pauli-map-normalization} then implies that every \(\mathcal{D}_{T, U}^{\mathcal{N}}\) is a subnormalized Pauli map.  
Combining Eq.~\eqref{eq:twirled-readout-map} with Theorem~\ref{prop:leakage-transition-Pauli-map-decomposition} leads to
\begin{equation}
    \mathcal{N}^{\mathrm{tw}}
    =
    \sum_{U,T\subseteq\mathsf{T}}
    \mathcal{E}_T^{\mathsf{T}}
    \circ
    \mathcal{L}_T^{\mathsf{T}}
    \circ
    \mathcal{D}_{T, U}^{\mathcal{N}}
    \circ
    \mathcal{C}^{\vpp}_{\mathrm{synd}}
    \circ
    \mathcal{F}_U^{\mathsf{T}}
    \circ
    \mathcal{P}_U^{\mathsf{T}},
    \label{eq:twirled-readout-leakage-transition-Pauli-map-decomposition}
\end{equation}
for the twirled version of the channel \(\mathcal{N}\) that captures noisy syndrome readout. 

For each \(R\subseteq\mathsf{D}\), denote by \(\mathcal{P}_R^{\mathsf{D}}\) and \(\mathcal{F}_R^{\mathsf{D}}\), respectively, the leakage-subspace projection map and filling map, as in Definitions~\ref{defn:leakage-subspace-projection-map} and \ref{defn:filling-map}, specialized to \(\mathsf{D}\). 
For each \(R\subseteq\mathsf{D}\), define \(U_R\subseteq\mathsf{T}\) to be the image of \(R\) under the canonical inclusion \(\mathsf{D}\hookrightarrow\mathsf{T}\). 

\begin{prop}
\label{prop:twirled complete-cycle-constituent-Pauli-map-expansion}
The twirled complete-cycle map \(\Phi^{\mathrm{tw}}\) 
satisfies
\begin{align}
    \Phi^{\mathrm{tw}}
    &= \sum_{\substack{R\subseteq\mathsf{D}\\T\subseteq\mathsf{T}}}
    \sum_{\theta\in\Theta}
    \mathcal{R}_{K(\theta)}
    \circ
    \mathcal{W}^{\mathsf{D}}(D(\theta))
    \circ
    \mathcal{M}_\theta \circ
    \mathcal{E}_T^{\mathsf{T}}
    \circ
    \mathcal{L}_T^{\mathsf{T}}
    \circ
    \mathcal{D}_{T, U_R}^{\mathcal{N}}
    \circ
    \mathcal{C}_{\mathrm{synd}}^{\vpp}
    \circ
    \mathcal{I}_{\mathsf{A}\mathsf{O}}
    \circ
    \mathcal{F}_R^{\mathsf{D}}
    \circ
    \mathcal{P}_R^{\mathsf{D}}.
    \label{eq:twirled complete-cycle-constituent-Pauli-map-expansion}
\end{align}
\end{prop}

\begin{proof}
Combining Eqs.~\eqref{eq:ancilla-outcome-initialization},
\eqref{eq:leakage-subspace-projection-map}, and~\eqref{eq:fill-map}
gives
\begin{equation}
    \mathcal{F}_U^{\mathsf{T}}
    \circ
    \mathcal{P}_U^{\mathsf{T}}
    \circ
    \mathcal{I}_{\mathsf{A}\mathsf{O}}
    =
    \begin{cases}
        \displaystyle
        \mathcal{I}_{\mathsf{A}\mathsf{O}}
        \circ
        \mathcal{F}_R^{\mathsf{D}}
        \circ
        \mathcal{P}_R^{\mathsf{D}},
        &U=U_R
        \text{ for some }R\subseteq\mathsf{D}
        \\[4pt]
        0
        &U\cap(\mathsf{A}\sqcup\mathsf{O})\neq\varnothing
    \end{cases} \, ,
    \label{eq:initialized-leakage-subspace-filling}
\end{equation}
for all \( U\subseteq\mathsf{T}\). 
Substituting Eq.~\eqref{eq:twirled-readout-leakage-transition-Pauli-map-decomposition} into Eq.~\eqref{eq:twirled complete-cycle} and applying Eq.~\eqref{eq:initialized-leakage-subspace-filling} termwise gives Eq.~\eqref{eq:twirled complete-cycle-constituent-Pauli-map-expansion}, completing the proof.
\end{proof}

\section{Syndrome-resolved logical structure of the QEC cycle}
\label{sec:leakage-aware-cycle}
\subsection{Setup}

For the remainder of this document, assume that the complete-cycle map \(\Phi\) in Eq.~\eqref{eq:complete-cycle} can be expressed as follows: there exists a family \(\{\mathcal{D}_{T, U}\}_{U,T\subseteq\mathsf{T}}\) of subnormalized Pauli maps on \(\operatorname{End}(\mathcal{K}_{\mathsf{T}})\) such that
\begin{align}
    \Phi
    &={}
    \sum_{\substack{
        R\subseteq\mathsf{D},\\
        T\subseteq\mathsf{T}
    }}
    \sum_{\theta\in\Theta}
    \mathcal{R}_{K(\theta)}
    \circ
    \mathcal{W}^{\mathsf{D}}(D(\theta))
    \circ
    \mathcal{M}_\theta \circ
    \mathcal{E}_T^{\mathsf{T}}
    \circ
    \mathcal{L}_T^{\mathsf{T}}
    \circ
    \mathcal{D}_{T, U_R}
    \circ
    \mathcal{C}_{\mathrm{synd}}^{\vpp}
    \circ
    \mathcal{I}_{\mathsf{A}\mathsf{O}}
    \circ
    \mathcal{F}_R^{\mathsf{D}}
    \circ
    \mathcal{P}_R^{\mathsf{D}},
    \label{eq:complete-cycle-projector-Markov-Pauli-form}
\end{align}
that further satisfies
\begin{equation}
    \sum_{T\subseteq\mathsf{T}}
    \operatorname{wt}\!\left(\mathcal{D}_{T, U}\right)= 1,
    \label{eq:complete-cycle-Pauli-map-family-normalization}
\end{equation}
for all \(U\subseteq\mathsf{T}\). 

Fix one such family.
Propositions~\ref{prop:twirled complete-cycle-constituent-Pauli-map-expansion} and~\ref{prop:leakage-transition-Pauli-map-normalization} show that \(\Phi^{\mathrm{tw}}\) and the family \(\{\mathcal{D}_{T, U}^{\mathcal{N}}\}_{U,T\subseteq\mathsf{T}}\) give one such realization.
The arguments below only invoke the fact that each \(\mathcal{D}_{T, U}\) is a subnormalized Pauli map and that Eqs.~\eqref{eq:complete-cycle-projector-Markov-Pauli-form} and \eqref{eq:complete-cycle-Pauli-map-family-normalization} hold; they do not depend on how these conditions are achieved.

Define \(\Delta^{\mathsf{D}}\) to be the leakage-pattern dephasing channel, as in Definition~\ref{defn:leakage-pattern-dephasing}, specialized to \(\mathsf{D}\). 
Equation~\eqref{eq:Delta-leakage-subspace-maps}, specialized to \(\mathsf{D}\), gives
\begin{equation}
    \Delta^{\mathsf{D}}
    =
    \sum_{R\subseteq\mathsf{D}}
    \mathcal{E}_R^{\mathsf{D}}
    \circ
    \mathcal{P}_R^{\mathsf{D}}.
    \label{eq:registerwise-leakage-dephasing}
\end{equation}
For \(T\subseteq\mathsf{T}\), define \(T_{\mathsf{D}}\coloneqq T\cap\mathsf{D}\).

\begin{prop}[Complete-cycle leakage-dephasing invariance]
\label{prop:complete-cycle-leakage-dephasing-from-Pauli-form}
The map \(\Phi\) satisfies
\begin{equation}
    \Phi
    =
    \Phi
    \circ
    \Delta^{\mathsf{D}}
    =
    \Delta^{\mathsf{D}}
    \circ
    \Phi
    =
    \Delta^{\mathsf{D}}
    \circ
    \Phi
    \circ
    \Delta^{\mathsf{D}}.
    \label{eq:complete-cycle-leakage-dephasing}
\end{equation}
\end{prop}

\begin{proof}
Equations~\eqref{eq:leakage-subspace}, \eqref{eq:leakage-subspace-extension-map}, \eqref{eq:leak-map}, and \eqref{eq:outcome-branch} give
\begin{equation}
    \operatorname{ran}\!\left(
        \mathcal{M}_\theta
        \circ
        \mathcal{E}_T^{\mathsf{T}}
        \circ
        \mathcal{L}_T^{\mathsf{T}}
    \right)
    \subseteq
    \operatorname{ran}(\mathcal{E}_{T_{\mathsf{D}}}^{\mathsf{D}}),
    \label{eq:outcome-branch-data-leakage-subspace-range}
\end{equation}
for all \(T\subseteq\mathsf{T},\ \theta\in\Theta\). 
Equations~\eqref{eq:leakage-extended-Pauli-subspace-commutation} and \eqref{eq:leakage-subspace-extension-map}, specialized to \(\mathsf{D}\), give
\begin{equation}
    \mathcal{W}^{\mathsf{D}}(v)
    \!\left(
        \operatorname{ran}(\mathcal{E}_S^{\mathsf{D}})
    \right)
    \subseteq
    \operatorname{ran}(\mathcal{E}_S^{\mathsf{D}}),
    \label{eq:data-leakage-subspace-Pauli-range-invariance}
\end{equation}
for all \(S\subseteq\mathsf{D}, \ v\in V_{\mathsf{K}_{\mathsf{D}}}\). 
Equations~\eqref{eq:return-map}, \eqref{eq:leakage-subspace}, and \eqref{eq:leakage-subspace-extension-map} give
\begin{equation}
    \operatorname{ran}\!\left(
        \mathcal{R}_{K(\theta)}
        \circ
        \mathcal{E}_{T_{\mathsf{D}}}^{\mathsf{D}}
    \right)
    \subseteq
    \operatorname{ran}\!\left(
        \mathcal{E}_{T_{\mathsf{D}}\setminus K(\theta)}^{\mathsf{D}}
    \right),
    \label{eq:return-map-data-leakage-subspace-range}
\end{equation}
for all \(T\subseteq\mathsf{T}, \ \theta\in\Theta\). 
Next, Eqs~\eqref{eq:outcome-branch-data-leakage-subspace-range}, \eqref{eq:data-leakage-subspace-Pauli-range-invariance}, and \eqref{eq:return-map-data-leakage-subspace-range} 
give
\begin{equation}
    \operatorname{ran}\!\left(
        \mathcal{R}_{K(\theta)}
        \circ
        \mathcal{W}^{\mathsf{D}}(D(\theta))
        \circ
        \mathcal{M}_\theta
        \circ
        \mathcal{E}_T^{\mathsf{T}}
        \circ
        \mathcal{L}_T^{\mathsf{T}}
    \right)
    \subseteq
    \operatorname{ran}\!\left(
        \mathcal{E}_{T_{\mathsf{D}}\setminus K(\theta)}^{\mathsf{D}}
    \right),
    \label{eq:complete-cycle-constituent-output-leakage-subspace-range}
\end{equation}
for all \(T\subseteq\mathsf{T}, \ \theta\in\Theta\). 
Substituting Eq.~\eqref{eq:registerwise-leakage-dephasing} and applying Eq.~\eqref{eq:cross-subspace-orthogonality} termwise, specialized to
\(\mathsf{D}\), gives
\begin{equation}
    \mathcal{P}_R^{\mathsf{D}}
    \circ
    \Delta^{\mathsf{D}}
    =
    \mathcal{P}_R^{\mathsf{D}},
    \qquad
    \Delta^{\mathsf{D}}
    \circ
    \mathcal{E}_R^{\mathsf{D}}
    =
    \mathcal{E}_R^{\mathsf{D}},
    \label{eq:data-leakage-subspace-dephasing-absorption}
\end{equation}
for all \(R\subseteq\mathsf{D}\). 
Right composing Eq.~\eqref{eq:complete-cycle-projector-Markov-Pauli-form} with \(\Delta^{\mathsf{D}}\) and applying
\(\mathcal{P}_R^{\mathsf{D}}\circ\Delta^{\mathsf{D}} =\mathcal{P}_R^{\mathsf{D}}\) from Eq.~\eqref{eq:data-leakage-subspace-dephasing-absorption} termwise gives \(\Phi\circ\Delta^{\mathsf{D}}=\Phi\).
Left composing Eq.~\eqref{eq:complete-cycle-projector-Markov-Pauli-form} with \(\Delta^{\mathsf{D}}\), using Eq.~\eqref{eq:complete-cycle-constituent-output-leakage-subspace-range}, and applying \(\Delta^{\mathsf{D}} \circ\mathcal{E}_{T_{\mathsf{D}}\setminus K(\theta)}^{\mathsf{D}} =\mathcal{E}_{T_{\mathsf{D}}\setminus K(\theta)}^{\mathsf{D}}\) from Eq.~\eqref{eq:data-leakage-subspace-dephasing-absorption} to each
term in the summand gives \(\Delta^{\mathsf{D}}\circ\Phi=\Phi\).
These identities imply that \(\Delta^{\mathsf{D}}\circ\Phi\circ\Delta^{\mathsf{D}}=\Phi\), proving Eq.~\eqref{eq:complete-cycle-leakage-dephasing}.
\end{proof}

\subsection{Clifford-frame representatives of complete-cycle leakage-subspace branches}
\label{subsec:filled-frame-representatives}

Denote by \(\{\Phi_{R_{\mathrm{out}}, R_{\mathrm{in}}}\}_{ R_{\mathrm{in}},R_{\mathrm{out}}\subseteq\mathsf{D}}\) the family of leakage-subspace branches associated with \(\Phi\), as in Definition~\ref{defn:leakage-subspace-branches}, specialized to \(\mathsf{D}\).

\begin{defn}[Clifford-frame representatives of complete-cycle
leakage-subspace branches]
\label{def:complete-cycle-filled-frame-representatives}
For every \(R_{\mathrm{in}},R_{\mathrm{out}}\subseteq\mathsf{D}\), define the \emph{Clifford-frame representative} \(\widetilde\Phi_{R_{\mathrm{out}}, R_{\mathrm{in}}}: \operatorname{End}(\mathcal{K}_{\mathsf{L}}\otimes\mathcal{K}_{\mathsf{S}}) \to \operatorname{End}(\mathcal{K}_{\mathsf{L}}\otimes\mathcal{K}_{\mathsf{S}})\) according to
\begin{equation}
    \widetilde\Phi_{R_{\mathrm{out}}, R_{\mathrm{in}}}
    \coloneqq 
    \mathcal{C}_{\mathrm{F}}^{\vpd}
    \circ
    \mathcal{F}_{R_{\mathrm{out}}}^{\mathsf{D}}
    \circ
    \Phi_{R_{\mathrm{out}}, R_{\mathrm{in}}}
    \circ
    \mathcal{L}_{R_{\mathrm{in}}}^{\mathsf{D}}
    \circ
    \mathcal{C}_{\mathrm{F}}^{-1}.
    \label{eq:complete-cycle-filled-frame-representative}
\end{equation}
\end{defn}
Lemma~\ref{lem:leakage-subspace-branch-properties} establishes that every \(\Phi_{R_{\mathrm{out}}, R_{\mathrm{in}}}\)  is completely
positive and trace nonincreasing.  
Unitarity of \(\CFrame\) implies that \(\mathcal{C}_{\mathrm{F}}\) and its inverse are CPTP.
Lemma~\ref{lem:filling-leakage-map-properties} and Eq.~\eqref{eq:complete-cycle-filled-frame-representative} therefore establish that every \(\widetilde\Phi_{R_{\mathrm{out}}, R_{\mathrm{in}}}\) is completely positive and trace nonincreasing.

\begin{prop}[Reconstruction from Clifford-frame representatives]
\label{prop:complete-cycle-filled-frame-reconstruction}
For every \(R_{\mathrm{in}},R_{\mathrm{out}}\subseteq\mathsf{D}\),
\begin{equation}
    \Phi_{R_{\mathrm{out}}, R_{\mathrm{in}}}
    =
    \mathcal{L}_{R_{\mathrm{out}}}^{\mathsf{D}}
    \circ
    \mathcal{C}_{\mathrm F}^{-1}
    \circ
    \widetilde\Phi_{R_{\mathrm{out}}, R_{\mathrm{in}}}
    \circ
    \mathcal{C}_{\mathrm F}^{\vpd}
    \circ
    \mathcal{F}_{R_{\mathrm{in}}}^{\mathsf{D}}.
    \label{eq:complete-cycle-filled-frame-representative-inversion}
\end{equation}
Moreover, 
\begin{align}
    \Phi
    &={}
    \sum_{R_{\mathrm{in}},R_{\mathrm{out}}\subseteq\mathsf{D}}
    \mathcal{E}_{R_{\mathrm{out}}}^{\mathsf{D}}
    \circ
    \Phi_{R_{\mathrm{out}}, R_{\mathrm{in}}}
    \circ
    \mathcal{P}_{R_{\mathrm{in}}}^{\mathsf{D}}
    \nonumber\\
    &={}
    \sum_{R_{\mathrm{in}},R_{\mathrm{out}}\subseteq\mathsf{D}}
    \mathcal{E}_{R_{\mathrm{out}}}^{\mathsf{D}}
    \circ
    \mathcal{L}_{R_{\mathrm{out}}}^{\mathsf{D}}
    \circ
    \mathcal{C}_{\mathrm{F}}^{-1}
    \circ
    \widetilde\Phi_{R_{\mathrm{out}}, R_{\mathrm{in}}}
    \circ
    \mathcal{C}_{\mathrm{F}}
    \circ
    \mathcal{F}_{R_{\mathrm{in}}}^{\mathsf{D}}
    \circ
    \mathcal{P}_{R_{\mathrm{in}}}^{\mathsf{D}}.
    \label{eq:complete-cycle-filled-frame-reconstruction}
\end{align}
\end{prop}

\begin{proof}
Unitarity of \(\CFrame\) gives
\begin{equation}
    \mathcal{C}_{\mathrm F}^{-1}
    \circ
    \mathcal{C}_{\mathrm F}^{\vpd}
    =
    \operatorname{id}_{
        \operatorname{End}(\mathcal{K}_{\mathsf{D}} )
    } \qquad \text{and} \qquad
    \mathcal{C}_{\mathrm F}^{\vpd}
    \circ
    \mathcal{C}_{\mathrm F}^{-1}
    =
    \operatorname{id}_{
        \operatorname{End}(\mathcal{K}_{\mathsf{L}}\otimes
        \mathcal{K}_{\mathsf{S}})
    }.
    \label{eq:Clifford-frame-channel-inverses}
\end{equation}
Substituting Eq.~\eqref{eq:complete-cycle-filled-frame-representative} into the right-hand side of Eq.~\eqref{eq:complete-cycle-filled-frame-representative-inversion} and applying Eqs.~\eqref{eq:Clifford-frame-channel-inverses} and \eqref{eq:leak-fill-identity} reduces that side to \(\Phi_{R_{\mathrm{out}}, R_{\mathrm{in}}}\).

Substituting Eq.~\eqref{eq:registerwise-leakage-dephasing} into both occurrences of \(\Delta^{\mathsf{D}}\) in the equality \(\Phi=\Delta^{\mathsf{D}}\circ\Phi\circ\Delta^{\mathsf{D}}\)
, distributing the compositions over the finite sums, and applying Eq.~\eqref{eq:leakage-subspace-branch} termwise gives the first equality in Eq.~\eqref{eq:complete-cycle-filled-frame-reconstruction}.
Substituting Eq.~\eqref{eq:complete-cycle-filled-frame-representative-inversion} termwise gives the second equality.
\end{proof}

For each \(R_{\mathrm{in}}\subseteq\mathsf{D}\), define \(U_{R_{\mathrm{in}}}\subseteq\mathsf{T}\) to be the image of \(R_{\mathrm{in}}\) under the canonical inclusion \(\mathsf{D}\hookrightarrow\mathsf{T}\).
For each such \(R_{\mathrm{in}}\), denote by \(\mathcal{G}_{U_{R_{\mathrm{in}}}}^{\mathsf{T}}\)  the computational replacement channel,  as in Definition~\ref{defn:computational-replacement-channel}, specialized to \(\mathsf{T}\) and \(U_{R_{\mathrm{in}}}\).

\begin{lem}[Explicit form of the Clifford-frame representatives]
\label{lem:complete-cycle-filled-frame-representative-form}
For every \(R_{\mathrm{in}},R_{\mathrm{out}}\subseteq\mathsf{D}\), 
\begin{align}
    \widetilde\Phi_{R_{\mathrm{out}}, R_{\mathrm{in}}}
    &={}
    \sum_{T\subseteq\mathsf{T}}
    \sum_{\theta\in\Theta}
    \mathcal{C}_{\mathrm F}
    \circ
    \mathcal{F}_{R_{\mathrm{out}}}^{\mathsf{D}}
    \circ
    \mathcal{P}_{R_{\mathrm{out}}}^{\mathsf{D}}
    \circ
    \mathcal{R}_{K(\theta)}
    \circ
    \mathcal{W}^{\mathsf{D}}(D(\theta))
    \circ
    \mathcal{M}_\theta
    \nonumber\\
    &\quad\circ
    \mathcal{E}_T^{\mathsf{T}}
    \circ
    \mathcal{L}_T^{\mathsf{T}}
    \circ
    \mathcal{D}_{T, U_{R_{\mathrm{in}}}}
    \circ
    \mathcal{C}_{\mathrm{synd}}^{\vpd}
    \circ
    \mathcal{G}_{U_{R_{\mathrm{in}}}}^{\mathsf{T}}
    \circ
    \mathcal{I}_{\mathsf{A}\mathsf{O}}
    \circ
    \mathcal{C}_{\mathrm F}^{-1}.
    \label{eq:complete-cycle-filled-frame-projector-form}
\end{align}
\end{lem}
\begin{proof}
Fix \(R_{\mathrm{in}},R_{\mathrm{out}}\subseteq\mathsf{D}\).
For \(R\neq R_{\mathrm{in}}\), Eq.~\eqref{eq:cross-subspace-orthogonality}, specialized to \(\mathsf{D}\), gives \( \mathcal{P}_R^{\mathsf{D}} \circ\mathcal{E}_{R_{\mathrm{in}}}^{\mathsf{D}}=0 \).
Left composing this equality with \(\mathcal{F}_R^{\mathsf{D}}\) and right composing with \(\mathcal{L}_{R_{\mathrm{in}}}^{\mathsf{D}}\) gives the
\(R\neq R_{\mathrm{in}}\) case below.
For \(R=R_{\mathrm{in}}\), Eq.~\eqref{eq:projection-extension-identity}, specialized to \(\mathsf{D}\), reduces the map  \( \mathcal{F}_R^{\mathsf{D}} \circ\mathcal{P}_R^{\mathsf{D}} \circ\mathcal{E}_{R_{\mathrm{in}}}^{\mathsf{D}} \circ\mathcal{L}_{R_{\mathrm{in}}}^{\mathsf{D}} \) to \( \mathcal{F}_{R_{\mathrm{in}}}^{\mathsf{D}} \circ\mathcal{L}_{R_{\mathrm{in}}}^{\mathsf{D}} \).
Let \(\mathcal{G}_{R_{\mathrm{in}}}^{\mathsf{D}}\) denote the computational replacement channel, as in Definition~\ref{defn:computational-replacement-channel}, specialized to \(\mathsf{D}\) and \(R_{\mathrm{in}}\).
Equation~\eqref{eq:computational-replacement-map}, specialized to \(\mathsf{D}\), gives \( \mathcal{F}_{R_{\mathrm{in}}}^{\mathsf{D}} \circ\mathcal{L}_{R_{\mathrm{in}}}^{\mathsf{D}} = \mathcal{G}_{R_{\mathrm{in}}}^{\mathsf{D}} \).
Hence, for every \(R\subseteq\mathsf{D}\), we find that
\begin{equation}
    \mathcal{F}_R^{\mathsf{D}}
    \circ
    \mathcal{P}_R^{\mathsf{D}}
    \circ
    \mathcal{E}_{R_{\mathrm{in}}}^{\mathsf{D}}
    \circ
    \mathcal{L}_{R_{\mathrm{in}}}^{\mathsf{D}}
    =
    \begin{cases}
        \mathcal{G}_{R_{\mathrm{in}}}^{\mathsf{D}} 
        &R=R_{\mathrm{in}},
        \\[2pt]
        0_{\operatorname{End}(\mathcal{K}_{\mathsf{D}} )
        \to\operatorname{End}(\mathcal{K}_{\mathsf{D}} )} 
        &R\neq R_{\mathrm{in}}.
    \end{cases}
    \label{eq:complete-cycle-input-subspace-replacement}
\end{equation}
Right composing Eq.~\eqref{eq:complete-cycle-projector-Markov-Pauli-form} with \(\mathcal{E}_{R_{\mathrm{in}}}^{\mathsf{D}} \circ\mathcal{L}_{R_{\mathrm{in}}}^{\mathsf{D}}\) and applying Eq.~\eqref{eq:complete-cycle-input-subspace-replacement} termwise annihilates the summands with \(R\neq R_{\mathrm{in}}\).
The remaining summands lead to
\begin{align}
    \Phi
    \circ
    \mathcal{E}_{R_{\mathrm{in}}}^{\mathsf{D}}
    \circ
    \mathcal{L}_{R_{\mathrm{in}}}^{\mathsf{D}}
    &={}
    \sum_{T\subseteq\mathsf{T}}
    \sum_{\theta\in\Theta}
    \mathcal{R}_{K(\theta)}
    \circ
    \mathcal{W}^{\mathsf{D}}(D(\theta))
    \circ
    \mathcal{M}_\theta
    \nonumber\\
    &\quad\circ
    \mathcal{E}_T^{\mathsf{T}}
    \circ
    \mathcal{L}_T^{\mathsf{T}}
    \circ
    \mathcal{D}_{T, U_{R_{\mathrm{in}}}}
    \circ
    \mathcal{C}_{\mathrm{synd}}^{\vpp}
    \circ
    \mathcal{I}_{\mathsf{A}\mathsf{O}}
    \circ
    \mathcal{G}_{R_{\mathrm{in}}}^{\mathsf{D}}.
    \label{eq:complete-cycle-fixed-input-before-embedding}
\end{align}
For every \(Y\in\operatorname{End}(\mathcal{K}_{\mathsf{D}} )\), Eq.~\eqref{eq:computational-replacement-action}, specialized to \(\mathsf{D}\) with \(S=R_{\mathrm{in}}\) and to \(\mathsf{T}\) with \(S=U_{R_{\mathrm{in}}}\), together with Eq.~\eqref{eq:ancilla-outcome-initialization}, gives
\begin{align}
    \bigl(
        \mathcal{I}_{\mathsf{A}\mathsf{O}}
        \circ
        \mathcal{G}_{R_{\mathrm{in}}}^{\mathsf{D}}
    \bigr)(Y)
    &={}
    \mathcal{G}_{R_{\mathrm{in}}}^{\mathsf{D}}(Y)
    \otimes
    \ZProj{0}_{\mathsf{A}}
    \otimes
    \ZProj{0}_{\mathsf{O}}
    \nonumber\\
    &={}
    \mathcal{G}_{U_{R_{\mathrm{in}}}}^{\mathsf{T}}
    \!\left(
        Y
        \otimes
        \ZProj{0}_{\mathsf{A}}
        \otimes
        \ZProj{0}_{\mathsf{O}}
    \right)
    \nonumber\\
    &={}
    \bigl(
        \mathcal{G}_{U_{R_{\mathrm{in}}}}^{\mathsf{T}}
        \circ
        \mathcal{I}_{\mathsf{A}\mathsf{O}}
    \bigr)(Y).
    \label{eq:complete-cycle-embedded-input-replacement-action}
\end{align}
Since the foregoing expression holds for every \(Y\in\operatorname{End}(\mathcal{K}_{\mathsf{D}} )\), it follows that
\begin{equation}
    \mathcal{I}_{\mathsf{A}\mathsf{O}}
    \circ
    \mathcal{G}_{R_{\mathrm{in}}}^{\mathsf{D}}
    =
    \mathcal{G}_{U_{R_{\mathrm{in}}}}^{\mathsf{T}}
    \circ
    \mathcal{I}_{\mathsf{A}\mathsf{O}}.
    \label{eq:complete-cycle-embedded-input-replacement}
\end{equation}
Substituting the relation above into each summand of Eq.~\eqref{eq:complete-cycle-fixed-input-before-embedding}, left composing with \(\mathcal{C}_{\mathrm F} \circ\mathcal{F}_{R_{\mathrm{out}}}^{\mathsf{D}} \circ\mathcal{P}_{R_{\mathrm{out}}}^{\mathsf{D}}\), and right composing with \(\mathcal{C}_{\mathrm F}^{-1}\) gives Eq.~\eqref{eq:complete-cycle-filled-frame-projector-form}, by Eqs.~\eqref{eq:leakage-subspace-branch} and~\eqref{eq:complete-cycle-filled-frame-representative}.
\end{proof}

\subsection{Syndrome-conditioned Pauli reduction of sector branches}
\label{subsec:conditioned-Pauli-reduction}

Denote by \(V_{\mathsf{L}}\) and \(V_{\mathsf{S}}\) the Pauli-label spaces of \(\mathsf{L}\) and \(\mathsf{S}\), respectively, as in
Definition~\ref{defn:Pauli-label-space}.

\begin{defn}[Syndrome-subspace selection maps]
\label{def:syndrome-block-projection-maps}
For each \(s\in\mathbb{Z}_2^{\mathsf{S}}\), define the map \(\mathcal{Z}_s: \operatorname{End}(\mathcal{K}_{\mathsf{L}}\otimes\mathcal{K}_{\mathsf{S}}) \to \operatorname{End}(\mathcal{K}_{\mathsf{L}}\otimes\mathcal{K}_{\mathsf{S}})\) according to
\begin{equation}
    \mathcal{Z}_s(Y)
    \coloneqq 
    \left(
        \id_{\mathsf{L}}\otimes\ZProj{s}_{\mathsf{S}}
    \right)
    Y
    \left(
        \id_{\mathsf{L}}\otimes\ZProj{s}_{\mathsf{S}}
    \right) ,
    \label{eq:syndrome-block-selection-map}
\end{equation}
for all \( Y \in \operatorname{End}(\mathcal{K}_{\mathsf{L}}\otimes\mathcal{K}_{\mathsf{S}})\), where \(\id_{\mathsf{L}}\) denotes the identity on \(\mathcal{K}_{\mathsf{L}}\).
\end{defn}

\begin{defn}[Syndrome-dephasing map]
\label{def:syndrome-dephasing-channel}
Define the map \(\mathcal{Z}_{\mathsf{S}}: \operatorname{End}(\mathcal{K}_{\mathsf{L}}\otimes\mathcal{K}_{\mathsf{S}}) \to \operatorname{End}(\mathcal{K}_{\mathsf{L}}\otimes\mathcal{K}_{\mathsf{S}})\) according to
\begin{equation}
    \mathcal{Z}_{\mathsf{S}}
    \coloneqq 
    \sum_{s\in\mathbb{Z}_2^{\mathsf{S}}}\mathcal{Z}_s,
    \label{eq:syndrome-dephasing-channel}
\end{equation}
with \(\mathcal{Z}_s\) as defined in Definition~\ref{def:syndrome-block-projection-maps}. 
\end{defn}

\begin{lem}
\label{lem:syndrome-dephasing-matrix-units}
For every \(X\in\operatorname{End}(\mathcal{K}_{\mathsf{L}})\) and \(s,s'\in\mathbb{Z}_2^{\mathsf{S}}\),
\begin{equation}
    \mathcal{Z}_{\mathsf{S}}
    \left(
        X\otimes\KetBra{s}{s'}_{\mathsf{S}}
    \right)
    =
    \delta_{s,s'}
    X\otimes\ZProj{s}_{\mathsf{S}}.
    \label{eq:syndrome-dephasing-matrix-units}
\end{equation}
\end{lem}

\begin{proof}
For every \(t,s,s'\in\mathbb{Z}_2^{\mathsf{S}}\) and
\(X\in\operatorname{End}(\mathcal{K}_{\mathsf{L}})\),
Eq.~\eqref{eq:syndrome-block-selection-map} gives
\begin{equation*}
    \mathcal{Z}_t
    \left(
        X\otimes \KetBra{s}{s'}_{\mathsf{S}}
    \right)
    =
    \delta_{t,s}\delta_{t,s'}
    X\otimes\ZProj{t}_{\mathsf{S}}.
\end{equation*}
Summing this equality over \(t\) and invoking Eq.~\eqref{eq:syndrome-dephasing-channel} recovers Eq.~\eqref{eq:syndrome-dephasing-matrix-units}.
\end{proof}

\begin{lem}
\label{lem:syndrome-dephasing-Pauli-channel}
The map \(\mathcal{Z}_{\mathsf{S}}\) satisfies
\begin{equation}
    \mathcal{Z}_{\mathsf{S}}
    =
    \frac{1}{2^{|\mathsf{S}|}}
    \sum_{z\in\mathbb{Z}_2^{\mathsf{S}}}
    \operatorname{Ad}_{
        \id_{\mathsf{L}}\otimes W^{\mathsf{S}}(0,z)
    },
    \label{eq:syndrome-dephasing-Pauli-channel}
\end{equation}
and is a CPTP Pauli channel.
\end{lem}

\begin{proof}
For every \(s,s'\in\mathbb{Z}_2^{\mathsf{S}}\),
\begin{align}
    &\frac{1}{2^{|\mathsf{S}|}}
    \sum_{z\in\mathbb{Z}_2^{\mathsf{S}}}
    (-1)^{z\mathbin{\cdot}(s+s')}
    =
    \prod_{i\in\mathsf{S}}
    \left[
        \frac{1}{2}
        \sum_{z_i\in\mathbb{Z}_2}
        (-1)^{z_i(s_i+s'_i)}
    \right]
    =
    \prod_{i\in\mathsf{S}}
    \frac{1+(-1)^{s_i+s'_i}}{2}
    =
    \delta_{s,s'}.
    \label{eq:syndrome-character-sum}
\end{align}
Combining the foregoing relation with Eqs.~\eqref{eq:computational-system-Pauli-generators}, \eqref{eq:app-Pauli-definition},
\eqref{eq:syndrome-character-sum}, and \eqref{eq:syndrome-dephasing-matrix-units} gives, for every \(X\in\operatorname{End}(\mathcal{K}_{\mathsf{L}})\) and \(s,s'\in\mathbb{Z}_2^{\mathsf{S}}\),
\begin{align}
    &\left[
        \frac{1}{2^{|\mathsf{S}|}}
        \sum_{z\in\mathbb{Z}_2^{\mathsf{S}}}
        \operatorname{Ad}_{
            \id_{\mathsf{L}}\otimes W^{\mathsf{S}}(0,z)
        }
    \right]
    \left(
        X\otimes\KetBra{s}{s'}_{\mathsf{S}}
    \right)
    \nonumber\\
    &\quad={}
    \frac{1}{2^{|\mathsf{S}|}}
    \sum_{z\in\mathbb{Z}_2^{\mathsf{S}}}
    (-1)^{z\mathbin{\cdot}(s+s')}
    X\otimes\KetBra{s}{s'}_{\mathsf{S}}
    \nonumber\\
    &\quad={}
    \delta_{s,s'}
    X\otimes\ZProj{s}_{\mathsf{S}}
    =
    \mathcal{Z}_{\mathsf{S}}
    \left(
        X\otimes\KetBra{s}{s'}_{\mathsf{S}}
    \right).
    \label{eq:syndrome-dephasing-Pauli-average-matrix-units}
\end{align}
The operators \(X\otimes\KetBra{s}{s'}_{\mathsf{S}} \) (with \(X\in\operatorname{End}(\mathcal{K}_{\mathsf{L}})\) and
\(s,s'\in\mathbb{Z}_2^{\mathsf{S}}\)) span \(\operatorname{End}(\mathcal{K}_{\mathsf{L}}\otimes\mathcal{K}_{\mathsf{S}})\), so Eq.~\eqref{eq:syndrome-dephasing-Pauli-channel} follows from Eq.~\eqref{eq:syndrome-dephasing-Pauli-average-matrix-units}.
Equation~\eqref{eq:syndrome-dephasing-Pauli-channel} and Definition~\ref{defn:pauli-map-normalization} establish that \(\mathcal{Z}_{\mathsf{S}}\) is a Pauli channel.
Lemma~\ref{lem:pauli-map-positivity-and-normalization} therefore implies that \(\mathcal{Z}_{\mathsf{S}}\) is CPTP.
\end{proof}

\begin{lem}
\label{lem:syndrome-dephasing-Pauli-commutation}
For every Pauli map \(\mathcal{D}\) on
\(\operatorname{End}(\mathcal{K}_{\mathsf{L}}\otimes\mathcal{K}_{\mathsf{S}})\),
\begin{equation}
    \mathcal{D}\circ\mathcal{Z}_{\mathsf{S}}
    =
    \mathcal{Z}_{\mathsf{S}}\circ\mathcal{D}.
    \label{eq:syndrome-dephasing-Pauli-commutation}
\end{equation}
\end{lem}

\begin{proof}
Becasue \(V_{\mathsf{L}}\times V_{\mathsf{S}}\) is abelian, applying  Eq.~\eqref{eq:app-Pauli-conjugation-action} and Lemma~\ref{lem:app-Pauli-representation} to \(\mathcal{K}_{\mathsf{L}}\otimes\mathcal{K}_{\mathsf{S}}\) shows that the maps \( \operatorname{Ad}_{W^{\mathsf{L}}(\ell)\otimes W^{\mathsf{S}}(r)}\), with \((\ell,r)\in V_{\mathsf{L}}\times V_{\mathsf{S}}\), commute with one another under composition.
Combining Eqs.~\eqref{eq:pauli-map-def} and \eqref{eq:syndrome-dephasing-Pauli-channel} proves Eq.~\eqref{eq:syndrome-dephasing-Pauli-commutation}.
\end{proof}

For every leakage pattern \(R\subseteq\mathsf{D}\), denote by \(V_R^{\mathsf{D}}\subseteq V_{\mathsf{K}_{\mathsf{D}}}\) the subgroup of Pauli labels supported on \(R\)
, specialized to \(\mathsf{K}_{\mathsf{D}}\).  
For every \(T\subseteq\mathsf{T}\), denote by \(V_T^{\mathsf{T}}\subseteq V_{\mathsf{K}_{\mathsf{T}}}\) the corresponding subgroup specialized to \(\mathsf{K}_{\mathsf{T}}\).

\begin{prop}
\label{prop:complete-cycle-conditioned-Pauli-form}
There exists a family of Pauli maps on \(\operatorname{End}(\mathcal{K}_{\mathsf{D}} )\), denoted 
\begin{equation*}
    \left\{
        \mathcal{B}^{R_{\mathrm{out}}, R_{\mathrm{in}}}_s
    \right\}^{R_{\mathrm{in}},R_{\mathrm{out}}\subseteq\mathsf{D}}_{s\in\mathbb{Z}_2^{\mathsf{S}}}, 
\end{equation*}
such that, for every \(R_{\mathrm{in}},R_{\mathrm{out}}\subseteq\mathsf{D}\), we have that
\begin{equation}
    \widetilde\Phi_{R_{\mathrm{out}}, R_{\mathrm{in}}}
    =
    \sum_{s\in\mathbb{Z}_2^{\mathsf{S}}}
    \mathcal{C}_{\mathrm F}^{\vpd}
    \circ
    \mathcal{B}^{R_{\mathrm{out}}, R_{\mathrm{in}}}_s
    \circ
    \mathcal{C}_{\mathrm F}^{-1}
    \circ
    \mathcal{Z}_s.
    \label{eq:complete-cycle-conditioned-Pauli-form}
\end{equation}
\end{prop}

\begin{proof}

Fix \(R_{\mathrm{in}},R_{\mathrm{out}}\subseteq\mathsf{D}\), \(X\in\operatorname{End}(\mathcal{K}_{\mathsf{L}})\), and \(s,s'\in\mathbb{Z}_2^{\mathsf{S}}\).
For \(T\subseteq\mathsf{T}\), write
\begin{equation*}
    T_{\mathsf{D}}\coloneqq T\cap\mathsf{D},
    \qquad
    T_{\mathsf{A}}\coloneqq T\cap\mathsf{A},
    \qquad \text{and} \qquad
    T_{\mathsf{O}}\coloneqq T\cap\mathsf{O}.
\end{equation*}
For \(v\in V_{\mathsf{K}_{\mathsf{T}}}\), denote by \(v_{\mathsf{D}}\in V_{\mathsf{K}_{\mathsf{D}}}\), \(v_{\mathsf{A}}\in V_{\mathsf{K}_{\mathsf{A}}}\), and \(v_{\mathsf{O}}\in V_{\mathsf{K}_{\mathsf{O}}}\), respectively, the restriction of \(v\) to \(\mathsf{D}\), \(\mathsf{A}\), and \(\mathsf{O}\). 
Evaluating Eq.~\eqref{eq:complete-cycle-filled-frame-projector-form} on \(X\otimes\KetBra{s}{s'}_{\mathsf{S}}\) gives
\begin{align}
    &\widetilde\Phi_{R_{\mathrm{out}}, R_{\mathrm{in}}}
    \left(
        X\otimes\KetBra{s}{s'}_{\mathsf{S}}
    \right)
    \nonumber\\
    &\quad={}
    \sum_{T\subseteq\mathsf{T}}
    \sum_{\theta\in\Theta}
    \left[
        \begin{aligned}
        &\mathcal{C}_{\mathrm F}^{\vpd}
        \circ
        \mathcal{F}_{R_{\mathrm{out}}}^{\mathsf{D}}
        \circ
        \mathcal{P}_{R_{\mathrm{out}}}^{\mathsf{D}}
        \circ
        \mathcal{R}_{K(\theta)}
        \\
        &\quad\circ
        \mathcal{W}^{\mathsf{D}}(D(\theta))
        \circ
        \mathcal{M}_\theta
        \circ
        \mathcal{E}_T^{\mathsf{T}}
        \circ
        \mathcal{L}_T^{\mathsf{T}}
        \\
        &\quad\circ
        \mathcal{D}_{T, U_{R_{\mathrm{in}}}}
        \circ
        \mathcal{C}^{\vpd}_{\mathrm{synd}}
        \circ
        \mathcal{G}_{U_{R_{\mathrm{in}}}}^{\mathsf{T}}
        \circ
        \mathcal{I}_{\mathsf{A}\mathsf{O}}
        \circ
        \mathcal{C}_{\mathrm F}^{-1}
        \end{aligned}
    \right]
    \left(
        X\otimes\KetBra{s}{s'}_{\mathsf{S}}
    \right).
    \label{eq:conditioned-starting-point}
\end{align}
For every \(T\subseteq\mathsf{T}\) and \(e\in V_{\mathsf{K}_{\mathsf{T}}}\), define
\begin{equation}
    q_e^{T, U_{R_{\mathrm{in}}}}
    \coloneqq 
    \frac{1}{
        \bigl(\dim\mathcal{K}_{\mathsf{T}}\bigr)^2
    }
    \left\langle
        \mathcal{W}^{\mathsf{K}_{\mathsf{T}}}(e),
        \mathcal{D}_{T, U_{R_{\mathrm{in}}}}
    \right\rangle_{\mathcal{S}}.
    \label{eq:conditioned-Pauli-coefficient}
\end{equation}
By the standing assumption on \(\{\mathcal{D}_{T, U}\}_{U,T\subseteq\mathsf{T}}\) in Appendix~\ref{sec:leakage-aware-cycle}, every \(\mathcal{D}_{T, U_{R_{\mathrm{in}}}}\) is a Pauli map.
Thus, Lemma~\ref{lem:pauli-map-structure}, specialized to \(\mathsf{K}_{\mathsf{T}}\), gives
\begin{equation}
    \mathcal{D}_{T, U_{R_{\mathrm{in}}}}
    =
    \sum_{e\in V_{\mathsf{K}_{\mathsf{T}}}}
    q_e^{T, U_{R_{\mathrm{in}}}}
    \mathcal{W}^{\mathsf{K}_{\mathsf{T}}}(e).
    \label{eq:conditioned-Pauli-expansion}
\end{equation}
Equation~\eqref{eq:computational-replacement-Pauli-channel}, specialized to \(\mathsf{T}\) with \(S=U_{R_{\mathrm{in}}}\), gives
\begin{equation}
    \mathcal{G}_{U_{R_{\mathrm{in}}}}^{\mathsf{T}}
    =
    \frac{1}{4^{|U_{R_{\mathrm{in}}}|}}
    \sum_{u\in V_{U_{R_{\mathrm{in}}}}^{\mathsf{T}}}
    \mathcal{W}^{\mathsf{K}_{\mathsf{T}}}(u).
    \label{eq:conditioned-input-replacement-expansion}
\end{equation}
For every \(e\in V_{\mathsf{K}_{\mathsf{T}}}\) and \(u\in V_{U_{R_{\mathrm{in}}}}^{\mathsf{T}}\), define
\begin{equation}
    v_{e,u}
    \coloneqq 
    e+F_{\mathrm{synd}}u.
    \label{eq:conditioned-combined-Pauli-label}
\end{equation}
Combining Eq.~\eqref{eq:app-Clifford-action}, specialized to \(C_{\mathrm{synd}}\) and \(F_{\mathrm{synd}}\), with Eqs.~\eqref{eq:app-Pauli-product} and \eqref{eq:app-adjoint-action-composition} gives
\begin{align}
    & \mathcal{W}^{\mathsf{K}_{\mathsf{T}}}(e)
    \circ
    \mathcal{C}_{\mathrm{synd}}^{\vpd}
    \circ
    \mathcal{W}^{\mathsf{K}_{\mathsf{T}}}(u)
    \nonumber\\
    &\quad={}
    \mathcal{W}^{\mathsf{K}_{\mathsf{T}}}(e)
    \circ
    \mathcal{W}^{\mathsf{K}_{\mathsf{T}}}(F_{\mathrm{synd}}u)
    \circ
    \mathcal{C}_{\mathrm{synd}}^{\vpd}
    \nonumber\\
    &\quad={}
    \mathcal{W}^{\mathsf{K}_{\mathsf{T}}}(e+F_{\mathrm{synd}}u)
    \circ
    \mathcal{C}_{\mathrm{synd}}^{\vpd}
    =
    \mathcal{W}^{\mathsf{K}_{\mathsf{T}}}(v_{e,u})
    \circ
    \mathcal{C}_{\mathrm{synd}}^{\vpd}.
    \label{eq:conditioned-Clifford-Pauli composition}
\end{align}
Substituting Eqs.~\eqref{eq:conditioned-Pauli-expansion} and \eqref{eq:conditioned-input-replacement-expansion} into Eq.~\eqref{eq:conditioned-starting-point} and applying Eq.~\eqref{eq:conditioned-Clifford-Pauli composition} termwise gives
\begin{align}
    &\widetilde\Phi_{R_{\mathrm{out}}, R_{\mathrm{in}}}
    \left(
        X\otimes\KetBra{s}{s'}_{\mathsf{S}}
    \right)
    \nonumber\\
    &\quad={}
    \sum_{T\subseteq\mathsf{T}}
    \sum_{\theta\in\Theta}
    \sum_{e\in V_{\mathsf{K}_{\mathsf{T}}}}
    \sum_{u\in V_{U_{R_{\mathrm{in}}}}^{\mathsf{T}}}
    \frac{
        q_e^{T, U_{R_{\mathrm{in}}}}
    }{
        4^{|U_{R_{\mathrm{in}}}|}
    }
    \left[
        \begin{aligned}
        &\mathcal{C}_{\mathrm F}
        \circ
        \mathcal{F}_{R_{\mathrm{out}}}^{\mathsf{D}}
        \circ
        \mathcal{P}_{R_{\mathrm{out}}}^{\mathsf{D}}
        \circ
        \mathcal{R}_{K(\theta)}
        \\
        &\quad\circ
        \mathcal{W}^{\mathsf{D}}(D(\theta))
        \circ
        \mathcal{M}_\theta
        \circ
        \mathcal{E}_T^{\mathsf{T}}
        \circ
        \mathcal{L}_T^{\mathsf{T}}
        \\
        &\quad\circ
        \mathcal{W}^{\mathsf{K}_{\mathsf{T}}}(v_{e,u})
        \circ
        \mathcal{C}_{\mathrm{synd}}^{\vpd}
        \circ
        \mathcal{I}_{\mathsf{A}\mathsf{O}}
        \circ
        \mathcal{C}_{\mathrm F}^{-1}
        \end{aligned}
    \right]
    \left(
        X\otimes\KetBra{s}{s'}_{\mathsf{S}}
    \right).
    \label{eq:conditioned-expanded}
\end{align}
For every \(e\in V_{\mathsf{K}_{\mathsf{T}}}\) and \(u\in V_{U_{R_{\mathrm{in}}}}^{\mathsf{T}}\), write
\begin{equation*}
    (v_{e,u})_{\mathsf{O}}
    =
    (x_{e,u,\mathsf{O}},z_{e,u,\mathsf{O}}),
\end{equation*}
where \(x_{e,u,\mathsf{O}},z_{e,u,\mathsf{O}} \in\mathbb{Z}_2^{\mathsf{O}}\).
For every \(e\in V_{\mathsf{K}_{\mathsf{T}}}\) and \(u\in V_{U_{R_{\mathrm{in}}}}^{\mathsf{T}}\), define the map \(r_{e,u}:\mathbb{Z}_2^{\mathsf{S}} \to\mathbb{Z}_2^{\mathsf{O}}\) via
\begin{equation}
    r_{e,u}(t)
    \coloneqq 
    E(t)+x_{e,u,\mathsf{O}} , 
    \label{eq:conditioned-shifted-outcome-record}
\end{equation}
for all \(t\in\mathbb{Z}_2^{\mathsf{S}}\). 
For every \(T\subseteq\mathsf{T}\), define the map \(\theta_T:\mathbb{Z}_2^{\mathsf{O}}\to\Theta\) according to
\begin{equation}
    [\theta_T(r)]_i
    \coloneqq 
    \begin{cases}
        \mathrm L,
        &i\in T_{\mathsf{O}},\\
        r_i,
        &i\in\mathsf{O}\setminus T_{\mathsf{O}},
    \end{cases} \, , 
    \label{eq:conditioned-outcome-record}
\end{equation}
for all \( r\in\mathbb{Z}_2^{\mathsf{O}},\; i\in\mathsf{O} \). 
For every \(e\in V_{\mathsf{K}_{\mathsf{T}}}\) and \(u\in V_{U_{R_{\mathrm{in}}}}^{\mathsf{T}}\), Eq.~\eqref{eq:app-Pauli-definition} gives
\begin{equation}
    W^{\mathsf{K}_{\mathsf{T}}}(v_{e,u})
    =
    W^{\mathsf{K}_{\mathsf{D}}}((v_{e,u})_{\mathsf{D}})
    \otimes
    W^{\mathsf{K}_{\mathsf{A}}}((v_{e,u})_{\mathsf{A}})
    \otimes
    W^{\mathsf{K}_{\mathsf{O}}}((v_{e,u})_{\mathsf{O}}).
    \label{eq:conditioned-full-register-Pauli-factorization}
\end{equation}
For every \(e\in V_{\mathsf{K}_{\mathsf{T}}}\) and \(u\in V_{U_{R_{\mathrm{in}}}}^{\mathsf{T}}\), taking outer products in Eq.~\eqref{eq:nominal-readout-vector-action}, and applying Eqs.~\eqref{eq:syndrome-readout-frame-conjugation}, \eqref{eq:ancilla-outcome-initialization}, \eqref{eq:app-Pauli-definition}, and \eqref{eq:conditioned-full-register-Pauli-factorization}, gives
\begin{align}
    &\left(
        \mathcal{W}^{\mathsf{K}_{\mathsf{T}}}(v_{e,u})
        \circ
        \mathcal{C}_{\mathrm{synd}}^{\vpd}
        \circ
        \mathcal{I}_{\mathsf{A}\mathsf{O}}
        \circ
        \mathcal{C}_{\mathrm F}^{-1}
    \right)
    \left(
        X\otimes\KetBra{s}{s'}_{\mathsf{S}}
    \right)
    \nonumber\\
    &\quad={}
    \left(
        \mathcal{W}^{\mathsf{K}_{\mathsf{D}}}(
            (v_{e,u})_{\mathsf{D}}
        )
        \circ
        \mathcal{C}_{\mathrm F}^{-1}
    \right)
    \left(
        X\otimes\KetBra{s}{s'}_{\mathsf{S}}
    \right)
    \nonumber\\
    &\qquad\otimes
    \mathcal{W}^{\mathsf{K}_{\mathsf{A}}}((v_{e,u})_{\mathsf{A}})
    \left( \KetBra{\alpha(s)}{\alpha(s')}_{\mathsf{A}}
    \right)
    \nonumber\\
    &\qquad\otimes
    (-1)^{z_{e,u,\mathsf{O}}\mathbin{\cdot}(E(s)+E(s'))}
    \KetBra{r_{e,u}(s)}{r_{e,u}(s')}_{\mathsf{O}} . 
    \label{eq:conditioned-registerwise-Pauli-action}
\end{align}
Substituting Eq.~\eqref{eq:conditioned-registerwise-Pauli-action} termwise into Eq.~\eqref{eq:conditioned-expanded} leads to
\begin{align}
    &\widetilde\Phi_{R_{\mathrm{out}}, R_{\mathrm{in}}}
    \left(
        X\otimes\KetBra{s}{s'}_{\mathsf{S}}
    \right)
    \nonumber\\
    &\quad={}
    \sum_{T\subseteq\mathsf{T}}
    \sum_{\theta\in\Theta}
    \sum_{e\in V_{\mathsf{K}_{\mathsf{T}}}}
    \sum_{u\in V_{U_{R_{\mathrm{in}}}}^{\mathsf{T}}}
    \frac{
        q_e^{T, U_{R_{\mathrm{in}}}}
    }{
        4^{|U_{R_{\mathrm{in}}}|}
    }
    \left[
        \begin{aligned}
        &\mathcal{C}_{\mathrm F}^{\vpd}
        \circ
        \mathcal{F}_{R_{\mathrm{out}}}^{\mathsf{D}}
        \circ
        \mathcal{P}_{R_{\mathrm{out}}}^{\mathsf{D}}
        \circ
        \mathcal{R}_{K(\theta)}
        \\
        &\quad\circ
        \mathcal{W}^{\mathsf{D}}(D(\theta))
        \circ
        \mathcal{M}_\theta
        \circ
        \mathcal{E}_T^{\mathsf{T}}
        \circ
        \mathcal{L}_T^{\mathsf{T}}
        \end{aligned}
    \right]
    \nonumber\\
    &\qquad\left(
        \begin{aligned}
        &\left(
            \mathcal{W}^{\mathsf{K}_{\mathsf{D}}}(
                (v_{e,u})_{\mathsf{D}}
            )
            \circ
            \mathcal{C}_{\mathrm F}^{-1}
        \right)
        \left(
            X\otimes\KetBra{s}{s'}_{\mathsf{S}}
        \right)
        \\
        &\quad\otimes
        \mathcal{W}^{\mathsf{K}_{\mathsf{A}}}(
            (v_{e,u})_{\mathsf{A}}
        )
        \left(
            \KetBra{\alpha(s)}{\alpha(s')}_{\mathsf{A}}
        \right)
        \\
        &\quad\otimes
        (-1)^{z_{e,u,\mathsf{O}}\mathbin{\cdot}(E(s)+E(s'))}
        \KetBra{r_{e,u}(s)}{r_{e,u}(s')}_{\mathsf{O}}
        \end{aligned}
    \right).
    \label{eq:conditioned-before-leakage-extension-factorization}
\end{align}
Writing \(T=T_{\mathsf{D}}\sqcup T_{\mathsf{A}}\sqcup T_{\mathsf{O}}\) and applying Eqs.~\eqref{eq:leak-map} and \eqref{eq:leakage-subspace-extension-map} to each tensor factor under the canonical labeled tensor-product identifications leads to
\begin{equation}
    \mathcal{E}_T^{\mathsf{T}}
    \circ
    \mathcal{L}_T^{\mathsf{T}}
    =
    \left(
        \mathcal{E}_{T_{\mathsf{D}}}^{\mathsf{D}}
        \circ
        \mathcal{L}_{T_{\mathsf{D}}}^{\mathsf{D}}
    \right)
    \otimes
    \left(
        \mathcal{E}_{T_{\mathsf{A}}}^{\mathsf{A}}
        \circ
        \mathcal{L}_{T_{\mathsf{A}}}^{\mathsf{A}}
    \right)
    \otimes
    \left(
        \mathcal{E}_{T_{\mathsf{O}}}^{\mathsf{O}}
        \circ
        \mathcal{L}_{T_{\mathsf{O}}}^{\mathsf{O}}
    \right).
    \label{eq:conditioned-leakage-extension-factorization}
\end{equation}
Next, for \(a\in\mathbb{Z}_2^{\mathsf{O}}\) and \(I\subseteq\mathsf{O}\), write \(a|_I\coloneqq (a_i)_{i\in I}\in\mathbb{Z}_2^I\). 
For every \(T\subseteq\mathsf{T}\) and \(r,r'\in\mathbb{Z}_2^{\mathsf{O}}\), Eqs.~\eqref{eq:leak-map} and \eqref{eq:leakage-subspace-extension-map}, specialized to \(\mathsf{O}\), give
\begin{equation}
    \left(
        \mathcal{E}_{T_{\mathsf{O}}}^{\mathsf{O}}
        \circ
        \mathcal{L}_{T_{\mathsf{O}}}^{\mathsf{O}}
    \right)
    \left(
        \KetBra{r}{r'}_{\mathsf{O}}
    \right)
    =
    \delta_{
        r|_{T_{\mathsf{O}}},
        r'|_{T_{\mathsf{O}}}
    }
    \KetBra{\theta_T(r)}{\theta_T(r')}_{\mathsf{O}}.
    \label{eq:conditioned-outcome-register-action}
\end{equation}
Suppose that \(s\neq s'\).  
Equation~\eqref{eq:conditioned-shifted-outcome-record} and injectivity of \(E\) give \(r_{e,u}(s)\neq r_{e,u}(s')\) for every \(e\in V_{\mathsf{K}_{\mathsf{T}}}\) and \(u\in V_{U_{R_{\mathrm{in}}}}^{\mathsf{T}}\).
If their  restrictions to \(T_{\mathsf{O}}\) differ, then the Kronecker delta in Eq.~\eqref{eq:conditioned-outcome-register-action} vanishes.  
Otherwise, their restrictions to \(\mathsf{O}\setminus T_{\mathsf{O}}\) differ, so that \(\theta_T(r_{e,u}(s))\neq\theta_T(r_{e,u}(s'))\).
Equations~\eqref{eq:outcome-projectors} and \eqref{eq:outcome-branch} therefore show that \(\mathcal{M}_\theta\) annihilates the corresponding term for every \(\theta\in\Theta\).  
Thus, every summand in Eq.~\eqref{eq:conditioned-before-leakage-extension-factorization} vanishes when \(s\neq s'\).

Now, suppose \(s=s'\).  
In this case, \(r_{e,u}(s)=r_{e,u}(s')\), the phase in Eq.~\eqref{eq:conditioned-before-leakage-extension-factorization} is one, and Eqs.~\eqref{eq:conditioned-outcome-register-action}, \eqref{eq:outcome-projectors}, and \eqref{eq:outcome-branch} contribute the factor \(\delta_{\theta,\theta_T(r_{e,u}(s))}\). 
Lemma~\ref{lem:filling-leakage-map-properties}, specialized to \(\mathsf{A}\), shows that \(\mathcal{L}_{T_{\mathsf{A}}}^{\mathsf{A}}\) is trace preserving, and Eq.~\eqref{eq:leakage-subspace-map-traces}, specialized to \(\mathsf{A}\), shows that \(\mathcal{E}_{T_{\mathsf{A}}}^{\mathsf{A}}\) is trace preserving.
Since \(W^{\mathsf{K}_{\mathsf{A}}}((v_{e,u})_{\mathsf{A}})\) is unitary and \(\ket{\alpha(s)}_{\mathsf{A}}\) is a unit vector, the traced \(\mathsf{A}\) factor is \(\trace_{\mathsf{A}} (\ZProj{\alpha(s)}_{\mathsf{A}})=1\).
Combining the cases \(s\neq s'\) and \(s=s'\) and substituting Eqs.~\eqref{eq:conditioned-leakage-extension-factorization} and \eqref{eq:conditioned-outcome-register-action} into Eq.~\eqref{eq:conditioned-before-leakage-extension-factorization}, with Eq.~\eqref{eq:outcome-branch}, gives
\begin{align}
    &\widetilde\Phi_{R_{\mathrm{out}}, R_{\mathrm{in}}}
    \left(
        X\otimes\KetBra{s}{s'}_{\mathsf{S}}
    \right)
    \nonumber\\
    &\quad={}
    \delta_{s,s'}
    \sum_{T\subseteq\mathsf{T}}
    \sum_{\theta\in\Theta}
    \sum_{e\in V_{\mathsf{K}_{\mathsf{T}}}}
    \sum_{u\in V_{U_{R_{\mathrm{in}}}}^{\mathsf{T}}}
    \frac{
        q_e^{T, U_{R_{\mathrm{in}}}}
    }{
        4^{|U_{R_{\mathrm{in}}}|}
    }
    \delta_{\theta,\theta_T(r_{e,u}(s))}
    \nonumber\\
    &\qquad\cdot
    \left[
        \begin{aligned}
        &\mathcal{C}_{\mathrm F}^{\vpd}
        \circ
        \mathcal{F}_{R_{\mathrm{out}}}^{\mathsf{D}}
        \circ
        \mathcal{P}_{R_{\mathrm{out}}}^{\mathsf{D}}
        \circ
        \mathcal{R}_{K(\theta)}
        \\
        &\quad\circ
        \mathcal{W}^{\mathsf{D}}(D(\theta))
        \circ
        \mathcal{E}_{T_{\mathsf{D}}}^{\mathsf{D}}
        \circ
        \mathcal{L}_{T_{\mathsf{D}}}^{\mathsf{D}}
        \\
        &\quad\circ
        \mathcal{W}^{\mathsf{K}_{\mathsf{D}}}(
            (v_{e,u})_{\mathsf{D}}
        )
        \circ
        \mathcal{C}_{\mathrm F}^{-1}
        \end{aligned}
    \right]
    \left(
        X\otimes\ZProj{s}_{\mathsf{S}}
    \right).
    \label{eq:conditioned-readout-selection}
\end{align}
For every \(T\subseteq\mathsf{T}\), \(\theta\in\Theta\), \(e\in V_{\mathsf{K}_{\mathsf{T}}}\), and \(u\in V_{U_{R_{\mathrm{in}}}}^{\mathsf{T}}\), we have that
\begin{align}
    &\mathcal{F}_{R_{\mathrm{out}}}^{\mathsf{D}}
    \circ
    \mathcal{P}_{R_{\mathrm{out}}}^{\mathsf{D}}
    \circ
    \mathcal{R}_{K(\theta)}
    \circ
    \mathcal{W}^{\mathsf{D}}(D(\theta))
    \circ
    \mathcal{E}_{T_{\mathsf{D}}}^{\mathsf{D}}
    \circ
    \mathcal{L}_{T_{\mathsf{D}}}^{\mathsf{D}}
    \circ
    \mathcal{W}^{\mathsf{K}_{\mathsf{D}}}(
        (v_{e,u})_{\mathsf{D}}
    )
    \nonumber\\
    &\quad={}
    \mathcal{F}_{R_{\mathrm{out}}}^{\mathsf{D}}
    \circ
    \mathcal{P}_{R_{\mathrm{out}}}^{\mathsf{D}}
    \circ
    \mathcal{R}_{K(\theta)}
    \circ
    \mathcal{E}_{T_{\mathsf{D}}}^{\mathsf{D}}
    \circ
    \mathcal{L}_{T_{\mathsf{D}}}^{\mathsf{D}}
    \circ
    \mathcal{W}^{\mathsf{K}_{\mathsf{D}}}(
        (v_{e,u})_{\mathsf{D}}+D(\theta)
    )
    \nonumber\\
    &\quad={}
    \delta_{
        R_{\mathrm{out}},
        T_{\mathsf{D}}\setminus K(\theta)
    }
    \mathcal{G}_{T_{\mathsf{D}}\cup K(\theta)}^{\mathsf{D}}
    \circ
    \mathcal{W}^{\mathsf{K}_{\mathsf{D}}}(
        (v_{e,u})_{\mathsf{D}}+D(\theta)
    ),
    \label{eq:conditioned-data-factor-reduction}
\end{align}
where Eqs.~\eqref{eq:leakage-subspace-extension-Pauli-intertwining}, \eqref{eq:leak-covariance}, \eqref{eq:subspace-Pauli-factorization}, \eqref{eq:app-Pauli-product}, and \eqref{eq:app-adjoint-action-composition}, specialized to \(\mathsf{D}\), give the first equality, while Eqs.~\eqref{eq:return-map}, \eqref{eq:leak-map}, \eqref{eq:leakage-subspace-extension-map}, \eqref{eq:cross-subspace-orthogonality}, \eqref{eq:computational-replacement-map}, and \eqref{eq:computational-replacement-action} give the second.

For \(T\subseteq\mathsf{T}\), \(e\in V_{\mathsf{K}_{\mathsf{T}}}\), \(u\in V_{U_{R_{\mathrm{in}}}}^{\mathsf{T}}\), and \(t\in\mathbb{Z}_2^{\mathsf{S}}\), define
\begin{subequations}
\label{eq:conditioned-data-parameters}
\begin{align}
    R_{T,v_{e,u}}(t)
    &\coloneqq {}
    T_{\mathsf{D}}\setminus K(\theta_T(r_{e,u}(t))) \\
    \Lambda_{T,v_{e,u}}(t)
    &\coloneqq 
    T_{\mathsf{D}}\cup K(\theta_T(r_{e,u}(t))), \\
    \nu_{T,v_{e,u}}(t)
    &\coloneqq 
    (v_{e,u})_{\mathsf{D}}+D(\theta_T(r_{e,u}(t))).
\end{align} 
\end{subequations}
Substituting Eq.~\eqref{eq:conditioned-data-factor-reduction} termwise into Eq.~\eqref{eq:conditioned-readout-selection}, the factor
\(\delta_{\theta,\theta_T(r_{e,u}(s))}\) restricts the sum over \(\theta\) to \(\theta=\theta_T(r_{e,u}(s))\).
Equation~\eqref{eq:conditioned-data-parameters} therefore gives
\begin{align}
    &\widetilde\Phi_{R_{\mathrm{out}}, R_{\mathrm{in}}}
    \left(
        X\otimes\KetBra{s}{s'}_{\mathsf{S}}
    \right)
    \nonumber\\
    &\quad={}
    \delta_{s,s'}
    \sum_{\substack{
        T\subseteq\mathsf{T},\;
        e\in V_{\mathsf{K}_{\mathsf{T}}},\;
        u\in V_{U_{R_{\mathrm{in}}}}^{\mathsf{T}}\\
        R_{T,v_{e,u}}(s)=R_{\mathrm{out}}
    }}
    \frac{
        q_e^{T, U_{R_{\mathrm{in}}}}
    }{
        4^{|U_{R_{\mathrm{in}}}|}
    }
    \left(
        \mathcal{C}_{\mathrm F}^{\vpd}
        \circ
        \mathcal{G}_{\Lambda_{T,v_{e,u}}(s)}^{\mathsf{D}}
        \circ
        \mathcal{W}^{\mathsf{K}_{\mathsf{D}}}(
            \nu_{T,v_{e,u}}(s)
        )
        \circ
        \mathcal{C}_{\mathrm F}^{-1}
    \right)
    \left(
        X\otimes\ZProj{s}_{\mathsf{S}}
    \right).
    \label{eq:conditioned-postprocessed-expansion}
\end{align}
For every \(t\in\mathbb{Z}_2^{\mathsf{S}}\), define the map
\begin{equation*}
    \mathcal{B}^{R_{\mathrm{out}}, R_{\mathrm{in}}}_t:
    \operatorname{End}(\mathcal{K}_{\mathsf{D}} )
    \to
    \operatorname{End}(\mathcal{K}_{\mathsf{D}} )
\end{equation*}
according to 
\begin{align}
    \mathcal{B}^{R_{\mathrm{out}}, R_{\mathrm{in}}}_t
    &\coloneqq {}
    \sum_{\substack{
        T\subseteq\mathsf{T},\;
        e\in V_{\mathsf{K}_{\mathsf{T}}},\;
        u\in V_{U_{R_{\mathrm{in}}}}^{\mathsf{T}}\\
        R_{T,v_{e,u}}(t)=R_{\mathrm{out}}
    }}
    \frac{
        q_e^{T, U_{R_{\mathrm{in}}}}
    }{
        4^{|U_{R_{\mathrm{in}}}|}
    }
    \mathcal{G}_{\Lambda_{T,v_{e,u}}(t)}^{\mathsf{D}}
    \circ
    \mathcal{W}^{\mathsf{K}_{\mathsf{D}}}(
        \nu_{T,v_{e,u}}(t)
    ).
    \label{eq:conditioned-physical-Pauli-map}
\end{align}
For every \(\Lambda\subseteq\mathsf{D}\) and \(\nu\in V_{\mathsf{K}_{\mathsf{D}}}\), Eq.~\eqref{eq:computational-replacement-Pauli-channel}, specialized to \(\mathsf{D}\), and Eqs.~\eqref{eq:app-Pauli-product} and \eqref{eq:app-adjoint-action-composition} give
\begin{align}
    \mathcal{G}_\Lambda^{\mathsf{D}}
    \circ
    \mathcal{W}^{\mathsf{K}_{\mathsf{D}}}(\nu)
    &=
    \frac{1}{4^{|\Lambda|}}
    \sum_{m\in V_\Lambda^{\mathsf{D}}}
    \mathcal{W}^{\mathsf{K}_{\mathsf{D}}}(m)
    \circ
    \mathcal{W}^{\mathsf{K}_{\mathsf{D}}}(\nu)
    =
    \frac{1}{4^{|\Lambda|}}
    \sum_{m\in V_\Lambda^{\mathsf{D}}}
    \mathcal{W}^{\mathsf{K}_{\mathsf{D}}}(m+\nu).
    \label{eq:conditioned-physical-branch-Pauli-channel}
\end{align}
Since \(|V_\Lambda^{\mathsf{D}}|=4^{|\Lambda|}\), Definition~\ref{defn:pauli-map-normalization} implies that \(\mathcal{G}_\Lambda^{\mathsf{D}} \circ \mathcal{W}^{\mathsf{K}_{\mathsf{D}}}(\nu)\) is a Pauli channel. 
Thus, Eq.~\eqref{eq:conditioned-physical-Pauli-map} shows that every \(\mathcal{B}^{R_{\mathrm{out}}, R_{\mathrm{in}}}_t\) is a Pauli map.  
Equations~\eqref{eq:conditioned-postprocessed-expansion} and \eqref{eq:conditioned-physical-Pauli-map} give
\begin{align}
    &\widetilde\Phi_{R_{\mathrm{out}}, R_{\mathrm{in}}}
    \left(
        X\otimes\KetBra{s}{s'}_{\mathsf{S}}
    \right)
    =
    \delta_{s,s'}
    \left(
        \mathcal{C}_{\mathrm F}^{\vpd}
        \circ
        \mathcal{B}^{R_{\mathrm{out}}, R_{\mathrm{in}}}_s
        \circ
        \mathcal{C}_{\mathrm F}^{-1}
    \right)
    \left(
        X\otimes\ZProj{s}_{\mathsf{S}}
    \right).
    \label{eq:conditioned-matrix-unit-branch-form}
\end{align}
For every \(t\in\mathbb{Z}_2^{\mathsf{S}}\),
Definition~\ref{def:syndrome-block-projection-maps} gives
\begin{equation}
    \mathcal{Z}_t
    \left(
        X\otimes\KetBra{s}{s'}_{\mathsf{S}}
    \right)
    =
    \delta_{s,s'}\delta_{t,s}
    X\otimes\ZProj{s}_{\mathsf{S}}.
    \label{eq:conditioned-syndrome-block-matrix-unit-action}
\end{equation}
Applying Eq.~\eqref{eq:conditioned-syndrome-block-matrix-unit-action} termwise gives
\begin{align}
    &\left[
        \sum_{t\in\mathbb{Z}_2^{\mathsf{S}}}
        \mathcal{C}_{\mathrm F}^{\vpd}
        \circ
        \mathcal{B}^{R_{\mathrm{out}}, R_{\mathrm{in}}}_t
        \circ
        \mathcal{C}_{\mathrm F}^{-1}
        \circ
        \mathcal{Z}_t
    \right]
    \left(
        X\otimes\KetBra{s}{s'}_{\mathsf{S}}
    \right)
    \nonumber\\
    &\quad={}
    \delta_{s,s'}
    \left(
        \mathcal{C}_{\mathrm F}
        \circ
        \mathcal{B}^{R_{\mathrm{out}}, R_{\mathrm{in}}}_s
        \circ
        \mathcal{C}_{\mathrm F}^{-1}
    \right)
    \left(
        X\otimes\ZProj{s}_{\mathsf{S}}
    \right)
    \nonumber\\
    &\quad={}
    \widetilde\Phi_{R_{\mathrm{out}}, R_{\mathrm{in}}}
    \left(
        X\otimes\KetBra{s}{s'}_{\mathsf{S}}
    \right),
    \label{eq:conditioned-matrix-unit-reconstruction}
\end{align}
where the last equality is Eq.~\eqref{eq:conditioned-matrix-unit-branch-form}.  
The operators \( X\otimes\KetBra{s}{s'}_{\mathsf{S}} \), with \(X\in\operatorname{End}(\mathcal{K}_{\mathsf{L}})\) and \(s,s'\in\mathbb{Z}_2^{\mathsf{S}}\), span \(\operatorname{End}(\mathcal{K}_{\mathsf{L}}\otimes\mathcal{K}_{\mathsf{S}})\).
Since \(R_{\mathrm{in}}\) and \(R_{\mathrm{out}}\) were arbitrary, Eq.~\eqref{eq:conditioned-matrix-unit-reconstruction} proves Eq.~\eqref{eq:complete-cycle-conditioned-Pauli-form} for every \(R_{\mathrm{in}},R_{\mathrm{out}}\subseteq\mathsf{D}\).
\end{proof}

An explicit construction of the family of Pauli maps
\(\left\{ \mathcal{B}^{R_{\mathrm{out}}, R_{\mathrm{in}}}_s \right\}^{R_{\mathrm{in}},R_{\mathrm{out}}\subseteq\mathsf{D}}_{s\in\mathbb{Z}_2^{\mathsf{S}}} \) appears in Eq.~\eqref{eq:conditioned-physical-Pauli-map}.

\begin{cor}[Syndrome dephasing of Clifford-frame sector branches]
\label{cor:complete-cycle-filled-frame-syndrome-dephasing}
For every \(R_{\mathrm{in}},R_{\mathrm{out}}\subseteq\mathsf{D}\),
\begin{equation}
    \widetilde\Phi_{R_{\mathrm{out}}, R_{\mathrm{in}}}
    =
    \widetilde\Phi_{R_{\mathrm{out}}, R_{\mathrm{in}}}
    \circ
    \mathcal{Z}_{\mathsf{S}}.
    \label{eq:complete-cycle-filled-frame-syndrome-dephasing}
\end{equation}
\end{cor}

\begin{proof}
For every \(s,t\in\mathbb{Z}_2^{\mathsf{S}}\),
Definition~\ref{def:syndrome-block-projection-maps} gives
\begin{equation*}
    \mathcal{Z}_s\circ\mathcal{Z}_t
    =
    \delta_{s,t}\mathcal{Z}_t.
\end{equation*}
Right composing Eq.~\eqref{eq:complete-cycle-conditioned-Pauli-form} by \(\mathcal{Z}_t\) therefore gives
\begin{equation*}
    \widetilde\Phi_{R_{\mathrm{out}}, R_{\mathrm{in}}}
    \circ
    \mathcal{Z}_t
    =
    \mathcal{C}_{\mathrm F}^{\vpd}
    \circ
    \mathcal{B}^{R_{\mathrm{out}}, R_{\mathrm{in}}}_t
    \circ
    \mathcal{C}_{\mathrm F}^{-1}
    \circ
    \mathcal{Z}_t.
\end{equation*}
Summing this equality over \(t\), and applying Eq.~\eqref{eq:complete-cycle-conditioned-Pauli-form} and Definition~\ref{def:syndrome-dephasing-channel}, gives Eq.~\eqref{eq:complete-cycle-filled-frame-syndrome-dephasing}.
\end{proof}

\begin{cor}
\label{cor:complete-cycle-conditioned-Pauli-normalization}
Let \(\left\{ \mathcal{B}^{R_{\mathrm{out}}, R_{\mathrm{in}}}_s \right\}^{R_{\mathrm{in}},R_{\mathrm{out}}\subseteq\mathsf{D}}_{s\in\mathbb{Z}_2^{\mathsf{S}}} \) 
be the family of Pauli maps constructed in the proof of Proposition~\ref{prop:complete-cycle-conditioned-Pauli-form}.
Every \(\mathcal{B}^{R_{\mathrm{out}}, R_{\mathrm{in}}}_s\) is a subnormalized Pauli map, and for every \(R_{\mathrm{in}}\subseteq\mathsf{D}\) and
\(s\in\mathbb{Z}_2^{\mathsf{S}}\), we have that
\begin{equation}
    \sum_{R_{\mathrm{out}}\subseteq\mathsf{D}}
    \operatorname{wt}\!\left(
        \mathcal{B}^{R_{\mathrm{out}}, R_{\mathrm{in}}}_s
    \right)
    =
    1.
    \label{eq:complete-cycle-conditioned-Pauli-normalization}
\end{equation}
\end{cor}

\begin{proof}
Fix \(R_{\mathrm{in}}\subseteq\mathsf{D}\) and \(s\in\mathbb{Z}_2^{\mathsf{S}}\). 
Use the coefficients \(q_e^{T, U_{R_{\mathrm{in}}}}\), the labels \(v_{e,u}\), and the sets and labels \(R_{T,v_{e,u}}(s)\), \(\Lambda_{T,v_{e,u}}(s)\), and \(\nu_{T,v_{e,u}}(s)\) defined in Eqs.~\eqref{eq:conditioned-Pauli-coefficient}, \eqref{eq:conditioned-combined-Pauli-label}, and
\eqref{eq:conditioned-data-parameters}. 
Each \(\mathcal{D}_{T, U_{R_{\mathrm{in}}}}\) is a subnormalized Pauli map by the standing assumption in Appendix~\ref{sec:leakage-aware-cycle}.  
Thus,  Eq.~\eqref{eq:subnormalized-pauli-map-coefficients} gives \(q_e^{T, U_{R_{\mathrm{in}}}}\geq0 \) for every \(T\subseteq\mathsf{T}\) and
\(e\in V_{\mathsf{K}_{\mathsf{T}}}\).  
Moreover, Eqs.~\eqref{eq:pauli-map-weight-sum} and \eqref{eq:complete-cycle-Pauli-map-family-normalization}---together with\(U=U_{R_{\mathrm{in}}}\)---give
\begin{equation}
    \sum_{T\subseteq\mathsf{T}}
    \sum_{e\in V_{\mathsf{K}_{\mathsf{T}}}}
    q_e^{T, U_{R_{\mathrm{in}}}}
    =
    \sum_{T\subseteq\mathsf{T}}
    \operatorname{wt}\!\left(
        \mathcal{D}_{T, U_{R_{\mathrm{in}}}}
    \right)
    =
    1.
    \label{eq:conditioned-coefficient-normalization}
\end{equation}
These inequalities, along with Eqs.~\eqref{eq:conditioned-physical-Pauli-map} and \eqref{eq:conditioned-physical-branch-Pauli-channel}, establish that every \(\mathcal{B}^{R_{\mathrm{out}}, R_{\mathrm{in}}}_s\) has nonnegative Pauli coefficients.  
Combining Eqs.~\eqref{eq:conditioned-physical-Pauli-map} and \eqref{eq:conditioned-physical-branch-Pauli-channel} with linearity of the weight functional \( \operatorname{wt}(\cdot)\) leads to
\begin{equation}
    \operatorname{wt}\!\left(
        \mathcal{B}^{R_{\mathrm{out}}, R_{\mathrm{in}}}_s
    \right)
    =
    \sum_{\substack{
        T\subseteq\mathsf{T},\;
        e\in V_{\mathsf{K}_{\mathsf{T}}},\;
        u\in V_{U_{R_{\mathrm{in}}}}^{\mathsf{T}}\\
        R_{T,v_{e,u}}(s)=R_{\mathrm{out}}
    }}
    \frac{
        q_e^{T, U_{R_{\mathrm{in}}}}
    }{
        4^{|U_{R_{\mathrm{in}}}|}
    }.
    \label{eq:conditioned-sector-branch-weight}
\end{equation}
Each triple \((T,e,u)\) determines exactly one value of \(R_{T,v_{e,u}}(s)\).  
Since \( |V_{U_{R_{\mathrm{in}}}}^{\mathsf{T}}| = 4^{|U_{R_{\mathrm{in}}}|} \), summing Eq.~\eqref{eq:conditioned-sector-branch-weight} over \(R_{\mathrm{out}}\subseteq\mathsf{D}\) gives
\begin{align*}
    &\sum_{R_{\mathrm{out}}\subseteq\mathsf{D}}
    \operatorname{wt}\!\left(
        \mathcal{B}^{R_{\mathrm{out}}, R_{\mathrm{in}}}_s
    \right)
    =
    \sum_{T\subseteq\mathsf{T}}
    \sum_{e\in V_{\mathsf{K}_{\mathsf{T}}}}
    \sum_{u\in V_{U_{R_{\mathrm{in}}}}^{\mathsf{T}}}
    \frac{
        q_e^{T, U_{R_{\mathrm{in}}}}
    }{
        4^{|U_{R_{\mathrm{in}}}|}
    }
    =
    \sum_{T\subseteq\mathsf{T}}
    \sum_{e\in V_{\mathsf{K}_{\mathsf{T}}}}
    q_e^{T, U_{R_{\mathrm{in}}}}
    =
    1,
\end{align*}
where the last equality is Eq.~\eqref{eq:conditioned-coefficient-normalization}.  
By Eq.~\eqref{eq:pauli-map-weight-sum}, the sum of the Pauli coefficients of each map \(\mathcal{B}^{R_{\mathrm{out}}, R_{\mathrm{in}}}_s\) equals its weight.  
Since those coefficients are nonnegative and each weight is at most one, every \(\mathcal{B}^{R_{\mathrm{out}}, R_{\mathrm{in}}}_s\) is a subnormalized Pauli map, by Definition~\ref{defn:pauli-map-normalization}.
\end{proof}

\subsection{Complete-cycle logical transition form}
\label{subsec:complete-cycle-transition-form}

\begin{defn}[Syndrome transition maps]
\label{def:syndrome-transition-map}
For every \(s_{\mathrm{in}},s_{\mathrm{out}} \in\mathbb{Z}_2^{\mathsf{S}}\), define the map \( \mathcal{S}_{s_{\mathrm{out}}, s_{\mathrm{in}}}: \operatorname{End}(\mathcal{K}_{\mathsf{S}}) \to \operatorname{End}(\mathcal{K}_{\mathsf{S}}) \) by
\begin{equation}
    \mathcal{S}_{s_{\mathrm{out}}, s_{\mathrm{in}}}(Y) \coloneqq \MatEl{s_{\mathrm{in}}}{Y}{s_{\mathrm{in}}}_{\mathsf{S}}
    \ZProj{s_{\mathrm{out}}}_{\mathsf{S}} \, ,
    \label{eq:syndrome-transition-map}
\end{equation}
for all \( Y\in\operatorname{End}(\mathcal{K}_{\mathsf{S}})\).
\end{defn}

\begin{prop}
\label{prop:complete-cycle-transition-form}
There exists a family
\begin{equation*}
    \left\{
        \mathcal{J}^{R_{\mathrm{out}}, R_{\mathrm{in}}}
        _{s_{\mathrm{out}},s_{\mathrm{in}}}
    \right\}^{R_{\mathrm{in}},R_{\mathrm{out}}\subseteq\mathsf{D}}_{s_{\mathrm{in}},s_{\mathrm{out}}\in\mathbb{Z}_2^{\mathsf{S}}}
\end{equation*}
of subnormalized Pauli maps on \(\operatorname{End}(\mathcal{K}_{\mathsf{L}})\) such that, for every \(R_{\mathrm{in}},R_{\mathrm{out}}\subseteq\mathsf{D}\),
\begin{equation}
    \widetilde\Phi_{R_{\mathrm{out}}, R_{\mathrm{in}}}
    =
    \sum_{s_{\mathrm{in}},s_{\mathrm{out}}
    \in\mathbb{Z}_2^{\mathsf{S}}}
    \mathcal{J}^{R_{\mathrm{out}}, R_{\mathrm{in}}}
    _{s_{\mathrm{out}},s_{\mathrm{in}}}
    \otimes
    \mathcal{S}_{s_{\mathrm{out}}, s_{\mathrm{in}}}.
    \label{eq:complete-cycle-sector-transition-decomposition}
\end{equation}
\end{prop}

\begin{proof}
Fix \(R_{\mathrm{in}},R_{\mathrm{out}}\subseteq\mathsf{D}\).
For every \(s_{\mathrm{in}}\in\mathbb{Z}_2^{\mathsf{S}}\), denote by
\begin{equation*}
    \left\{
        b_{q,s_{\mathrm{in}}}^{R_{\mathrm{out}}, R_{\mathrm{in}}}
    \right\}_{q\in V_{\mathsf{K}_{\mathsf{D}}}}
\end{equation*}
the unique coefficient family of \(\mathcal{B}^{R_{\mathrm{out}}, R_{\mathrm{in}}}_{s_{\mathrm{in}}}\) from Lemma~\ref{lem:pauli-map-structure}, specialized to \(\mathsf{K}_{\mathsf{D}}\). 
Corollary~\ref{cor:complete-cycle-conditioned-Pauli-normalization} and Eq.~\eqref{eq:subnormalized-pauli-map-coefficients} imply that \( b_{q,s_{\mathrm{in}}}^{R_{\mathrm{out}}, R_{\mathrm{in}}} \geq 0 \) for every \(q\in V_{\mathsf{K}_{\mathsf{D}}}\). 
Then, the relations
\begin{align}
    \mathcal{B}^{R_{\mathrm{out}}, R_{\mathrm{in}}}
    _{s_{\mathrm{in}}}
    &={}
    \sum_{q\in V_{\mathsf{K}_{\mathsf{D}}}}
    b_{q,s_{\mathrm{in}}}^{R_{\mathrm{out}}, R_{\mathrm{in}}}
    \mathcal{W}^{\mathsf{K}_{\mathsf{D}}}(q),
    \label{eq:complete-cycle-conditioned-Pauli-coefficients}
    \\
    \sum_{q\in V_{\mathsf{K}_{\mathsf{D}}}}
    b_{q,s_{\mathrm{in}}}^{R_{\mathrm{out}}, R_{\mathrm{in}}}
    &\leq
    1
    \label{eq:complete-cycle-conditioned-Pauli-coefficient-bound}
\end{align}
hold.
Renaming the summation index \(s\) in Eq.~\eqref{eq:complete-cycle-conditioned-Pauli-form} as \(s_{\mathrm{in}}\), substituting Eq.~\eqref{eq:complete-cycle-conditioned-Pauli-coefficients} for \(\mathcal{B}^{R_{\mathrm{out}}, R_{\mathrm{in}}}_{s_{\mathrm{in}}}\) in each summand, and distributing the composition over the resulting sum give
\begin{equation}
    \widetilde\Phi_{R_{\mathrm{out}}, R_{\mathrm{in}}}
    =
    \sum_{s_{\mathrm{in}}\in\mathbb{Z}_2^{\mathsf{S}}}
    \sum_{q\in V_{\mathsf{K}_{\mathsf{D}}}}
    b_{q,s_{\mathrm{in}}}^{R_{\mathrm{out}}, R_{\mathrm{in}}}
    \mathcal{C}_{\mathrm F}^{\vpd}
    \circ
    \mathcal{W}^{\mathsf{K}_{\mathsf{D}}}(q)
    \circ
    \mathcal{C}_{\mathrm F}^{-1}
    \circ
    \mathcal{Z}_{s_{\mathrm{in}}}.
    \label{eq:proof complete-cycle-framed-Pauli-expansion}
\end{equation}
By the Clifford assumption on \(\CFrame\), fix a symplectic isomorphism
\(
    \kappa=(\kappa_{\mathsf{L}},\kappa_{\mathsf{S}}):
    V_{\mathsf{K}_{\mathsf{D}}}
    \to
    V_{\mathsf{L}}\times V_{\mathsf{S}}
\)
such that, for every \(q\in V_{\mathsf{K}_{\mathsf{D}}}\),
\begin{equation}
    \mathcal{C}_{\mathrm F}^{\vpd}
    \circ
    \mathcal{W}^{\mathsf{K}_{\mathsf{D}}}(q)
    \circ
    \, \mathcal{C}_{\mathrm F}^{-1}
    =
    \mathcal{W}^{\mathsf{L}}(\kappa_{\mathsf{L}}(q))
        \otimes \mathcal{W}^{\mathsf{S}}(\kappa_{\mathsf{S}}(q))
    .
    \label{eq:Clifford-frame-Pauli-labels}
\end{equation}
Next, define the map
\(
    \mathrm{x}:
    V_{\mathsf{S}}
    \to
    \mathbb{Z}_2^{\mathsf{S}}
\)
by \(\mathrm{x}(x,z)\coloneqq x\).
For every \(s_{\mathrm{in}}\in\mathbb{Z}_2^{\mathsf{S}}\) and \(q\in V_{\mathsf{K}_{\mathsf{D}}}\),
\begin{equation}
    \mathcal{C}_{\mathrm F}^{\vpd}
    \circ
    \mathcal{W}^{\mathsf{K}_{\mathsf{D}}}(q)
    \circ
    \, \mathcal{C}_{\mathrm F}^{-1}
    \circ
    \mathcal{Z}_{s_{\mathrm{in}}}
    =
    \mathcal{W}^{\mathsf{L}}(\kappa_{\mathsf{L}}(q))
    \otimes
    \mathcal{S}_{
        s_{\mathrm{in}}
        +\mathrm{x}(\kappa_{\mathsf{S}}(q))
        ,
        s_{\mathrm{in}}
    }.
    \label{eq:complete-cycle-single-Pauli-transition}
\end{equation}
To verify Eq.~\eqref{eq:complete-cycle-single-Pauli-transition}, Eqs.~\eqref{eq:computational-system-Pauli-generators} and \eqref{eq:app-Pauli-definition}, specialized to \(\mathsf{S}\), give
\(
    \mathcal{W}^{\mathsf{S}}(r)
    (\ZProj{s}_{\mathsf{S}})
    =
    \ZProj{s+\mathrm{x}(r)}_{\mathsf{S}}
\)
for every \(r\in V_{\mathsf{S}}\) and \(s\in\mathbb{Z}_2^{\mathsf{S}}\).  
This equality and Eqs.~\eqref{eq:Clifford-frame-Pauli-labels}, \eqref{eq:syndrome-block-selection-map}, and \eqref{eq:syndrome-transition-map} give, for every \(X\in\operatorname{End}(\mathcal{K}_{\mathsf{L}})\) and \(Y\in\operatorname{End}(\mathcal{K}_{\mathsf{S}})\),
\begin{align*}
    &\left(
        \mathcal{C}_{\mathrm F}^{\vpd}
        \circ
        \mathcal{W}^{\mathsf{K}_{\mathsf{D}}}(q)
        \circ
        \mathcal{C}_{\mathrm F}^{-1}
        \circ
        \mathcal{Z}_{s_{\mathrm{in}}}
    \right)(X\otimes Y)
    \\
    &\quad={}
    \MatEl{s_{\mathrm{in}}}{Y}{s_{\mathrm{in}}}_{\mathsf{S}}
    \mathcal{W}^{\mathsf{L}} (\kappa_{\mathsf{L}}(q))(X)
    \otimes
    \ZProj{
        s_{\mathrm{in}}
        +\mathrm{x}(\kappa_{\mathsf{S}}(q))
    }_{\mathsf{S}}
    \\
    &\quad={}
    \left(
        \mathcal{W}^{\mathsf{L}}(\kappa_{\mathsf{L}}(q))
        \otimes
        \mathcal{S}_{
            s_{\mathrm{in}}
            +\mathrm{x}(\kappa_{\mathsf{S}}(q))
            ,
            s_{\mathrm{in}}
        }
    \right)(X\otimes Y).
\end{align*}
Since simple tensors span \(\operatorname{End}(\mathcal{K}_{\mathsf{L}}\otimes \mathcal{K}_{\mathsf{S}})\), this proves Eq.~\eqref{eq:complete-cycle-single-Pauli-transition}.

For every \(s_{\mathrm{in}},s_{\mathrm{out}} \in\mathbb{Z}_2^{\mathsf{S}}\),  define the map
\(
    \mathcal{J}^{R_{\mathrm{out}}, R_{\mathrm{in}}}
    _{s_{\mathrm{out}},s_{\mathrm{in}}}:
    \operatorname{End}(\mathcal{K}_{\mathsf{L}})
    \to
    \operatorname{End}(\mathcal{K}_{\mathsf{L}})
\)
by
\begin{equation}
    \mathcal{J}^{R_{\mathrm{out}}, R_{\mathrm{in}}}
    _{s_{\mathrm{out}},s_{\mathrm{in}}}
    \coloneqq {}
    \hspace{-6mm} \sum_{\substack{
        q\in V_{\mathsf{K}_{\mathsf{D}}}\\
        s_{\mathrm{in}}
        +\mathrm{x}(\kappa_{\mathsf{S}}(q))
        =s_{\mathrm{out}}
    }} \hspace{-4mm} 
    b_{q,s_{\mathrm{in}}}^{R_{\mathrm{out}}, R_{\mathrm{in}}}
    \mathcal{W}^{\mathsf{L}}(\kappa_{\mathsf{L}}(q)).
    \label{eq:complete-cycle-logical-Pauli-form}
\end{equation}
Grouping the terms in Eq.~\eqref{eq:complete-cycle-logical-Pauli-form} by \(\kappa_{\mathsf{L}}(q)\) gives Pauli coefficients that are sums of the nonnegative coefficients
\(
    b_{q,s_{\mathrm{in}}}^{R_{\mathrm{out}}, R_{\mathrm{in}}}
\).  
Their sum is at most
\(
    \sum_{q\in V_{\mathsf{K}_{\mathsf{D}}}}
    b_{q,s_{\mathrm{in}}}^{R_{\mathrm{out}}, R_{\mathrm{in}}}
\), 
which is at most one, by Eq.~\eqref{eq:complete-cycle-conditioned-Pauli-coefficient-bound}.
Henc, every \(\mathcal{J}^{R_{\mathrm{out}}, R_{\mathrm{in}}}_{s_{\mathrm{out}},s_{\mathrm{in}}}\) is a subnormalized Pauli map, by Definition~\ref{defn:pauli-map-normalization}.

Substituting Eq.~\eqref{eq:complete-cycle-single-Pauli-transition} termwise into Eq.~\eqref{eq:proof complete-cycle-framed-Pauli-expansion}, grouping by the value of \(s_{\mathrm{in}}+\mathrm{x}(\kappa_{\mathsf{S}}(q))\), and applying Eq.~\eqref{eq:complete-cycle-logical-Pauli-form} give
Eq.~\eqref{eq:complete-cycle-sector-transition-decomposition}.
\end{proof}

The members of the family
\(
    \left\{
        \mathcal{J}^{R_{\mathrm{out}}, R_{\mathrm{in}}}
        _{s_{\mathrm{out}},s_{\mathrm{in}}}
    \right\}^{R_{\mathrm{in}},R_{\mathrm{out}}\subseteq\mathsf{D}}_{s_{\mathrm{in}},s_{\mathrm{out}}\in\mathbb{Z}_2^{\mathsf{S}}}
\)
constructed in the foregoing proof are given explicitly by Eq.~\eqref{eq:complete-cycle-logical-Pauli-form}.

\begin{cor}
\label{cor:leakage-resolved-transition-normalization}
Denote by 
\(
    \left\{
        \mathcal{J}^{R_{\mathrm{out}}, R_{\mathrm{in}}}
        _{s_{\mathrm{out}},s_{\mathrm{in}}}
    \right\}^{R_{\mathrm{in}},R_{\mathrm{out}}\subseteq\mathsf{D}}_{s_{\mathrm{in}},s_{\mathrm{out}}\in\mathbb{Z}_2^{\mathsf{S}}}
\)
the family constructed in the proof of Proposition~\ref{prop:complete-cycle-transition-form}.
For every \(R_{\mathrm{in}}\subseteq\mathsf{D}\) and \(s_{\mathrm{in}}\in\mathbb{Z}_2^{\mathsf{S}}\),
\begin{equation}
    \sum_{R_{\mathrm{out}}\subseteq\mathsf{D}}
    \sum_{s_{\mathrm{out}}\in\mathbb{Z}_2^{\mathsf{S}}}
    \operatorname{wt}\!\left(
        \mathcal{J}^{R_{\mathrm{out}}, R_{\mathrm{in}}}
        _{s_{\mathrm{out}},s_{\mathrm{in}}}
    \right)
    =
    1.
    \label{eq:leakage-resolved-transition-normalization}
\end{equation}
\end{cor}

\begin{proof}
Fix
\(R_{\mathrm{in}}\subseteq\mathsf{D}\) and
\(s_{\mathrm{in}}\in\mathbb{Z}_2^{\mathsf{S}}\).
For each
\(R_{\mathrm{out}}\subseteq\mathsf{D}\), as
\(s_{\mathrm{out}}\) varies over \(\mathbb{Z}_2^{\mathsf{S}}\), the
conditions in Eq.~\eqref{eq:complete-cycle-logical-Pauli-form}
partition the summation indices in
Eq.~\eqref{eq:complete-cycle-conditioned-Pauli-coefficients}.
Grouping terms with the same adjoint map in
Eq.~\eqref{eq:complete-cycle-logical-Pauli-form} does not change the
total coefficient sum.  Combining Eqs.~\eqref{eq:pauli-map-weight-sum},
\eqref{eq:complete-cycle-conditioned-Pauli-coefficients}, and
\eqref{eq:complete-cycle-logical-Pauli-form} and then applying
Eq.~\eqref{eq:complete-cycle-conditioned-Pauli-normalization} gives
\begin{align*}
    &\sum_{R_{\mathrm{out}}\subseteq\mathsf{D}}
    \sum_{s_{\mathrm{out}}\in\mathbb{Z}_2^{\mathsf{S}}}
    \operatorname{wt}\!\left(
        \mathcal{J}^{R_{\mathrm{out}}, R_{\mathrm{in}}}
        _{s_{\mathrm{out}},s_{\mathrm{in}}}
    \right) =
    \sum_{R_{\mathrm{out}}\subseteq\mathsf{D}}
    \operatorname{wt}\!\left(
        \mathcal{B}^{R_{\mathrm{out}}, R_{\mathrm{in}}}
        _{s_{\mathrm{in}}}
    \right) = 1.
\end{align*}
\end{proof}

For every linear map
\(
    \mathcal{A}:
    \operatorname{End}(\mathcal{K}_{\mathsf{L}})
    \to
    \operatorname{End}(\mathcal{K}_{\mathsf{L}})
\)
and every
\(s_{\mathrm{in}},s_{\mathrm{out}}
\in\mathbb{Z}_2^{\mathsf{S}}\),
Eqs.~\eqref{eq:syndrome-block-selection-map},
\eqref{eq:syndrome-dephasing-channel}, and
\eqref{eq:syndrome-transition-map} give
\begin{equation*}
    \mathcal{Z}_{\mathsf{S}}
    \circ
    \left(
        \mathcal{A}
        \otimes
        \mathcal{S}_{s_{\mathrm{out}}, s_{\mathrm{in}}}
    \right)
    =
    \mathcal{A}
    \otimes
    \mathcal{S}_{s_{\mathrm{out}}, s_{\mathrm{in}}}
    =
    \left(
        \mathcal{A}
        \otimes
        \mathcal{S}_{s_{\mathrm{out}}, s_{\mathrm{in}}}
    \right)
    \circ
    \mathcal{Z}_{\mathsf{S}}.
\end{equation*}
For every
\(R_{\mathrm{in}},R_{\mathrm{out}}\subseteq\mathsf{D}\), applying
this identity termwise to
Eq.~\eqref{eq:complete-cycle-sector-transition-decomposition} gives
\begin{equation}
    \widetilde\Phi_{R_{\mathrm{out}}, R_{\mathrm{in}}}
    =
    \mathcal{Z}_{\mathsf{S}}
    \circ
    \widetilde\Phi_{R_{\mathrm{out}}, R_{\mathrm{in}}}
    =
    \widetilde\Phi_{R_{\mathrm{out}}, R_{\mathrm{in}}}
    \circ
    \mathcal{Z}_{\mathsf{S}}
    =
    \mathcal{Z}_{\mathsf{S}}
    \circ
    \widetilde\Phi_{R_{\mathrm{out}}, R_{\mathrm{in}}}
    \circ
    \mathcal{Z}_{\mathsf{S}}.
    \label{eq:complete-cycle-filled-frame-double-syndrome-dephasing}
\end{equation}

\begin{thm}
\label{thm:single-cycle}
Let \(\Phi\) be the complete QEC-cycle channel, assumed to satisfy Eq.~\eqref{eq:complete-cycle-projector-Markov-Pauli-form}, and let \(
    \left\{
        \mathcal{J}^{R_{\mathrm{out}}, R_{\mathrm{in}}}
        _{s_{\mathrm{out}},s_{\mathrm{in}}}
    \right\}^{R_{\mathrm{in}},R_{\mathrm{out}}\subseteq\mathsf{D}}_{s_{\mathrm{in}},s_{\mathrm{out}}\in\mathbb{Z}_2^{\mathsf{S}}}
\)
be the family constructed in the proof of Proposition~\ref{prop:complete-cycle-transition-form}.  
Then
\begin{equation}
    \Phi = 
    \hspace{-2.5mm}
    \sum_{\substack{R_{\mathrm{in}},R_{\mathrm{out}}\subseteq\mathsf{D} \\ s_{\mathrm{in}},s_{\mathrm{out}} \in\mathbb{Z}_2^{\mathsf{S}}}}
    \hspace{-1mm}
        \mathcal{E}_{R_{\mathrm{out}}}^{\mathsf{D}}
        \circ
        \mathcal{L}_{R_{\mathrm{out}}}^{\mathsf{D}}
        \circ
        \mathcal{C}_{\mathrm F}^{-1}\circ
        \left( 
        \mathcal{J}^{R_{\mathrm{out}}, R_{\mathrm{in}}}
        _{s_{\mathrm{out}},s_{\mathrm{in}}}
        \otimes
        \mathcal{S}_{s_{\mathrm{out}}, s_{\mathrm{in}}}
        \right)
    \circ
        \mathcal{C}_{\mathrm F}^{\vpd}
        \circ
        \mathcal{F}_{R_{\mathrm{in}}}^{\mathsf{D}}
        \circ
        \mathcal{P}_{R_{\mathrm{in}}}^{\mathsf{D}}
        .
    \label{eq:physical-single-cycle-decomposition}
\end{equation}
\end{thm}

\begin{proof}
Substituting
Eq.~\eqref{eq:complete-cycle-sector-transition-decomposition} into
Eq.~\eqref{eq:complete-cycle-filled-frame-reconstruction} gives
Eq.~\eqref{eq:physical-single-cycle-decomposition}.
\end{proof}

\section{Leakage--syndrome-dephased operators and hidden-memory dynamics}
\label{sec:single-cycle}

\subsection{Leakage--syndrome dephasing}
\label{subsec:canonical-leakage--syndrome-dephasing}

\begin{defn}[Leakage--syndrome-dephasing map]
\label{def:leakage--syndrome-dephasing-map}
Define the map
\(
    \mathfrak{D}_{\mathrm{LS}}:
    \operatorname{End}(\mathcal{H}_{\mathsf{D}})
    \to
    \operatorname{End}(\mathcal{H}_{\mathsf{D}})
\)
according to
\begin{equation}
    \mathfrak{D}_{\mathrm{LS}}
    \coloneqq 
    \sum_{R\subseteq\mathsf{D}}
    \mathcal{E}_R^{\mathsf{D}}
    \circ
    \mathcal{L}_R^{\mathsf{D}}
    \circ
    \mathcal{C}_{\mathrm F}^{-1}
    \circ
    \mathcal{Z}_{\mathsf{S}}
    \circ
    \mathcal{C}_{\mathrm F}^{\vpd}
    \circ
    \mathcal{F}_R^{\mathsf{D}}
    \circ
    \mathcal{P}_R^{\mathsf{D}}.
    \label{eq:leakage--syndrome-dephasing-map}
\end{equation}
\end{defn}

\begin{lem}
\label{lem:leakage--syndrome-dephasing-channel}
The map \(\mathfrak{D}_{\mathrm{LS}}\) 
s completely positive and trace preserving (CPTP).
\end{lem}

\begin{proof}
For every \(R\subseteq\mathsf{D}\), the maps \(\mathcal{P}_R^{\mathsf{D}}\) and \(\mathcal{E}_R^{\mathsf{D}}\) are completely positive.
Lemma~\ref{lem:filling-leakage-map-properties}, unitarity of \(\CFrame\), and Lemma~\ref{lem:syndrome-dephasing-Pauli-channel} show that the remaining factors in every summand of Eq.~\eqref{eq:leakage--syndrome-dephasing-map} are completely positive. 
Hence \(\mathfrak{D}_{\mathrm{LS}}\) is completely positive.

Equation~\eqref{eq:leakage-subspace-map-traces}, specialized to \(\mathsf{D}\), Lemma~\ref{lem:filling-leakage-map-properties}, together with unitarity of \(\CFrame\) and Lemma~\ref{lem:syndrome-dephasing-Pauli-channel}, show that every factor in each summand other than \(\mathcal{P}_R^{\mathsf{D}}\) is trace preserving.  
Therefore, for every \(X\in\operatorname{End}(\mathcal{H}_{\mathsf{D}})\), Eq.~\eqref{eq:registerwise-leakage-dephasing} and trace preservation of \(\Delta^{\mathsf{D}}\) give
\begin{align*}
    \trace\!\left[
        \mathfrak{D}_{\mathrm{LS}}(X)
    \right]
    &=
    \sum_{R\subseteq\mathsf{D}}
    \trace\!\left[
        \mathcal{P}_R^{\mathsf{D}}(X)
    \right]
    =
    \trace\!\left[
        \Delta^{\mathsf{D}}(X)
    \right]
    =
    \trace(X).
\end{align*}
\end{proof}

\begin{lem}
For every \(R\subseteq\mathsf{D}\), we have that
\begin{equation}
    \left(
        \mathcal{C}_{\mathrm F}^{\vpd}
        \circ
        \mathcal{F}_R^{\mathsf{D}}
        \circ
        \mathcal{P}_R^{\mathsf{D}}
    \right)
    \circ
    \mathfrak{D}_{\mathrm{LS}}
    =
    \mathcal{Z}_{\mathsf{S}}
    \circ
    \left(
        \mathcal{C}_{\mathrm F}^{\vpd}
        \circ
        \mathcal{F}_R^{\mathsf{D}}
        \circ
        \mathcal{P}_R^{\mathsf{D}}
    \right).
    \label{eq:canonical-dephasing-filled-representative}
\end{equation}
\end{lem}

\begin{proof}
Fix \(R\subseteq\mathsf{D}\) and let \(\mathcal{G}_R^{\mathsf{D}}\) be the computational replacement channel, as in Definition~\ref{defn:computational-replacement-channel}, specialized to \(\mathsf{D}\) and \(R\).
Substituting Eq.~\eqref{eq:leakage--syndrome-dephasing-map} into
\(
    (\mathcal{C}_{\mathrm F}^{\vpp}
    \circ\mathcal{F}_R^{\mathsf{D}}
    \circ\mathcal{P}_R^{\mathsf{D}})
    \circ\mathfrak{D}_{\mathrm{LS}}
\)
and renaming its summation variable at \(T\), we obtain a sum in which Eq.~\eqref{eq:complete-cycle-input-subspace-replacement} applies with \(R_{\mathrm{in}}=T\).  
All summands with \(T\neq R\) vanish, and the summand with \(T=R\) gives
\begin{align}
    \left(
        \mathcal{C}_{\mathrm F}^{\vpd}
        \circ
        \mathcal{F}_R^{\mathsf{D}}
        \circ
        \mathcal{P}_R^{\mathsf{D}}
    \right)
    \circ
    \mathfrak{D}_{\mathrm{LS}}
    =
    \left(
        \mathcal{C}_{\mathrm F}^{\vpd}
        \circ
        \mathcal{G}_R^{\mathsf{D}}
        \circ
        \mathcal{C}_{\mathrm F}^{-1}
    \right)
    \circ
    \mathcal{Z}_{\mathsf{S}}
    \circ
    \left(
        \mathcal{C}_{\mathrm F}^{\vpd}
        \circ
        \mathcal{F}_R^{\mathsf{D}}
        \circ
        \mathcal{P}_R^{\mathsf{D}}
    \right).
    \label{eq:canonical-dephasing-sector-reduction}
\end{align}
Conjugating the expansion in Eq.~\eqref{eq:computational-replacement-Pauli-channel} by \(\mathcal{C}_{\mathrm F}^{\vpd}\) and applying Eq.~\eqref{eq:Clifford-frame-Pauli-labels} show that
\(
    \mathcal{C}_{\mathrm F}^{\vpp}
    \circ
    \mathcal{G}_R^{\mathsf{D}}
    \circ
    \mathcal{C}_{\mathrm F}^{-1}
\)
is a Pauli channel on \(\operatorname{End}(\mathcal{K}_{\mathsf{L}}\otimes \mathcal{K}_{\mathsf{S}})\).
Equation \eqref{eq:syndrome-dephasing-Pauli-commutation} therefore gives
\begin{align}
    &\left(
        \mathcal{C}_{\mathrm F}^{\vpd}
        \circ
        \mathcal{G}_R^{\mathsf{D}}
        \circ
        \mathcal{C}_{\mathrm F}^{-1}
    \right)
    \circ
    \mathcal{Z}_{\mathsf{S}}
    =
    \mathcal{Z}_{\mathsf{S}}
    \circ
    \left(
        \mathcal{C}_{\mathrm F}^{\vpd}
        \circ
        \mathcal{G}_R^{\mathsf{D}}
        \circ
        \mathcal{C}_{\mathrm F}^{-1}
    \right).
    \label{eq:canonical-dephasing-replacement-commutation}
\end{align}
Equations~\eqref{eq:Clifford-frame-channel-inverses}, \eqref{eq:computational-replacement-map}, and \eqref{eq:leak-fill-identity} give
\begin{align}
    &\left(
        \mathcal{C}_{\mathrm F}^{\vpd}
        \circ
        \mathcal{G}_R^{\mathsf{D}}
        \circ
        \mathcal{C}_{\mathrm F}^{-1}
    \right)
    \circ
    \left(
        \mathcal{C}_{\mathrm F}^{\vpd}
        \circ
        \mathcal{F}_R^{\mathsf{D}}
        \circ
        \mathcal{P}_R^{\mathsf{D}}
    \right)
    =
    \mathcal{C}_{\mathrm F}^{\vpd}
    \circ
    \mathcal{F}_R^{\mathsf{D}}
    \circ
    \mathcal{L}_R^{\mathsf{D}}
    \circ
    \mathcal{F}_R^{\mathsf{D}}
    \circ
    \mathcal{P}_R^{\mathsf{D}}
    =
    \mathcal{C}_{\mathrm F}^{\vpd}
    \circ
    \mathcal{F}_R^{\mathsf{D}}
    \circ
    \mathcal{P}_R^{\mathsf{D}}.
    \label{eq:canonical-dephasing-replacement-reduction}
\end{align}
Combining Eqs.~\eqref{eq:canonical-dephasing-sector-reduction}, \eqref{eq:canonical-dephasing-replacement-commutation}, and \eqref{eq:canonical-dephasing-replacement-reduction} gives Eq.~\eqref{eq:canonical-dephasing-filled-representative}.
\end{proof}

\begin{cor}
\label{cor:leakage--syndrome-dephasing-idempotence}
The map \(\mathfrak{D}_{\mathrm{LS}}\) is idempotent:
\begin{equation*}
    \mathfrak{D}_{\mathrm{LS}}
    \circ
    \mathfrak{D}_{\mathrm{LS}}
    =
    \mathfrak{D}_{\mathrm{LS}}.
\end{equation*}
\end{cor}

\begin{proof}
For every \(s,t\in\mathbb{Z}_2^{\mathsf{S}}\), Definition~\ref{def:syndrome-block-projection-maps} gives
\(\mathcal{Z}_s\circ\mathcal{Z}_t=\delta_{s,t}\mathcal{Z}_t\).
Summing this equality over \(s,t\in\mathbb{Z}_2^{\mathsf{S}}\) and applying Eq.~\eqref{eq:syndrome-dephasing-channel} give
\(\mathcal{Z}_{\mathsf{S}}^2=\mathcal{Z}_{\mathsf{S}}\).
Substituting Eq.~\eqref{eq:canonical-dephasing-filled-representative} into Eq.~\eqref{eq:leakage--syndrome-dephasing-map} and using \(\mathcal{Z}_{\mathsf{S}}^2=\mathcal{Z}_{\mathsf{S}}\) give
\begin{align*}
    \mathfrak{D}_{\mathrm{LS}}
    \circ
    \mathfrak{D}_{\mathrm{LS}}
    &={}
    \sum_{R\subseteq\mathsf{D}}
    \mathcal{E}_R^{\mathsf{D}}
    \circ
    \mathcal{L}_R^{\mathsf{D}}
    \circ
    \mathcal{C}_{\mathrm F}^{-1}
    \circ
    \mathcal{Z}_{\mathsf{S}}^2
    \circ
    \mathcal{C}_{\mathrm F}^{\vpd}
    \circ
    \mathcal{F}_R^{\mathsf{D}}
    \circ
    \mathcal{P}_R^{\mathsf{D}}
    =
    \mathfrak{D}_{\mathrm{LS}}.
\end{align*}
\end{proof}

\begin{defn}[Leakage--syndrome-dephased operators]
\label{def:leakage--syndrome-dephased-operator}
An operator \(X\in\operatorname{End}(\mathcal{H}_{\mathsf{D}})\) is \emph{leakage--syndrome dephased} if \(\mathfrak{D}_{\mathrm{LS}}(X)=X\).
\end{defn}

Corollary~\ref{cor:leakage--syndrome-dephasing-idempotence} and Definition~\ref{def:leakage--syndrome-dephased-operator} imply that \(\mathfrak{D}_{\mathrm{LS}}(X)\) is leakage--syndrome dephased for every \(X\in\operatorname{End}(\mathcal{H}_{\mathsf{D}})\).

\begin{prop}
\label{prop:complete-cycle-leakage--syndrome-dephasing}
The complete physical QEC-cycle channel \(\Phi\) 
satisfies
\begin{equation}
    \Phi
    =
    \Phi\circ\mathfrak{D}_{\mathrm{LS}}
    =
    \mathfrak{D}_{\mathrm{LS}}\circ\Phi.
    \label{eq:complete-cycle-leakage--syndrome-factorization}
\end{equation}
Moreover, \(\Phi(X)\) is leakage--syndrome dephased for every \(X\in\operatorname{End}(\mathcal{H}_{\mathsf{D}})\).
\end{prop}

\begin{proof}
Right composing Eq.~\eqref{eq:complete-cycle-filled-frame-reconstruction} by \(\mathfrak{D}_{\mathrm{LS}}\) and substituting Eq.~\eqref{eq:canonical-dephasing-filled-representative} give
\begin{align}
    \Phi\circ\mathfrak{D}_{\mathrm{LS}}
    &={}
    \sum_{R_{\mathrm{in}},R_{\mathrm{out}}\subseteq\mathsf{D}}
    \mathcal{E}_{R_{\mathrm{out}}}^{\mathsf{D}}
    \circ
    \mathcal{L}_{R_{\mathrm{out}}}^{\mathsf{D}}
    \circ
    \mathcal{C}_{\mathrm F}^{-1}
    \circ
    \widetilde\Phi_{R_{\mathrm{out}}, R_{\mathrm{in}}}
    \circ
    \mathcal{Z}_{\mathsf{S}}
    \circ
    \mathcal{C}_{\mathrm F}^{\vpd}
    \circ
    \mathcal{F}_{R_{\mathrm{in}}}^{\mathsf{D}}
    \circ
    \mathcal{P}_{R_{\mathrm{in}}}^{\mathsf{D}}.
    \label{eq:complete-cycle-right-factorization-expansion}
\end{align}
Substituting Eq.~\eqref{eq:complete-cycle-filled-frame-syndrome-dephasing} into Eq.~\eqref{eq:complete-cycle-right-factorization-expansion} and applying Eq.~\eqref{eq:complete-cycle-filled-frame-reconstruction} give \(\Phi\circ\mathfrak{D}_{\mathrm{LS}}=\Phi\).

Equations~\eqref{eq:leakage-subspace-branch}, \eqref{eq:complete-cycle-filled-frame-representative}, \eqref{eq:Clifford-frame-channel-inverses}, and \eqref{eq:leak-fill-identity} give, for every \(R_{\mathrm{in}},R_{\mathrm{out}}\subseteq\mathsf{D}\),
\begin{align}
    &\widetilde\Phi_{R_{\mathrm{out}}, R_{\mathrm{in}}}
    \circ
    \left(
        \mathcal{C}_{\mathrm F}^{\vpd}
        \circ
        \mathcal{F}_{R_{\mathrm{in}}}^{\mathsf{D}}
        \circ
        \mathcal{P}_{R_{\mathrm{in}}}^{\mathsf{D}}
    \right)
    =
    \left(
        \mathcal{C}_{\mathrm F}^{\vpd}
        \circ
        \mathcal{F}_{R_{\mathrm{out}}}^{\mathsf{D}}
        \circ
        \mathcal{P}_{R_{\mathrm{out}}}^{\mathsf{D}}
    \right)
    \circ
    \Phi
    \circ
    \mathcal{E}_{R_{\mathrm{in}}}^{\mathsf{D}}
    \circ
    \mathcal{P}_{R_{\mathrm{in}}}^{\mathsf{D}}.
    \label{eq:complete-cycle-filled-frame-sector composition}
\end{align}
Fix \(R\subseteq\mathsf{D}\).
Summing Eq.~\eqref{eq:complete-cycle-filled-frame-sector composition} with \(R_{\mathrm{out}}=R\) over \(R_{\mathrm{in}}\subseteq\mathsf{D}\), applying Eq.~\eqref{eq:registerwise-leakage-dephasing}, and using \(\Phi\circ\Delta^{\mathsf{D}}=\Phi\) from Eq.~\eqref{eq:complete-cycle-leakage-dephasing} give
\begin{align}
    &\left(
        \mathcal{C}_{\mathrm F}^{\vpd}
        \circ
        \mathcal{F}_R^{\mathsf{D}}
        \circ
        \mathcal{P}_R^{\mathsf{D}}
    \right)
    \circ
    \Phi
    =
    \sum_{R_{\mathrm{in}}\subseteq\mathsf{D}}
    \widetilde\Phi_{R, R_{\mathrm{in}}}
    \circ
    \left(
        \mathcal{C}_{\mathrm F}^{\vpd}
        \circ
        \mathcal{F}_{R_{\mathrm{in}}}^{\mathsf{D}}
        \circ
        \mathcal{P}_{R_{\mathrm{in}}}^{\mathsf{D}}
    \right).
    \label{eq:complete-cycle-output-filled-frame-expansion}
\end{align}
Left composing Eq.~\eqref{eq:complete-cycle-output-filled-frame-expansion} by \(\mathcal{Z}_{\mathsf{S}}\) and applying Eq.~\eqref{eq:complete-cycle-filled-frame-double-syndrome-dephasing} give
\begin{equation}
    \mathcal{Z}_{\mathsf{S}}
    \circ
    \left(
        \mathcal{C}_{\mathrm F}^{\vpd}
        \circ
        \mathcal{F}_R^{\mathsf{D}}
        \circ
        \mathcal{P}_R^{\mathsf{D}}
    \right)
    \circ
    \Phi
    =
    \left(
        \mathcal{C}_{\mathrm F}^{\vpd}
        \circ
        \mathcal{F}_R^{\mathsf{D}}
        \circ
        \mathcal{P}_R^{\mathsf{D}}
    \right)
    \circ
    \Phi,
    \label{eq:complete-cycle-output-filled-frame-dephasing}
\end{equation}
for every \(R\subseteq\mathsf{D}\). 
Substituting Eq.~\eqref{eq:complete-cycle-output-filled-frame-dephasing} into Eq.~\eqref{eq:leakage--syndrome-dephasing-map} and applying Eqs.~\eqref{eq:Clifford-frame-channel-inverses}, \eqref{eq:leak-fill-identity}, and \eqref{eq:registerwise-leakage-dephasing}, together with \(\Delta^{\mathsf{D}}\circ\Phi=\Phi\) from Eq.~\eqref{eq:complete-cycle-leakage-dephasing}, give
\begin{align}
    \mathfrak{D}_{\mathrm{LS}}
    \circ
    \Phi
    &={}
    \sum_{R\subseteq\mathsf{D}}
    \mathcal{E}_R^{\mathsf{D}}
    \circ
    \mathcal{L}_R^{\mathsf{D}}
    \circ
    \mathcal{C}_{\mathrm F}^{-1}
    \circ
    \mathcal{Z}_{\mathsf{S}}
    \circ
    \mathcal{C}_{\mathrm F}
    \circ
    \mathcal{F}_R^{\mathsf{D}}
    \circ
    \mathcal{P}_R^{\mathsf{D}}
    \circ
    \Phi
    \nonumber\\
    &=
    \sum_{R\subseteq\mathsf{D}}
    \mathcal{E}_R^{\mathsf{D}}
    \circ
    \mathcal{L}_R^{\mathsf{D}}
    \circ
    \mathcal{F}_R^{\mathsf{D}}
    \circ
    \mathcal{P}_R^{\mathsf{D}}
    \circ
    \Phi =
    \Delta^{\mathsf{D}}
    \circ
    \Phi =
    \Phi.
    \label{eq:complete-cycle-left-factorization}
\end{align}
Combining \(\Phi\circ\mathfrak{D}_{\mathrm{LS}}=\Phi\) with Eq.~\eqref{eq:complete-cycle-left-factorization} gives Eq.~\eqref{eq:complete-cycle-leakage--syndrome-factorization}.
Equation~\eqref{eq:complete-cycle-left-factorization} gives \(\mathfrak{D}_{\mathrm{LS}}(\Phi(X))=\Phi(X)\) for every \(X\in\operatorname{End}(\mathcal{H}_{\mathsf{D}})\).
Definition~\ref{def:leakage--syndrome-dephased-operator} therefore implies that \(\Phi(X)\) is leakage--syndrome dephased.
\end{proof}

\subsection{Logical blocks and standard form}
\label{subsec:logical-blocks-standard-form}

\begin{defn}[Clifford-frame representatives]
\label{def:filled-frame-sector-representatives}
For every \(X\in\operatorname{End}(\mathcal{H}_{\mathsf{D}})\) and for every \(R\subseteq\mathsf{D}\), define the operator \(\widetilde X_R\in \operatorname{End}(\mathcal{K}_{\mathsf{L}}\otimes\mathcal{K}_{\mathsf{S}})\) by
\begin{equation}
    \widetilde X_R
    \coloneqq 
    \left(
        \mathcal{C}_{\mathrm F}^{\vpd}
        \circ
        \mathcal{F}_R^{\mathsf{D}}
        \circ
        \mathcal{P}_R^{\mathsf{D}}
    \right)(X).
    \label{eq:filled-frame-representative}
\end{equation}
Define the \emph{Clifford-frame family associated with \(X\)} by \(\{\widetilde X_R\}_{R\subseteq\mathsf{D}}\).
\end{defn}

\begin{defn}[Hidden-memory index]
\label{def:hidden-memory-index}
Define the \emph{hidden-memory index set} by
\(
    \mathsf{M}
    \coloneqq 
    2^{\mathsf{D}}
    \times
    \mathbb{Z}_2^{\mathsf{S}}
\).
For \(m=(R,s)\in\mathsf{M}\), write
\(
    R(m)\coloneqq R
\)
and
\(
    s(m)\coloneqq s
\).
\end{defn}

\begin{defn}[Logical blocks]
\label{def:hidden-memory-logical-blocks}
For \(X\in\operatorname{End}(\mathcal{H}_{\mathsf{D}})\), let
\(
    \{\widetilde X_R\}_{R\subseteq\mathsf{D}}
\)
be the Clifford-frame family associated with \(X\) as in
Definition~\ref{def:filled-frame-sector-representatives}.
For every \(m\in\mathsf{M}\), define the
\emph{logical block}
\(X_m^{\mathsf{L}}\in\operatorname{End}(\mathcal{K}_{\mathsf{L}})\) by
\begin{equation}
    X_m^{\mathsf{L}}
    \coloneqq 
    \trace_{\mathsf{S}}
    \left[
        \mathcal{Z}_{s(m)}
        \left(
            \widetilde X_{R(m)}
        \right)
    \right].
    \label{eq:logical-block-definition}
\end{equation}
Define the \emph{logical-block family associated with \(X\)} to be
\(
    \{X_m^{\mathsf{L}}\}_{m\in\mathsf{M}}
\).
\end{defn}

\begin{lem}
\label{lem:canonical-dephasing-clifford-frame-expansion}
For \(X\in\operatorname{End}(\mathcal{H}_{\mathsf{D}})\), denote by \(\{X_m^{\mathsf{L}}\}_{m\in\mathsf{M}}\)  the logical-block family associated with \(X\), as in Definition~\ref{def:hidden-memory-logical-blocks}, and denote by \(\{\widetilde{\mathfrak{D}_{\mathrm{LS}}(X)}_R\}_{R\subseteq\mathsf{D}}\) the Clifford-frame family associated with \(\mathfrak{D}_{\mathrm{LS}}(X)\), as in Definition~\ref{def:filled-frame-sector-representatives}.
Then, for every \(R\subseteq\mathsf{D}\),
\begin{equation}
    \widetilde{\mathfrak{D}_{\mathrm{LS}}(X)}_R
    =
    \sum_{s\in\mathbb{Z}_2^{\mathsf{S}}}
    X_{(R,s)}^{\mathsf{L}}
    \otimes
    \ZProj{s}_{\mathsf{S}}.
    \label{eq:canonical-dephasing-filled-block-expansion}
\end{equation}
\end{lem}

\begin{proof}
Denote by \(\{\widetilde X_R\}_{R\subseteq\mathsf{D}}\) the Clifford-frame family associated with \(X\), as in Definition~\ref{def:filled-frame-sector-representatives}, and fix \(R\subseteq\mathsf{D}\).
Equations~\eqref{eq:canonical-dephasing-filled-representative}, \eqref{eq:filled-frame-representative}, and \eqref{eq:syndrome-dephasing-channel} give
\begin{equation*}
    \widetilde{\mathfrak{D}_{\mathrm{LS}}(X)}_R
    =
    \mathcal{Z}_{\mathsf{S}}
    \left(
        \widetilde X_R
    \right)
    =
    \sum_{s\in\mathbb{Z}_2^{\mathsf{S}}}
    \mathcal{Z}_s
    \left(
        \widetilde X_R
    \right).
\end{equation*}
For every \(s\in\mathbb{Z}_2^{\mathsf{S}}\), Eq.~\eqref{eq:syndrome-block-selection-map} gives
\begin{equation*}
    \mathcal{Z}_s
    \left(
        \widetilde X_R
    \right)
    =
    \left(
        \id_{\mathsf{L}}\otimes\ZProj{s}_{\mathsf{S}}
    \right)
    \mathcal{Z}_s
    \left(
        \widetilde X_R
    \right)
    \left(
        \id_{\mathsf{L}}\otimes\ZProj{s}_{\mathsf{S}}
    \right).
\end{equation*}
Since \(\ZProj{s}_{\mathsf{S}}\) has rank one and trace one, this equality implies that
\begin{equation*}
    \mathcal{Z}_s
    \left(
        \widetilde X_R
    \right)
    =
    \trace_{\mathsf{S}}
    \left[
        \mathcal{Z}_s
        \left(
            \widetilde X_R
        \right)
    \right]
    \otimes
    \ZProj{s}_{\mathsf{S}}.
\end{equation*}
Substituting these equalities into
\(
    \sum_{s\in\mathbb{Z}_2^{\mathsf{S}}}
    \mathcal{Z}_s(\widetilde X_R)
\)
and applying Eq.~\eqref{eq:logical-block-definition} give
Eq.~\eqref{eq:canonical-dephasing-filled-block-expansion}.
\end{proof}

For every \(A\in\operatorname{End}(\mathcal{K}_{\mathsf{L}})\) and
\(s,t\in\mathbb{Z}_2^{\mathsf{S}}\), one has
\[
    \ZProj{s}_{\mathsf{S}}
    \ZProj{t}_{\mathsf{S}}
    \ZProj{s}_{\mathsf{S}}
    =
    \delta_{s,t}\ZProj{s}_{\mathsf{S}},
    \qquad
    \trace_{\mathsf{S}}
    \left(
        \ZProj{s}_{\mathsf{S}}
    \right)
    =
    1.
\]
Equation~\eqref{eq:syndrome-block-selection-map} therefore gives
\begin{equation}
    \trace_{\mathsf{S}}
    \left[
        \mathcal{Z}_s
        \left(
            A\otimes\ZProj{t}_{\mathsf{S}}
        \right)
    \right]
    =
    \delta_{s,t}A.
    \label{eq:syndrome-block-extraction-identity}
\end{equation}
For \(X\in\operatorname{End}(\mathcal{H}_{\mathsf{D}})\), let \(\{X_m^{\mathsf{L}}\}_{m\in\mathsf{M}}\) and \(\{(\mathfrak{D}_{\mathrm{LS}}(X))_m^{\mathsf{L}}\}_{m\in\mathsf{M}}\) be the logical-block families associated with \(X\) and \(\mathfrak{D}_{\mathrm{LS}}(X)\), respectively, as in Definition~\ref{def:hidden-memory-logical-blocks}.
Fix \(m\in\mathsf{M}\). 
Equations~\eqref{eq:logical-block-definition}, \eqref{eq:canonical-dephasing-filled-block-expansion}, and \eqref{eq:syndrome-block-extraction-identity}, together with Definition~\ref{def:hidden-memory-index}, give
\begin{align*}
    \left(
        \mathfrak{D}_{\mathrm{LS}}(X)
    \right)_m^{\mathsf{L}}
    &=
    \trace_{\mathsf{S}}
    \left[
        \mathcal{Z}_{s(m)}
        \left(
            \widetilde{\mathfrak{D}_{\mathrm{LS}}(X)}_{R(m)}
        \right)
    \right]
    \\
    &=
    \sum_{s\in\mathbb{Z}_2^{\mathsf{S}}}
    \trace_{\mathsf{S}}
    \left[
        \mathcal{Z}_{s(m)}
        \left(
            X_{(R(m),s)}^{\mathsf{L}}
            \otimes
            \ZProj{s}_{\mathsf{S}}
        \right)
    \right]
    \\
    &=
    X_{(R(m),s(m))}^{\mathsf{L}}
    =
    X_m^{\mathsf{L}}.
\end{align*}

\begin{cor}
\label{lem:leakage--syndrome-standard-form}
For \(X\in\operatorname{End}(\mathcal{H}_{\mathsf{D}})\), let
\(
    \{X_m^{\mathsf{L}}\}_{m\in\mathsf{M}}
\)
be the logical-block family associated with \(X\) as in
Definition~\ref{def:hidden-memory-logical-blocks}.  Then
\begin{equation}
    \mathfrak{D}_{\mathrm{LS}}(X)
    =
    \sum_{m\in\mathsf{M}}
    \left(
        \mathcal{E}_{R(m)}^{\mathsf{D}}
        \circ
        \mathcal{L}_{R(m)}^{\mathsf{D}}
        \circ
        \mathcal{C}_{\mathrm F}^{-1}
    \right)
    \left(
        X_m^{\mathsf{L}}
        \otimes
        \ZProj{s(m)}_{\mathsf{S}}
    \right).
    \label{eq:leakage--syndrome-standard-form}
\end{equation}
\end{cor}

\begin{proof}
Let
\(
    \{\widetilde{\mathfrak{D}_{\mathrm{LS}}(X)}_R\}_
    {R\subseteq\mathsf{D}}
\)
be the Clifford-frame family associated with
\(\mathfrak{D}_{\mathrm{LS}}(X)\) as in
Definition~\ref{def:filled-frame-sector-representatives}.
Equations~\eqref{eq:leakage--syndrome-dephasing-map},
\eqref{eq:canonical-dephasing-filled-representative}, and
\eqref{eq:filled-frame-representative} give
\begin{equation*}
    \mathfrak{D}_{\mathrm{LS}}(X)
    =
    \sum_{R\subseteq\mathsf{D}}
    \left(
        \mathcal{E}_R^{\mathsf{D}}
        \circ
        \mathcal{L}_R^{\mathsf{D}}
        \circ
        \mathcal{C}_{\mathrm F}^{-1}
    \right)
    \left(
        \widetilde{\mathfrak{D}_{\mathrm{LS}}(X)}_R
    \right).
\end{equation*}
Substituting
Eq.~\eqref{eq:canonical-dephasing-filled-block-expansion} and
reindexing the resulting double sum by
\(m=(R,s)\in\mathsf{M}\), using
Definition~\ref{def:hidden-memory-index}, give
Eq.~\eqref{eq:leakage--syndrome-standard-form}.
\end{proof}

\subsection{Single-cycle hidden-memory update}
\label{subsec:single-cycle-hidden-memory-update}

For \(m_{\mathrm{in}},m_{\mathrm{out}}\in\mathsf{M}\), define
\(\mathcal{J}_{m_{\mathrm{out}}, m_{\mathrm{in}}}\) by
\begin{equation}
    \mathcal{J}_{m_{\mathrm{out}}, m_{\mathrm{in}}}
    \coloneqq 
    \mathcal{J}^{R(m_{\mathrm{out}}), R(m_{\mathrm{in}})}
    _{s(m_{\mathrm{out}}),s(m_{\mathrm{in}})}.
    \label{eq:memory-indexed-logical-transition-map}
\end{equation}

\begin{prop}
\label{prop:single-cycle-standard-form-update}
For every \(X\in\operatorname{End}(\mathcal{H}_{\mathsf{D}})\), let
\(
    \{X_m^{\mathsf{L}}\}_{m\in\mathsf{M}}
\)
and
\(
    \{(\Phi(X))_m^{\mathsf{L}}\}_{m\in\mathsf{M}}
\)
be the logical-block families associated with \(X\) and \(\Phi(X)\),
respectively, as in
Definition~\ref{def:hidden-memory-logical-blocks}.  Then, for every
\(m_{\mathrm{out}}\in\mathsf{M}\),
\begin{equation}
    \left(
        \Phi(X)
    \right)_{m_{\mathrm{out}}}^{\mathsf{L}}
    =
    \sum_{m_{\mathrm{in}}\in\mathsf{M}}
    \mathcal{J}_{m_{\mathrm{out}}, m_{\mathrm{in}}}
    \left(
        X_{m_{\mathrm{in}}}^{\mathsf{L}}
    \right).
    \label{eq:single-cycle-standard-form-update}
\end{equation}
\end{prop}

\begin{proof}
Let
\(
    \{\widetilde{\mathfrak{D}_{\mathrm{LS}}(X)}_R\}_
    {R\subseteq\mathsf{D}}
\)
and
\(
    \{\widetilde{\Phi(X)}_R\}_{R\subseteq\mathsf{D}}
\)
be the Clifford-frame families associated with
\(\mathfrak{D}_{\mathrm{LS}}(X)\) and \(\Phi(X)\), respectively, as in
Definition~\ref{def:filled-frame-sector-representatives}.
Evaluating Eq.~\eqref{eq:complete-cycle-output-filled-frame-expansion}
at \(\mathfrak{D}_{\mathrm{LS}}(X)\), using
\(\Phi\circ\mathfrak{D}_{\mathrm{LS}}=\Phi\) from
Eq.~\eqref{eq:complete-cycle-leakage--syndrome-factorization}, and
applying Eq.~\eqref{eq:filled-frame-representative} give, for every
\(R_{\mathrm{out}}\subseteq\mathsf{D}\),
\begin{equation}
    \widetilde{\Phi(X)}_{R_{\mathrm{out}}}
    =
    \sum_{R_{\mathrm{in}}\subseteq\mathsf{D}}
    \widetilde\Phi_{R_{\mathrm{out}}, R_{\mathrm{in}}}
    \left(
        \widetilde{\mathfrak{D}_{\mathrm{LS}}(X)}_{R_{\mathrm{in}}}
    \right).
    \label{eq:single-cycle-filled-frame-branch-sum}
\end{equation}
For
\(s_{\mathrm{in}},s_{\mathrm{out}},t
\in\mathbb{Z}_2^{\mathsf{S}}\),
Eq.~\eqref{eq:syndrome-transition-map} gives
\begin{equation}
    \mathcal{S}_{s_{\mathrm{out}}, s_{\mathrm{in}}}
    \left(
        \ZProj{t}_{\mathsf{S}}
    \right)
    =
    \delta_{s_{\mathrm{in}},t}
    \ZProj{s_{\mathrm{out}}}_{\mathsf{S}}.
    \label{eq:syndrome-transition-projector-action}
\end{equation}
Applying Eq.~\eqref{eq:complete-cycle-sector-transition-decomposition}
to
\(\widetilde{\mathfrak{D}_{\mathrm{LS}}(X)}_{R_{\mathrm{in}}}\)
and substituting
Eq.~\eqref{eq:canonical-dephasing-filled-block-expansion} termwise,
with its summation index renamed \(t\), yield a sum over
\(s_{\mathrm{in}},s_{\mathrm{out}},t\).
By Eq.~\eqref{eq:syndrome-transition-projector-action}, only the terms
with \(t=s_{\mathrm{in}}\) remain.  Thus, for every
\(R_{\mathrm{in}},R_{\mathrm{out}}\subseteq\mathsf{D}\),
\begin{align}
    &\widetilde\Phi_{R_{\mathrm{out}}, R_{\mathrm{in}}}
    \left(
        \widetilde{\mathfrak{D}_{\mathrm{LS}}(X)}_{R_{\mathrm{in}}}
    \right)
    =
    \sum_{s_{\mathrm{in}},s_{\mathrm{out}}
    \in\mathbb{Z}_2^{\mathsf{S}}}
    \mathcal{J}^{R_{\mathrm{out}}, R_{\mathrm{in}}}
    _{s_{\mathrm{out}},s_{\mathrm{in}}}
    \left(
        X_{(R_{\mathrm{in}},s_{\mathrm{in}})}^{\mathsf{L}}
    \right)
    \otimes
    \ZProj{s_{\mathrm{out}}}_{\mathsf{S}}.
    \label{eq:single-cycle-sector-branch-logical-update}
\end{align}
Substituting Eq.~\eqref{eq:single-cycle-sector-branch-logical-update} termwise into Eq.~\eqref{eq:single-cycle-filled-frame-branch-sum}, using Eq.~\eqref{eq:memory-indexed-logical-transition-map}, and reindexing the double sum over \(R_{\mathrm{in}}\) and \(s_{\mathrm{in}}\) by \(m_{\mathrm{in}}=(R_{\mathrm{in}},s_{\mathrm{in}})\) give
\begin{align}
    \widetilde{\Phi(X)}_{R_{\mathrm{out}}}
    &={}
    \sum_{s_{\mathrm{out}}\in\mathbb{Z}_2^{\mathsf{S}}}
    \left[
        \sum_{m_{\mathrm{in}}\in\mathsf{M}}
        \mathcal{J}_{(R_{\mathrm{out}},s_{\mathrm{out}})
        , m_{\mathrm{in}}}
        \left(
            X_{m_{\mathrm{in}}}^{\mathsf{L}}
        \right)
    \right]
    \otimes
    \ZProj{s_{\mathrm{out}}}_{\mathsf{S}}.
    \label{eq:single-cycle-filled-frame-update}
\end{align}
Fix \(m_{\mathrm{out}}\in\mathsf{M}\).
Substituting Eq.~\eqref{eq:single-cycle-filled-frame-update} with \(R_{\mathrm{out}}=R(m_{\mathrm{out}})\) into Eq.~\eqref{eq:logical-block-definition} gives a sum over \(s_{\mathrm{out}}\).
Termwise application of Eq.~\eqref{eq:syndrome-block-extraction-identity} leaves only the term with \(s_{\mathrm{out}}=s(m_{\mathrm{out}})\), giving Eq.~\eqref{eq:single-cycle-standard-form-update}.
\end{proof}

\begin{cor}
\label{cor:single-cycle-hidden-memory-standard-form}
For every \(X\in\operatorname{End}(\mathcal{H}_{\mathsf{D}})\), let \( \{X_m^{\mathsf{L}}\}_{m\in\mathsf{M}} \) be the logical-block family associated with \(X\), as in Definition~\ref{def:hidden-memory-logical-blocks}. 
Then
\begin{equation}
    \begin{aligned}
    \Phi(X)
    &=
    \hspace{-2.5mm} \sum_{m_{\mathrm{in}},m_{\mathrm{out}}\in\mathsf{M}} \hspace{-1mm}
    \left(
        \mathcal{E}_{R(m_{\mathrm{out}})}^{\mathsf{D}}
        \circ
        \mathcal{L}_{R(m_{\mathrm{out}})}^{\mathsf{D}}
        \circ
        \mathcal{C}_{\mathrm F}^{-1}
    \right)
    \left[
            \mathcal{J}_{m_{\mathrm{out}}, m_{\mathrm{in}}}
            \left(
                X_{m_{\mathrm{in}}}^{\mathsf{L}}
            \right)
        \otimes
        \ZProj{s(m_{\mathrm{out}})}_{\mathsf{S}} \right] \, .
    \end{aligned}
    \label{eq:single-cycle-hidden-memory-standard-form}
\end{equation}
\end{cor}

\begin{proof}
Let \( \{(\Phi(X))_m^{\mathsf{L}}\}_{m\in\mathsf{M}} \) be the logical-block family associated with \(\Phi(X)\), as in Definition~\ref{def:hidden-memory-logical-blocks}.
The factorization in Eq.~\eqref{eq:complete-cycle-leakage--syndrome-factorization} gives \(\Phi(X)=\mathfrak{D}_{\mathrm{LS}}(\Phi(X))\).
Applying Eq.~\eqref{eq:leakage--syndrome-standard-form} to \(\Phi(X)\), followed by termwise substitution of Eq.~\eqref{eq:single-cycle-standard-form-update}, gives Eq.~\eqref{eq:single-cycle-hidden-memory-standard-form}.
\end{proof}

\subsection{Repeated-cycle logical-block dynamics}
\label{subsec:multicycle}

\begin{prop}
\label{prop:repeated-cycle-logical-block composition}

For every integer \(t\geq1\), define \(\Phi^t\) to be the \(t\)-fold composition of \(\Phi\).  
For every such \(t\) and every \( X\in\operatorname{End}(\mathcal{H}_{\mathsf{D}})\), let \( \{X_m^{\mathsf{L}}\}_{m\in\mathsf{M}} \) and \( \{(\Phi^t(X))_m^{\mathsf{L}}\}_{m\in\mathsf{M}} \) be the logical-block families associated with \(X\) and \(\Phi^t(X)\), respectively, as in Definition~\ref{def:hidden-memory-logical-blocks}. 
Then, for every \(m_t\in\mathsf{M}\),
\begin{equation}
    \left(
        \Phi^t(X)
    \right)_{m_t}^{\mathsf{L}}
    =
    \sum_{m_0,\ldots,m_{t-1}\in\mathsf{M}}
    \left(
        \mathcal{J}_{m_t, m_{t-1}}
        \circ\cdots\circ
        \mathcal{J}_{m_1, m_0}
    \right)
    \left(
        X_{m_0}^{\mathsf{L}}
    \right).
    \label{eq:repeated-cycle-logical-block composition}
\end{equation}
For \(t=1\), the composition in Eq.~\eqref{eq:repeated-cycle-logical-block composition} is the single map \(\mathcal{J}_{m_1, m_0}\).
\end{prop}

\begin{proof}
When \(t=1\), Eq.~\eqref{eq:repeated-cycle-logical-block composition} reduces to Eq.~\eqref{eq:single-cycle-standard-form-update}.
Fix \(t\geq2\), and assume that Eq.~\eqref{eq:repeated-cycle-logical-block composition} holds with \(t\) replaced by \(t-1\). 
Let \( \{(\Phi^{t-1}(X))_m^{\mathsf{L}}\}_{m\in\mathsf{M}} \) be the logical-block family associated with \(\Phi^{t-1}(X)\), as in Definition~\ref{def:hidden-memory-logical-blocks}.
Fix \(m_t\in\mathsf{M}\).  
Applying Eq.~\eqref{eq:single-cycle-standard-form-update} with \(X\) replaced by \(\Phi^{t-1}(X)\) and substituting the induction hypothesis termwise give
\begin{align*}
    \left(
        \Phi^t(X)
    \right)_{m_t}^{\mathsf{L}}
    &={}
    \sum_{m_{t-1}\in\mathsf{M}}
    \mathcal{J}_{m_t, m_{t-1}}
    \left(
        \left(\Phi^{t-1}(X)\right)_{m_{t-1}}^{\mathsf{L}}
    \right)
    \\
    &={}
    \sum_{m_0,\ldots,m_{t-1}\in\mathsf{M}}
    \left(
        \mathcal{J}_{m_t, m_{t-1}}
        \circ\cdots\circ
        \mathcal{J}_{m_1, m_0}
    \right)
    \left(
        X_{m_0}^{\mathsf{L}}
    \right).
\end{align*}
\end{proof}

\subsection{Memory experiments from logical-Pauli transfer matrices}
\label{subsec:memory-experiment-transfer-matrices}

Define \( d_{\mathsf{L}}\coloneqq \dim(\mathcal{K}_{\mathsf{L}})=2^{|\mathsf{L}|}\) and denote by \(\trace_{\mathsf{L}}\) the trace on \(\mathcal{K}_{\mathsf{L}}\).
For \(\ell\in V_{\mathsf{L}}\), define \( P_\ell\coloneqq W^{\mathsf{L}}(\ell) \), using the specialization of the Pauli notation in Eq.~\eqref{eq:app-Pauli-definition} to \(\mathsf{L}\).
For \(\ell\in V_{\mathsf{L}}\), define the linear functional
\( c_\ell: \operatorname{End}(\mathcal{K}_{\mathsf{L}}) \to \mathbb{C} \) by
\begin{equation}
    c_\ell(A)
    \coloneqq 
    \frac{1}{d_{\mathsf{L}}}
    \trace_{\mathsf{L}}(P_\ell A),
    \label{eq:memory-logical-Pauli-coefficient}
\end{equation}
for all \(A\in\operatorname{End}(\mathcal{K}_{\mathsf{L}})\).
Combining Eqs.~\eqref{eq:memory-logical-Pauli-coefficient} and \eqref{eq:app-Pauli-orthogonality}, specialized to \(\mathsf{L}\), gives,
\begin{equation}
    A
    =
    \sum_{\ell\in V_{\mathsf{L}}}
    c_\ell(A)P_\ell ,
    \label{eq:memory-logical-Pauli-expansion}
\end{equation}
for all \(A\in\operatorname{End}(\mathcal{K}_{\mathsf{L}})\).

\begin{defn}[Pauli eigenvalues]
\label{def:memory-Pauli-map-eigenvalues}
Let
\(
    \mathcal{D}:
    \operatorname{End}(\mathcal{K}_{\mathsf{L}})
    \to
    \operatorname{End}(\mathcal{K}_{\mathsf{L}})
\)
be a Pauli map, and let
\(
    \{q_e\}_{e\in V_{\mathsf{L}}}
\)
be its unique Pauli-map coefficient family, as in Lemma~\ref{lem:pauli-map-structure}, specialized to \(\mathsf{L}\).
For \(\ell\in V_{\mathsf{L}}\), define the \emph{\(P_\ell\)-eigenvalue} of \(\mathcal{D}\) by
\begin{equation}
    \lambda_\ell(\mathcal{D})
    \coloneqq 
    \sum_{e\in V_{\mathsf{L}}}
    q_e(-1)^{[e,\ell]}.
    \label{eq:memory-Pauli-map-eigenvalue}
\end{equation}
\end{defn}

\begin{lem}
\label{lem:memory-Pauli-map-eigenoperators}
Let
\(
    \mathcal{D}:
    \operatorname{End}(\mathcal{K}_{\mathsf{L}})
    \to
    \operatorname{End}(\mathcal{K}_{\mathsf{L}})
\)
be a Pauli map.  
Then, for every \(\ell\in V_{\mathsf{L}}\),
\begin{equation}
    \mathcal{D}(P_\ell)
    =
    \lambda_\ell(\mathcal{D})P_\ell.
    \label{eq:memory-Pauli-map-eigenoperator}
\end{equation}
If \(\mathcal{D}\) is a subnormalized Pauli map, then \(\lambda_\ell(\mathcal{D})\in\mathbb R\) for every \(\ell\in V_{\mathsf{L}}\).
\end{lem}

\begin{proof}
Let \(\{q_e\}_{e\in V_{\mathsf{L}}}\) be the unique Pauli-map coefficient family of \(\mathcal{D}\), as in Lemma~\ref{lem:pauli-map-structure}.
Substituting Eq.~\eqref{eq:pauli-map-def} and using Eqs.~\eqref{eq:app-adjoint-action}, \eqref{eq:app-Pauli-Hermiticity}, and \eqref{eq:app-Pauli-conjugation}, specialized to \(\mathsf{L}\), give
\begin{align*}
    \mathcal{D}(P_\ell)
    =
    \sum_{e\in V_{\mathsf{L}}}
    q_e\operatorname{Ad}_{P_e}(P_\ell)
    =
    \sum_{e\in V_{\mathsf{L}}}
    q_e(-1)^{[e,\ell]}P_\ell
    =
    \lambda_\ell(\mathcal{D})P_\ell,
\end{align*}
which proves Eq.~\eqref{eq:memory-Pauli-map-eigenoperator}.
If \(\mathcal{D}\) is subnormalized, then Eq.~\eqref{eq:subnormalized-pauli-map-coefficients}, specialized to \(\mathsf{L}\), gives \(q_e\in[0,\infty)\) for every \(e\in V_{\mathsf{L}}\).
It follows from Eq.~\eqref{eq:memory-Pauli-map-eigenvalue} that \(\lambda_\ell(\mathcal{D})\in\mathbb R\) for every \(\ell\in V_{\mathsf{L}}\).
\end{proof}

\begin{defn}[Logical-Pauli transfer matrices]
\label{def:memory-logical-Pauli-transfer-matrices}
For \(\ell\in V_{\mathsf{L}}\), define \(T_\ell\in\mathbb C^{\mathsf{M}\times\mathsf{M}}\) by
\begin{equation}
    [T_\ell]_{m_{\mathrm{out}},m_{\mathrm{in}}}
    \coloneqq 
    \lambda_\ell\!\left(
        \mathcal{J}_{m_{\mathrm{out}}, m_{\mathrm{in}}}
    \right),
    \label{eq:memory-logical-Pauli-transfer-matrix}
\end{equation}
for every \(m_{\mathrm{in}},m_{\mathrm{out}}\in\mathsf{M}\).
\end{defn}

Proposition~\ref{prop:complete-cycle-transition-form}, Eq.~\eqref{eq:memory-indexed-logical-transition-map}, Lemma~\ref{lem:memory-Pauli-map-eigenoperators}, and Eq.~\eqref{eq:memory-logical-Pauli-transfer-matrix} imply that \(T_\ell\in\mathbb R^{\mathsf{M}\times\mathsf{M}}\) for every \(\ell\in V_{\mathsf{L}}\).

\begin{lem}
\label{lem:memory-logical-Pauli-transition-coefficient-scaling}
For every \(A\in\operatorname{End}(\mathcal{K}_{\mathsf{L}})\), \(\ell\in V_{\mathsf{L}}\), and \(m_{\mathrm{in}},m_{\mathrm{out}}\in\mathsf{M}\),
\begin{equation}
    c_\ell\!\left(
        \mathcal{J}_{m_{\mathrm{out}}, m_{\mathrm{in}}}(A)
    \right)
    =
    [T_\ell]_{m_{\mathrm{out}},m_{\mathrm{in}}}c_\ell(A).
    \label{eq:memory-logical-Pauli-transition-coefficient-scaling}
\end{equation}
\end{lem}

\begin{proof}
Fix \(A\in\operatorname{End}(\mathcal{K}_{\mathsf{L}})\), \(\ell\in V_{\mathsf{L}}\), and \(m_{\mathrm{in}},m_{\mathrm{out}}\in\mathsf{M}\).
By Proposition~\ref{prop:complete-cycle-transition-form} and Eq.~\eqref{eq:memory-indexed-logical-transition-map}, Lemma~\ref{lem:memory-Pauli-map-eigenoperators} applies to \(\mathcal{J}_{m_{\mathrm{out}}, m_{\mathrm{in}}}\).
Applying this map termwise to Eq.~\eqref{eq:memory-logical-Pauli-expansion} and using Eq.~\eqref{eq:memory-Pauli-map-eigenoperator} give
\begin{equation}
    \mathcal{J}_{m_{\mathrm{out}}, m_{\mathrm{in}}}(A)
    =
    \sum_{a\in V_{\mathsf{L}}}
    \lambda_a\!\left(
        \mathcal{J}_{m_{\mathrm{out}}, m_{\mathrm{in}}}
    \right)
    c_a(A)P_a.
    \label{eq:memory-transition-map-Pauli-expansion}
\end{equation}
Combining Eqs.~\eqref{eq:memory-logical-Pauli-coefficient} and \eqref{eq:app-Pauli-orthogonality}, specialized to \(\mathsf{L}\), gives \(c_\ell(P_a)=\delta_{\ell,a}\).
Applying \(c_\ell\) termwise to Eq.~\eqref{eq:memory-transition-map-Pauli-expansion} leaves only the term with \(a=\ell\).
Using Eq.~\eqref{eq:memory-logical-Pauli-transfer-matrix} then gives Eq.~\eqref{eq:memory-logical-Pauli-transition-coefficient-scaling}.
\end{proof}

\begin{defn}[Logical-Pauli state-preparation vectors]
\label{def:memory-logical-Pauli-block-coefficient-vectors}
For \(\rho\in\operatorname{End}(\mathcal{H}_{\mathsf{D}})\) with \(\rho\succeq0\) and \(\trace(\rho)=1\), let
\(
    \{\rho_m^{\mathsf{L}}\}_{m\in\mathsf{M}}
\)
be the logical-block family associated with \(\rho\), as in Definition~\ref{def:hidden-memory-logical-blocks}.
For \(\ell\in V_{\mathsf{L}}\), define the state-preparation vector \(\boldsymbol r_\ell(\rho)\in\mathbb C^{\mathsf{M}}\) by
\begin{equation}
    [\boldsymbol r_\ell(\rho)]_m
    \coloneqq 
    c_\ell(\rho_m^{\mathsf{L}}),
    \label{eq:memory-logical-Pauli-block-coefficient-vector}
\end{equation}
for all \(m\in\mathsf{M}\). 
\end{defn}

\begin{lem}
\label{lem:memory-single-cycle-logical-Pauli-dynamics}
For every \(\rho\in\operatorname{End}(\mathcal{H}_{\mathsf{D}})\) with \(\rho\succeq0\) and \(\trace(\rho)=1\), and every \(\ell\in V_{\mathsf{L}}\),
\begin{equation}
    \boldsymbol r_\ell\!\left(\Phi(\rho)\right)
    =
    T_\ell\boldsymbol r_\ell(\rho).
    \label{eq:memory-single-cycle-logical-Pauli-dynamics}
\end{equation}
\end{lem}

\begin{proof}
Fix \(\ell\in V_{\mathsf{L}}\) and \(m_{\mathrm{out}}\in\mathsf{M}\).
Applying \(c_\ell\) termwise to Eq.~\eqref{eq:single-cycle-standard-form-update} with \(X=\rho\), substituting Eq.~\eqref{eq:memory-logical-Pauli-transition-coefficient-scaling}, and using Eq.~\eqref{eq:memory-logical-Pauli-block-coefficient-vector} give
\begin{equation}
    [\boldsymbol r_\ell(\Phi(\rho))]_{m_{\mathrm{out}}}
    =
    \sum_{m_{\mathrm{in}}\in\mathsf{M}}
    [T_\ell]_{m_{\mathrm{out}},m_{\mathrm{in}}}
    [\boldsymbol r_\ell(\rho)]_{m_{\mathrm{in}}}
    =
    [T_\ell\boldsymbol r_\ell(\rho)]_{m_{\mathrm{out}}}.
    \label{eq:memory-single-cycle-logical-Pauli component-update}
\end{equation}
Since \(m_{\mathrm{out}}\) was arbitrary, Eq.~\eqref{eq:memory-single-cycle-logical-Pauli component-update} gives Eq.~\eqref{eq:memory-single-cycle-logical-Pauli-dynamics}.
\end{proof}

\begin{cor}
\label{cor:memory-repeated-cycle-logical-Pauli-dynamics}
For every \(\rho\in\operatorname{End}(\mathcal{H}_{\mathsf{D}})\) with \(\rho\succeq0\) and \(\trace(\rho)=1\), every \(\ell\in V_{\mathsf{L}}\), and every integer \(k\geq1\),
\begin{equation}
    \boldsymbol r_\ell\!\left(\Phi^k(\rho)\right)
    =
    T_\ell^k\boldsymbol r_\ell(\rho).
    \label{eq:memory-repeated-cycle-logical-Pauli-dynamics}
\end{equation}
\end{cor}

\begin{proof}
Fix \(\rho\in\operatorname{End}(\mathcal{H}_{\mathsf{D}})\) with \(\rho\succeq0\) and \(\trace(\rho)=1\), and fix \(\ell\in V_{\mathsf{L}}\).
The case \(k=1\) follows from Lemma~\ref{lem:memory-single-cycle-logical-Pauli-dynamics}.
Fix \(k\geq2\) and assume that Eq.~\eqref{eq:memory-repeated-cycle-logical-Pauli-dynamics} holds, with \(k\) replaced by \(k-1\).
Since \(\Phi\) is a channel, \(\Phi^{k-1}(\rho)\) is a state. 
Applying Lemma~\ref{lem:memory-single-cycle-logical-Pauli-dynamics} to \(\Phi^{k-1}(\rho)\) and substituting the induction hypothesis give
\begin{equation}
    \boldsymbol r_\ell\!\left(\Phi^k(\rho)\right)
    =
    T_\ell
    \boldsymbol r_\ell\!\left(\Phi^{k-1}(\rho)\right)
    =
    T_\ell T_\ell^{k-1}\boldsymbol r_\ell(\rho)
    =
    T_\ell^k\boldsymbol r_\ell(\rho).
    \label{eq:memory-repeated-cycle-logical-Pauli-induction-step}
\end{equation}
\end{proof}

\begin{defn}[Logical-Pauli measurement vectors]
\label{def:memory-logical-Pauli-measurement-vectors}
For \(E\in\operatorname{End}(\mathcal{H}_{\mathsf{D}})\) with \(0\preceq E\preceq\id_{\mathcal{H}_{\mathsf{D}}}\), let \(\{E_m^{\mathsf{L}}\}_{m\in\mathsf{M}}\) be the logical-block family associated with \(E\), as in Definition~\ref{def:hidden-memory-logical-blocks}.
For \(\ell\in V_{\mathsf{L}}\) and any \(m\in\mathsf{M}\), define the measurement vector \(\boldsymbol e_\ell(E)\in\mathbb C^{\mathsf{M}}\) as
\begin{equation}
    [\boldsymbol e_\ell(E)]_m
    \coloneqq 
    2^{|R(m)|}c_\ell(E_m^{\mathsf{L}}).
    \label{eq:memory-logical-Pauli-measurement-vector}
\end{equation}
\end{defn}

\begin{lem}
\label{lem:memory-logical-block-trace-pairing}
Let \(E,Y\in\operatorname{End}(\mathcal{H}_{\mathsf{D}})\) satisfy
\(0\preceq E\preceq\id_{\mathcal{H}_{\mathsf{D}}}\),
\(Y\succeq0\), and \(\trace(Y)=1\).
Let \(\{E_m^{\mathsf{L}}\}_{m\in\mathsf{M}}\) and
\(\{Y_m^{\mathsf{L}}\}_{m\in\mathsf{M}}\) be the logical-block
families associated with \(E\) and \(Y\), respectively, as in
Definition~\ref{def:hidden-memory-logical-blocks}, and assume that
\(Y\) is leakage--syndrome dephased.  Then
\begin{align}
    \trace(EY)
    &={}
    \sum_{m\in\mathsf{M}}
    2^{|R(m)|}
    \trace_{\mathsf{L}}\!\left(
        E_m^{\mathsf{L}}Y_m^{\mathsf{L}}
    \right)
    \label{eq:memory-logical-block-trace-pairing}
    \\
    &={}
    d_{\mathsf{L}}
    \sum_{\ell\in V_{\mathsf{L}}}
    \boldsymbol e_\ell(E)^{\mathsf{T}}
    \boldsymbol r_\ell(Y).
    \label{eq:memory-logical-Pauli-trace-pairing}
\end{align}
\end{lem}

\begin{proof}
Denote by \(\{\widetilde E_R\}_{R\subseteq\mathsf{D}}\) the Clifford-frame family associated with \(E\) as in Definition~\ref{def:filled-frame-sector-representatives}.
Since \(Y\) is leakage--syndrome dephased, Definition~\ref{def:leakage--syndrome-dephased-operator} gives \(Y=\mathfrak{D}_{\mathrm{LS}}(Y)\).
Substituting Eq.~\eqref{eq:leakage--syndrome-standard-form} with \(X=Y\) into \(\trace(EY)\) gives
\begin{align}
    \trace(EY)
    &={}
    \sum_{m\in\mathsf{M}}
    \trace\!\left[
        E
        \left(
            \mathcal{E}_{R(m)}^{\mathsf{D}}
            \circ
            \mathcal{L}_{R(m)}^{\mathsf{D}}
            \circ
            \mathcal{C}_{\mathrm F}^{-1}
        \right)
        \left(
            Y_m^{\mathsf{L}}
            \otimes
            \ZProj{s(m)}_{\mathsf{S}}
        \right)
    \right].
    \label{eq:memory-standard-form-trace-pairing}
\end{align}
Fix \(R\subseteq\mathsf{D}\) and \(C\in\operatorname{End}(\mathcal{H}_{\mathsf{D}}^{R})\).
Since \(\mathfrak{l}_R\) is one dimensional, there is a unique \(C_{\overline R}\in\operatorname{End}(\mathfrak{c}_{\overline R})\) such that \(C=\id_{\mathfrak{l}_R}\otimes C_{\overline R}\).
For \(B\in\operatorname{End}(\mathcal{K}_{\mathsf{D}} )\), Eqs.~\eqref{eq:fill-map} and~\eqref{eq:leak-map}, specialized to \(\mathsf{D}\) and \(R\), respectively, give
\begin{align}
    \trace\!\left[
        C\,\mathcal{L}_R^{\mathsf{D}}(B)
    \right]
    &={}
    \trace\!\left[
        C_{\overline R}
        \trace_{\mathfrak{c}_R}(B)
    \right]
    =
    \trace\!\left[
        \left(
            \id_{\mathfrak{c}_R}\otimes C_{\overline R}
        \right)B
    \right]
    =
    2^{|R|}
    \trace\!\left[
        \mathcal{F}_R^{\mathsf{D}}(C)B
    \right].
    \label{eq:memory-fill-leak-trace-duality}
\end{align}
Fix \(m\in\mathsf{M}\), and set \(R\coloneqq R(m)\) and \(s\coloneqq s(m)\).
Applying Eqs.~\eqref{eq:leakage-subspace-projection-map} and \eqref{eq:leakage-subspace-extension-map}, followed by Eq.~\eqref{eq:memory-fill-leak-trace-duality}, gives the first two transformations in the following calculation.  
Cyclicity of the trace, followed by Eq.~\eqref{eq:filled-frame-representative}, give the next two.
The remaining transformations follow from Eq.~\eqref{eq:syndrome-block-selection-map}, the defining property of the partial trace, and Eq.~\eqref{eq:logical-block-definition}:
\begin{align}
    &\trace\!\left[
        E
        \left(
            \mathcal{E}_R^{\mathsf{D}}
            \circ
            \mathcal{L}_R^{\mathsf{D}}
            \circ
            \mathcal{C}_{\mathrm F}^{-1}
        \right)
        \left(
            Y_m^{\mathsf{L}}\otimes\ZProj{s}_{\mathsf{S}}
        \right)
    \right]
    \nonumber\\
    &\quad={}
    \trace\!\left[
        \mathcal{P}_R^{\mathsf{D}}(E)
        \left(
            \mathcal{L}_R^{\mathsf{D}}
            \circ
            \mathcal{C}_{\mathrm F}^{-1}
        \right)
        \left(
            Y_m^{\mathsf{L}}\otimes\ZProj{s}_{\mathsf{S}}
        \right)
    \right]
    \nonumber\\
    &\quad={}
    2^{|R|}
    \trace\!\left[
        \left(
            \mathcal{F}_R^{\mathsf{D}}
            \circ
            \mathcal{P}_R^{\mathsf{D}}
        \right)(E)
        \, \mathcal{C}_{\mathrm F}^{-1}
        \left(
            Y_m^{\mathsf{L}}\otimes\ZProj{s}_{\mathsf{S}}
        \right)
    \right]
    \nonumber\\
    &\quad={}
    2^{|R|}
    \trace\!\left[
        \mathcal{C}_{\mathrm F}^{\vpd}\!\left(
            \left(
                \mathcal{F}_R^{\mathsf{D}}
                \circ
                \mathcal{P}_R^{\mathsf{D}}
            \right)(E)
        \right)
        \left(
            Y_m^{\mathsf{L}}\otimes\ZProj{s}_{\mathsf{S}}
        \right)
    \right]
    \nonumber\\
    &\quad={}
    2^{|R|}
    \trace\!\left[
        \widetilde E_R
        \left(
            Y_m^{\mathsf{L}}\otimes\ZProj{s}_{\mathsf{S}}
        \right)
    \right]
    \nonumber\\
    &\quad={}
    2^{|R|}
    \trace\!\left[
        \mathcal{Z}_s(\widetilde E_R)
        \left(
            Y_m^{\mathsf{L}}\otimes\id_{\mathcal{K}_{\mathsf{S}}}
        \right)
    \right]
    \nonumber\\
    &\quad={}
    2^{|R|}
    \trace_{\mathsf{L}}\!\left(
        \trace_{\mathsf{S}}\!\left[
            \mathcal{Z}_s(\widetilde E_R)
        \right]
        Y_m^{\mathsf{L}}
    \right)
    \nonumber\\
    &\quad={}
    2^{|R(m)|}
    \trace_{\mathsf{L}}\!\left(
        E_m^{\mathsf{L}}Y_m^{\mathsf{L}}
    \right).
    \label{eq:memory-standard-form-summand-pairing}
\end{align}
Substituting Eq.~\eqref{eq:memory-standard-form-summand-pairing} termwise into Eq.~\eqref{eq:memory-standard-form-trace-pairing} gives Eq.~\eqref{eq:memory-logical-block-trace-pairing}.

For \(m\in\mathsf{M}\), substituting Eq.~\eqref{eq:memory-logical-Pauli-expansion} with \(A=E_m^{\mathsf{L}}\) and \(A=Y_m^{\mathsf{L}}\) into \(\trace_{\mathsf{L}}(E_m^{\mathsf{L}}Y_m^{\mathsf{L}})\) and applying Eq.~\eqref{eq:app-Pauli-orthogonality} termwise, specialized to \(\mathsf{L}\), leads to
\begin{align}
    \trace_{\mathsf{L}}\!\left(
        E_m^{\mathsf{L}}Y_m^{\mathsf{L}}
    \right)
    &={}
    \sum_{\ell,\ell'\in V_{\mathsf{L}}}
    c_\ell(E_m^{\mathsf{L}})
    c_{\ell'}(Y_m^{\mathsf{L}})
    \trace_{\mathsf{L}}(P_\ell P_{\ell'})
    \nonumber\\
    &={}
    d_{\mathsf{L}}
    \sum_{\ell\in V_{\mathsf{L}}}
    c_\ell(E_m^{\mathsf{L}})
    c_\ell(Y_m^{\mathsf{L}}).
    \label{eq:memory-logical-block-Pauli-pairing}
\end{align}
Substituting Eq.~\eqref{eq:memory-logical-block-Pauli-pairing} termwise into Eq.~\eqref{eq:memory-logical-block-trace-pairing}, interchanging the finite sums, and using Eqs.~\eqref{eq:memory-logical-Pauli-block-coefficient-vector} and~\eqref{eq:memory-logical-Pauli-measurement-vector} give Eq.~\eqref{eq:memory-logical-Pauli-trace-pairing}.
\end{proof}

Fix \(\rho\in\operatorname{End}(\mathcal{H}_{\mathsf{D}})\) with \(\rho\succeq0\) and \(\trace(\rho)=1\), and fix \(E_y\in\operatorname{End}(\mathcal{H}_{\mathsf{D}})\) with \(0\preceq E_y\preceq\id_{\mathcal{H}_{\mathsf{D}}}\), where \(y\) is an outcome label.  
The operator \(\rho\) is intended to represent the prepared state, and \(E_y\) is intended to represent the POVM element associated with the outcome labeled by \(y\).
For every integer \(k\geq1\), define the probability of outcome \(y\) after \(k\) QEC cycles by
\begin{equation}
    p_y(k)
    \coloneqq 
    \trace\!\left[
        E_y\Phi^k(\rho)
    \right].
    \label{eq:memory-experiment-outcome-probability}
\end{equation}

\begin{thm}
\label{thm:memory-logical-Pauli-transfer-formula}
For every integer \(k\geq1\), the outcome probability \(p_y(k)\)
satisfies
\begin{equation}
    p_y(k)
    =
    d_{\mathsf{L}}
    \sum_{\ell\in V_{\mathsf{L}}}
    \boldsymbol e_\ell(E_y)^{\mathsf{T}}
    T_\ell^k
    \boldsymbol r_\ell(\rho).
    \label{eq:memory-experiment-logical-Pauli-transfer-formula}
\end{equation}
\end{thm}

\begin{proof}
Fix an integer \(k\geq1\), and set
\(Y\coloneqq \Phi^k(\rho)\).
Since \(\Phi\) is a channel, \(Y\) is a state.  By
Proposition~\ref{prop:complete-cycle-leakage--syndrome-dephasing},
\(Y\) is leakage--syndrome dephased.
Applying Eq.~\eqref{eq:memory-logical-Pauli-trace-pairing} with
\(E=E_y\) and using
Eq.~\eqref{eq:memory-experiment-outcome-probability} give
\begin{equation}
    p_y(k)
    =
    d_{\mathsf{L}}
    \sum_{\ell\in V_{\mathsf{L}}}
    \boldsymbol e_\ell(E_y)^{\mathsf{T}}
    \boldsymbol r_\ell(Y).
    \label{eq:memory-experiment-Pauli-vector-pairing}
\end{equation}
By Eq.~\eqref{eq:memory-repeated-cycle-logical-Pauli-dynamics},
\(\boldsymbol r_\ell(Y)=T_\ell^k\boldsymbol r_\ell(\rho)\) for every
\(\ell\in V_{\mathsf{L}}\).
Substituting these equalities termwise into
Eq.~\eqref{eq:memory-experiment-Pauli-vector-pairing} gives
Eq.~\eqref{eq:memory-experiment-logical-Pauli-transfer-formula}.
\end{proof}

\section{Elementary operator-space lemmas}
\label{app:elementary-operator-space-lemmas}

Let \(\mathcal{X}\) be a finite-dimensional Hilbert space.  Define
\(\mathcal{A}\) to be \(\operatorname{End}(\mathcal{X})\) equipped with
the Hilbert--Schmidt inner product
\begin{equation}
    \langle A,B\rangle_{\mathcal{A}}
    =
    \trace_{\mathcal{X}}(A^\dagger B),
    \label{eq:app-generic-HS-inner-product}
\end{equation}
for any \(A,B\in\mathcal{A}\). 
For \(\Phi\in\operatorname{End}(\mathcal{A})\), write \(\Phi^\dagger\) for the adjoint of \(\Phi\) with respect to the inner product on \(\mathcal{A}\).  
Define \(\mathcal{S}\) to be \(\operatorname{End}(\mathcal{A})\) equipped with the Hilbert--Schmidt inner product
\begin{equation}
    \langle\Phi,\Psi\rangle_{\mathcal{S}}
    =
    \trace_{\mathcal{A}}
    \!\left(\Phi^\dagger\circ\Psi\right),
    \label{eq:app-generic-superoperator-inner-product}
\end{equation}
for any \(\Phi,\Psi\in\mathcal{S}\). 
Next, for a unitary \(U\in\mathcal{A}\), define \( \operatorname{Ad}_U \in \mathcal{S}\) by
\begin{equation}
    \operatorname{Ad}_U(B)
    =
    UBU^\dagger \eqqcolon \mathcal{U}(B) ,
    \label{eq:app-adjoint-action}
\end{equation}
for any \( B \in \mathcal{A}\).

\begin{lem}[Unitary adjoint action]
\label{lem:app-unitary-adjoint-action}
Let \(U,V\in\mathcal{A}\) be unitary.  Then
\begin{align}
    \operatorname{Ad}_U\circ\operatorname{Ad}_V
    &=
    \operatorname{Ad}_{UV},
    \label{eq:app-adjoint-action-composition}
    \\
    \left(\operatorname{Ad}_U\right)^\dagger
    &=
    \operatorname{Ad}_{U^\dagger}
    =
    \left(\operatorname{Ad}_U\right)^{-1}.
    \label{eq:app-adjoint-action-adjoint}
\end{align}
Consequently, \(\operatorname{Ad}_U\) is a unitary element of
\(\mathcal{S}\).
\end{lem}

\begin{proof}
For \(B\in\mathcal{A}\), applying
Eq.~\eqref{eq:app-adjoint-action} first to
\(\operatorname{Ad}_V\) and then to \(\operatorname{Ad}_U\) gives
\begin{equation}
    \begin{aligned}
        \left(
            \operatorname{Ad}_U
            \circ
            \operatorname{Ad}_V
        \right)(B)
        &=
        U\left(VBV^\dagger\right)U^\dagger
        \\
        &=
        UVB(UV)^\dagger
        \\
        &=
        \operatorname{Ad}_{UV}(B).
    \end{aligned}
    \label{eq:app-adjoint-action composition-pointwise}
\end{equation}
Since \(B\in\mathcal{A}\) was arbitrary,
Eq.~\eqref{eq:app-adjoint-action composition-pointwise} proves
Eq.~\eqref{eq:app-adjoint-action-composition}.

For \(B,C\in\mathcal{A}\),
Eqs.~\eqref{eq:app-generic-HS-inner-product} and
\eqref{eq:app-adjoint-action}, together with cyclicity of the trace,
give
\begin{equation}
    \begin{aligned}
        \left\langle
            \operatorname{Ad}_U(B),
            C
        \right\rangle_{\mathcal{A}}
        &=
        \trace_{\mathcal{X}}
        \!\left(
            \left(UBU^\dagger\right)^\dagger C
        \right)
        \\
        &=
        \trace_{\mathcal{X}}
        \!\left(
            UB^\dagger U^\dagger C
        \right)
        \\
        &=
        \trace_{\mathcal{X}}
        \!\left(
            B^\dagger U^\dagger C U
        \right)
        \\
        &=
        \left\langle
            B,
            \operatorname{Ad}_{U^\dagger}(C)
        \right\rangle_{\mathcal{A}}.
    \end{aligned}
    \label{eq:app-adjoint-action-adjoint-calculation}
\end{equation}
By the defining property of the adjoint, Eq.~\eqref{eq:app-adjoint-action-adjoint-calculation} implies that
\[
    \left(\operatorname{Ad}_U\right)^\dagger
    =
    \operatorname{Ad}_{U^\dagger}.
\]
Applying Eq.~\eqref{eq:app-adjoint-action-composition}  with \((U,V)=(U^\dagger,U)\) and \((U,V)=(U,U^\dagger)\), together with Eq.~\eqref{eq:app-adjoint-action}, gives
\begin{equation}
    \operatorname{Ad}_{U^\dagger}
    \circ
    \operatorname{Ad}_U
    =
    \operatorname{Ad}_U
    \circ
    \operatorname{Ad}_{U^\dagger}
    =
    \operatorname{id}_{\mathcal{A}}.
    \label{eq:app-adjoint-action-inverse}
\end{equation}
Combining the relation \(\left(\operatorname{Ad}_U\right)^\dagger =\operatorname{Ad}_{U^\dagger}\) with Eq.~\eqref{eq:app-adjoint-action-inverse} proves Eq.~\eqref{eq:app-adjoint-action-adjoint} and shows that \(\operatorname{Ad}_U\) is a unitary element of \(\mathcal{S}\).
\end{proof}

Next, for all \(A,B,D\in\mathcal{A}\), define the multiplication map
\begin{equation}
    \mathcal{L}_{B,D}
    \in
    \mathcal{S},
    \qquad
    \mathcal{L}_{B,D}(A)=BAD .
    \label{eq:app-elementary-superoperator-definition}
\end{equation}
\begin{lem}[Trace of \(\mathcal{L}_{B,D}\)]
\label{lem:app-elementary-superoperator-trace}
For \(B,D\in\mathcal{A}\),
\begin{equation}
    \trace_{\mathcal{A}}
    \!\left(\mathcal{L}_{B,D}\right)
    =
    \trace_{\mathcal{X}}(B)
    \trace_{\mathcal{X}}(D).
    \label{eq:app-elementary-superoperator-trace}
\end{equation}
Consequently, for unitary \(U,U'\in\mathcal{A}\),
\begin{equation}
    \left\langle
        \operatorname{Ad}_U,
        \operatorname{Ad}_{U'}
    \right\rangle_{\mathcal{S}}
    =
    \left|
        \trace_{\mathcal{X}}(U^\dagger U')
    \right|^2.
    \label{eq:app-adjoint-action-inner-product}
\end{equation}
\end{lem}

\begin{proof}
Let \(n=\dim(\mathcal{X})\), let \(\{\ket{j}\}_{j=1}^{n}\) be an orthonormal basis for \(\mathcal{X}\), and set
\(E_{jk}=\ket{j}\!\bra{k}\).  
The family \(\{E_{jk}\}_{1\leq j,k\leq n}\) is an orthonormal basis for \(\mathcal{A}\), and
\[
    E_{jk}^\dagger B E_{jk}^{\,}
    = \MatEl{j}{B}{j} \ZProj{k}.
\]
The trace of an endomorphism of \(\mathcal{A}\) is the sum of its diagonal matrix elements in any orthonormal basis. 
Therefore, Eqs.~\eqref{eq:app-generic-HS-inner-product} and \eqref{eq:app-elementary-superoperator-definition} give
\begin{equation}
    \begin{aligned}
        \trace_{\mathcal{A}}
        \!\left(\mathcal{L}_{B,D}\right)
        &=
        \sum_{j,k=1}^{n}
        \left\langle
            E_{jk},
            \mathcal{L}_{B,D}(E_{jk})
        \right\rangle_{\mathcal{A}}
        \\
        &=
        \sum_{j,k=1}^{n}
        \trace_{\mathcal{X}}
        \!\left(
            E_{jk}^\dagger B E_{jk}D
        \right)
        \\
        &=
        \sum_{j,k=1}^{n}
        \bra{j}B\ket{j}
        \bra{k}D\ket{k}
        \\
        &=
        \trace_{\mathcal{X}}(B)
        \trace_{\mathcal{X}}(D).
    \end{aligned}
    \label{eq:app-elementary-superoperator-trace-calculation}
\end{equation}
Equation~\eqref{eq:app-elementary-superoperator-trace-calculation} proves Eq.~\eqref{eq:app-elementary-superoperator-trace}. 
For unitary \(U,U'\in\mathcal{A}\), Eqs.~\eqref{eq:app-adjoint-action} and \eqref{eq:app-elementary-superoperator-definition} give
\[
    \operatorname{Ad}_{U^\dagger U'}
    =
    \mathcal{L}_{U^\dagger U',\,U'^\dagger U} 
    \quad \text{and} \quad
    \trace_{\mathcal{X}}(U'^\dagger U)
    =
    \overline{
        \trace_{\mathcal{X}}(U^\dagger U')
    }.
\]
Equations~\eqref{eq:app-generic-superoperator-inner-product}, \eqref{eq:app-adjoint-action-composition}, \eqref{eq:app-adjoint-action-adjoint}, and \eqref{eq:app-elementary-superoperator-trace}, together with these equalities, give
\begin{equation}
    \begin{aligned}
        \left\langle
            \operatorname{Ad}_U,
            \operatorname{Ad}_{U'}
        \right\rangle_{\mathcal{S}}
        &=
        \trace_{\mathcal{A}}
        \!\left[
            \left(\operatorname{Ad}_U\right)^\dagger
            \circ
            \operatorname{Ad}_{U'}
        \right]
        \\
        &=
        \trace_{\mathcal{A}}
        \!\left(
            \operatorname{Ad}_{U^\dagger U'}
        \right)
        \\
        &=
        \trace_{\mathcal{A}}
        \!\left(
            \mathcal{L}_{U^\dagger U',\,U'^\dagger U}
        \right)
        \\
        &=
        \trace_{\mathcal{X}}(U^\dagger U')
        \trace_{\mathcal{X}}(U'^\dagger U)
        \\
        &=
        \left|
            \trace_{\mathcal{X}}(U^\dagger U')
        \right|^2.
    \end{aligned}
    \label{eq:app-adjoint-action-inner-product-calculation}
\end{equation}
Equation~\eqref{eq:app-adjoint-action-inner-product-calculation} proves Eq.~\eqref{eq:app-adjoint-action-inner-product}.
\end{proof}

\end{document}